\documentclass[fleqn,usenatbib]{mnras}

\usepackage{newtxtext,newtxmath}

\usepackage[T1]{fontenc}

\DeclareRobustCommand{\VAN}[3]{#2}
\let\VANthebibliography\thebibliography
\def\thebibliography{\DeclareRobustCommand{\VAN}[3]{##3}\VANthebibliography}

\usepackage{graphicx}	
\usepackage{adjustbox}  
\usepackage{amsmath}	
\usepackage{wasysym}    
\usepackage{bm}         
\usepackage{etoolbox}
\usepackage{float}
\usepackage{enumerate}
\makeatletter 
  \patchcmd{\NAT@citex}
    {\@citea\NAT@hyper@{%
      \NAT@nmfmt{\NAT@nm}%
      \hyper@natlinkbreak{\NAT@aysep\NAT@spacechar}{\@citeb\@extra@b@citeb}%
      \NAT@date}}
    {\@citea\NAT@nmfmt{\NAT@nm}%
    \NAT@aysep\NAT@spacechar\NAT@hyper@{\NAT@date}}{}{}

  \patchcmd{\NAT@citex}
    {\@citea\NAT@hyper@{%
      \NAT@nmfmt{\NAT@nm}%
      \hyper@natlinkbreak{\NAT@spacechar\NAT@@open\if*#1*\else#1\NAT@spacechar\fi}%
        {\@citeb\@extra@b@citeb}%
      \NAT@date}}
    {\@citea\NAT@nmfmt{\NAT@nm}%
    \NAT@spacechar\NAT@@open\if*#1*\else#1\NAT@spacechar\fi\NAT@hyper@{\NAT@date}}
    {}{}
\makeatother

\newtoggle{referee}            %
\togglefalse{referee}          %
\iftoggle{referee}{
    \usepackage{color,soul}    %
}{
}                              %

\newcommand\Msun{\text{M}_{\astrosun}} 
\newcommand\HI{\ion{H}{I}} 
\newcommand\HII{\ion{H}{II}} 
\newcommand\HeI{\ion{He}{I}} 
\newcommand\HeII{\ion{He}{II}} 
\newcommand\thesan{\mbox{\textsc{thesan}}}

\newcommand\orcid[1]{\href{http://orcid.org/#1}{\adjustbox{trim={-.15\width} {0\height} {-.15\width} {0\height},clip}{\includegraphics[height=12pt]{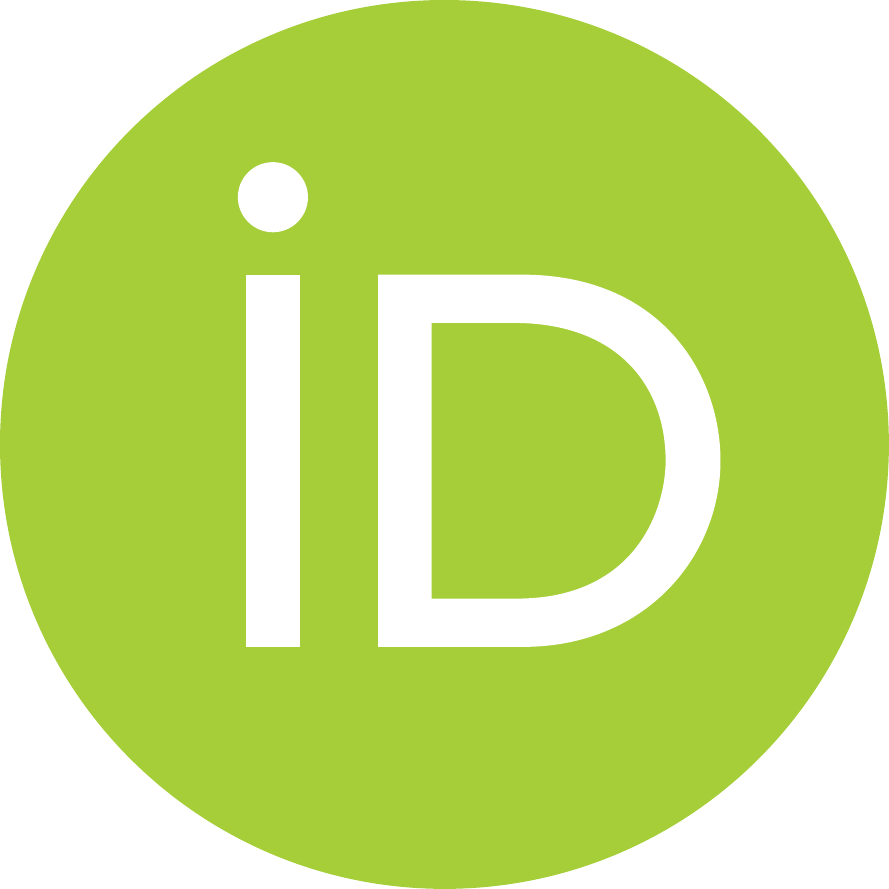}}}}

\title[Ly$\alpha$ transmission during the EoR]{The \thesan\ project: Lyman-$\bmath{\alpha}$ emission and transmission during the \mbox{Epoch of Reionization}}

\author[A.~Smith et al.]{%
A.~Smith,$^{1}$\thanks{E-mail: \href{mailto:arsmith@mit.edu}{arsmith@mit.edu}; NHFP Einstein Fellow.}
R.~Kannan,$^{2}$\thanks{E-mail: \href{mailto:rahul.kannan@cfa.harvard.edu}{rahul.kannan@cfa.harvard.edu}}
E.~Garaldi,$^{3}$\thanks{E-mail: \href{mailto:egaraldi@mpa-garching.mpg.de}{egaraldi@mpa-garching.mpg.de}}
M.~Vogelsberger,$^{1}$
R.~Pakmor,$^{3}$
V.~Springel$^{3}$
and L.~Hernquist$^{2}$
\\
$^{1}$Department of Physics, Massachusetts Institute of Technology, Cambridge, MA 02139, USA \\
$^{2}$Center for Astrophysics $\vert$ Harvard $\&$ Smithsonian, 60 Garden Street, Cambridge, MA 02138, USA \\
$^{3}$Max-Planck Institute for Astrophysics, Karl-Schwarzschild-Str.~1, D-85741 Garching, Germany 
}

\date{Accepted XXX. Received YYY; in original form ZZZ}

\pubyear{2022}

\begin{document}
\label{firstpage}
\pagerange{\pageref{firstpage}--\pageref{lastpage}}
\maketitle

\begin{abstract}
  The visibility of high-redshift Lyman-alpha emitting galaxies (LAEs) provides important constraints on galaxy formation processes and the Epoch of Reionization (EoR). However, predicting realistic and representative statistics for comparison with observations represents a significant challenge in the context of large-volume cosmological simulations. The \thesan\ project offers a unique framework for addressing such limitations by combining state-of-the-art galaxy formation (IllustrisTNG) and dust models with the \textsc{arepo-rt} radiation-magnetohydrodynamics solver. In this initial study we present Lyman-alpha centric analysis for the flagship simulation that resolves atomic cooling haloes throughout a $(95.5\,\text{cMpc})^3$ region of the Universe. To avoid numerical artefacts we devise a novel method for accurate frequency-dependent line radiative transfer in the presence of continuous Hubble flow, transferable to broader astrophysical applications as well. Our scalable approach highlights the utility of LAEs and red damping-wing transmission as probes of reionization, which reveal nontrivial trends across different galaxies, sightlines, and frequency bands that can be modelled in the framework of covering fractions. In fact, after accounting for environmental factors influencing large-scale ionized bubble formation such as redshift and UV magnitude, the variation across galaxies and sightlines mainly depends on random processes including peculiar velocities and self-shielded systems that strongly impact unfortunate rays more than others. Throughout the EoR local and cosmological optical depths are often greater than or less than unity such that the $\exp(-\tau)$ behavior leads to anisotropic and bimodal transmissivity. Future surveys will benefit by targeting both rare bright objects and Goldilocks zone LAEs to infer the presence of these (un)predictable (dis)advantages.
\end{abstract}




\section{Introduction}
\label{sec:intro}
The Epoch of Reionization (EoR) is the time period in the history of the Universe when the radiation from the first stars and galaxies initiated a cosmic phase transition throughout the intergalactic medium (IGM), which went from being cold and neutral to warm and ionized. In recent years we have witnessed significant progress in understanding the reionization process and advancing the current observational and computational frontiers. However, many of the central questions only have tentative answers that may be revised as the constraints from available data improve. For example, there is significant uncertainty about the role and contribution of low- and high-mass galaxies, the sources and escape of ionizing photons, and even the timing, duration, and morphology of reionization itself \citep{Barkana2001,LoebFurlanetto2013,Dayal2018,Wise2019}.

There is mounting evidence for the so-called late reionization in which most of the IGM is rapidly ionized around redshift $z \sim 7$--$8$ \citep{Naidu2020}. This understanding comes from the \textit{Planck} measurement of a low optical depth for electron scattering of CMB photons \citep{Planck2018}, the declining fraction of Lyman-alpha emitters (LAEs) among the galaxy population at $z \gtrsim 6$ \citep{Stark2011,Schenker2014,Mason2019}, the imprint of neutral hydrogen in the IGM as a damping wing absorption feature on the spectrum of high-redshift quasars \citep{Simcoe2012,Davies2018}, and the spatial fluctuations of the Ly$\alpha$ forest transmission \citep{Onorbe2017,Kulkarni2019}. We anticipate a more complete understanding of the EoR from upcoming facilities, including the \textit{James Webb Space Telescope} (\textit{JWST}) for the characterization of high-$z$ galaxies and the Low-Frequency Array (LOFAR), Hydrogen Epoch of Reionization Array (HERA), and Square Kilometer Array (SKA) for 21\,cm cosmology measurements to map out the distribution of neutral hydrogen in the Universe. On the theory side, radiation hydrodynamic (RHD) simulations have played an increasingly important role in capturing the non-linear, multiscale physics of reionization although there are still important computational challenges to overcome in the coming decades as well \citep{Ciardi2000,Iliev2006,Gnedin2014,Ocvirk2016,Pawlik2017,Rosdahl2018}.

Ultimately, the joint analysis of observational probes sensitive to unique aspects of galaxy formation and IGM properties will provide definitive answers to the main questions about the EoR. It is in this spirit that we pursue an initial study of Lyman-alpha (Ly$\alpha$) emission from atomic hydrogen gas during the EoR from the \thesan\ suite of large-volume cosmological reionization simulations (\citet{KannanThesan2022}, \citet{GaraldiThesan2022}, hereafter Papers~I and II). The \thesan\ project utilizes the adaptive moving mesh magneto-hydrodynamics code \textsc{arepo} \citep{Springel2010,Weinberger2020}, in combination with the state-of-the-art IllustrisTNG galaxy formation model \citep{Weinberger2017,Pillepich2018}, self-consistent radiation hydrodynamics \citep{Kannan2019}, and dust modelling \citep{McKinnon2017}. The flagship \thesan-1 run resolves atomic cooling haloes throughout a $(95.5\,\text{cMpc})^3$ region of the Universe, providing sufficient particle resolution and halo statistics to bring unique insights about galaxy and IGM properties. Of crucial importance for LAEs specifically, \thesan\ connects the production of Ly$\alpha$ photons to their subsequent transmission through the IGM. One of the main drawbacks is the subresolution treatment of the interstellar medium (ISM) as a two-phase gas where cold clumps are embedded in a smooth, hot phase produced by supernova explosions \citep{Springel2003}. However, such subgrid modelling comes with the territory of large-volume simulations with demonstrated agreement with observations down to $z = 0$ \citep{Vogelsberger2020}. Thus, we proceed with our current exploration emphasizing that the \thesan\ project will be followed up by high-resolution zoom-in resimulations of a wide range of galaxies from the flagship run for self-consistent Ly$\alpha$ radiative transfer modelling from ISM to IGM scales.

The phenomenological impact of the IGM on radiation in the vicinity of the Ly$\alpha$ line is well known \citep{GunnPeterson1965,Miralda-Escude1998,MadauRees2000}, as are the implications when leveraging LAEs as a probe of reionization \citep{MalhotraRhoads2004,McQuinn2007,Dijkstra2014,Kakiichi2016}. In essence, neutral hydrogen far from the source can remove Ly$\alpha$ photons with a single scattering out of the line of sight. For fortunate LAEs residing within ionized bubbles on the order of $\sim0.1$--$1$ physical Mpc (somewhat less stringent for peaks with large red velocity offsets), the light can redshift sufficiently far from resonance to avoid total suppression by the intervening IGM. However, numerical studies exhibit a complex landscape of sightline-to-sightline and galaxy-to-galaxy variations, which hints towards non-trivial dependence on the reionization history, environmental and proximity effects, and even galaxy properties such as the halo mass, specific star formation rate (SFR), and infall and peculiar velocities \citep{Laursen2011,Dayal2012,Jensen2014,Byrohl2020,Garel2021,Gronke2021,Park2021}. The growing number of studies working in the context of large-volume cosmological RHD simulations is indicative of the importance of providing higher accuracy predictions for current and upcoming LAE surveys extending into the EoR. Such detailed studies will be invaluable to interpreting the observational signatures of both high-$z$ galaxies (e.g. from the \textit{JWST}) and integrated diffuse emission (e.g. from the \textit{Spectro-Photometer for the history of the Universe, Epoch of Reionization and Ices Explorer: SPHEREx}).

In this study, we provide comprehensive galaxy Ly$\alpha$ emission and IGM transmission catalogues, which will be available with the public release of \thesan. The accessibility of such simulation-based surveys is timely given the current and forthcoming state of Ly$\alpha$ observation data. In fact, narrow-band surveys such as SILVERRUSH have already mapped over $2000$ LAEs at $z = 5.7$ and $6.6$, revealing target halo masses of $M_\text{halo} \sim 10^{11}\,\Msun$ with per cent level duty cycles \citep{Ouchi2018,Ouchi2020}. Likewise, surveys of $3 \lesssim z \lesssim 6$ LAEs observed with the Multi-Unit Spectroscopic Explorer (MUSE) reveal that Ly$\alpha$ haloes are ubiquitous and possibly associated with high-$z$ cosmic structure formation \citep{Leclercq2017,Gallego2018}. These windows correspond to the tail-end of reionization when most bubbles overlap but cosmic voids still harbour neutral islands capable of boosting the LAE clustering signal. Searches at $z \gtrsim 7$ are still impeded by low number statistics in constraining the reionization history at these epochs \citep{Jung2020}.

Our work is distinct from previous high-$z$ studies in several aspects. In particular, the \thesan\ project employs a realistic galaxy formation model that for example produces the correct stellar-to-halo mass relation over a wide range of halo masses, includes novel secondary physics such as dust processes and black hole radiation and feedback, and medium resolution physics variations within the suite. The simulations do not require post-processing ionizing radiative transfer as is done for the modern pioneering Ly$\alpha$ IGM transmission analysis by \citet{Laursen2011}. Furthermore, our volumes are large enough to avoid global cosmic variance that can bias transmission statistics. Recently, \citet{Garel2021} performed end-to-end Monte Carlo Ly$\alpha$ radiative transfer for thousands of galaxies within a $(10\,\text{cMpc})^3$ \textsc{sphinx} simulation, providing valuable insights about using LAEs to constrain the global reionization history. Still, the \thesan\ volumes are almost a thousand times larger, which is essential for capturing a representative number of bright LAEs for comparison with current and upcoming observations. Finally, other recent Ly$\alpha$ transmission studies with comparable volumes as ours are either focused on lower redshifts when the spatially uniform UV background approximation is valid \citep[e.g.][]{Behrens2018,Byrohl2020} or due to the poor spatial resolution within galaxies the astrophysical connections remain tenuous \citep[e.g.][]{Gronke2021,Park2021}. Self-consistent simulation-based LAE science represents a monumental challenge with encouraging progress from several fronts, including improving treatments of ISM-to-IGM scale radiative transfer modelling focused on the EoR \citep[e.g.][]{Behrens2019,Laursen2019,Smith2019,Garel2021}. Of course, our intuition and understanding of Ly$\alpha$ radiative transfer has been aided by analytical studies \citep[e.g.][]{Harrington1973,Neufeld1990,HansenOh2006,LaoSmith2020}, idealized setups \citep[e.g.][]{Dijkstra2006,Behrens2014,Gronke2017,Song2020,LiSteidel2021}, and isolated galaxy simulations \citep[e.g.][]{Verhamme2012,BehrensBraun2014,Smith2021}, which allow better resolution and control of the small-scale ISM physics.

The paper is organized as follows. In Section~\ref{sec:simulations}, we briefly describe the flagship \thesan\ simulation employed throughout this paper. In Section~\ref{sec:galaxies}, we introduce the Ly$\alpha$ emission catalogues that form the basis of statistical explorations of intrinsic luminosity based properties. In Section~\ref{sec:IGM}, we outline the procedure for calculating frequency-dependent transmission curves, including a novel integration scheme to account for continuous Hubble flow within the IGM. We also present our main results exploring the non-trivial dependence on frequency, redshift, and UV magnitude. Finally, in Section~\ref{sec:summary}, we provide a summary and brief perspective utilizing \thesan\ for Ly$\alpha$ science.

\begin{figure*}
  \centering
  \includegraphics[width=\textwidth]{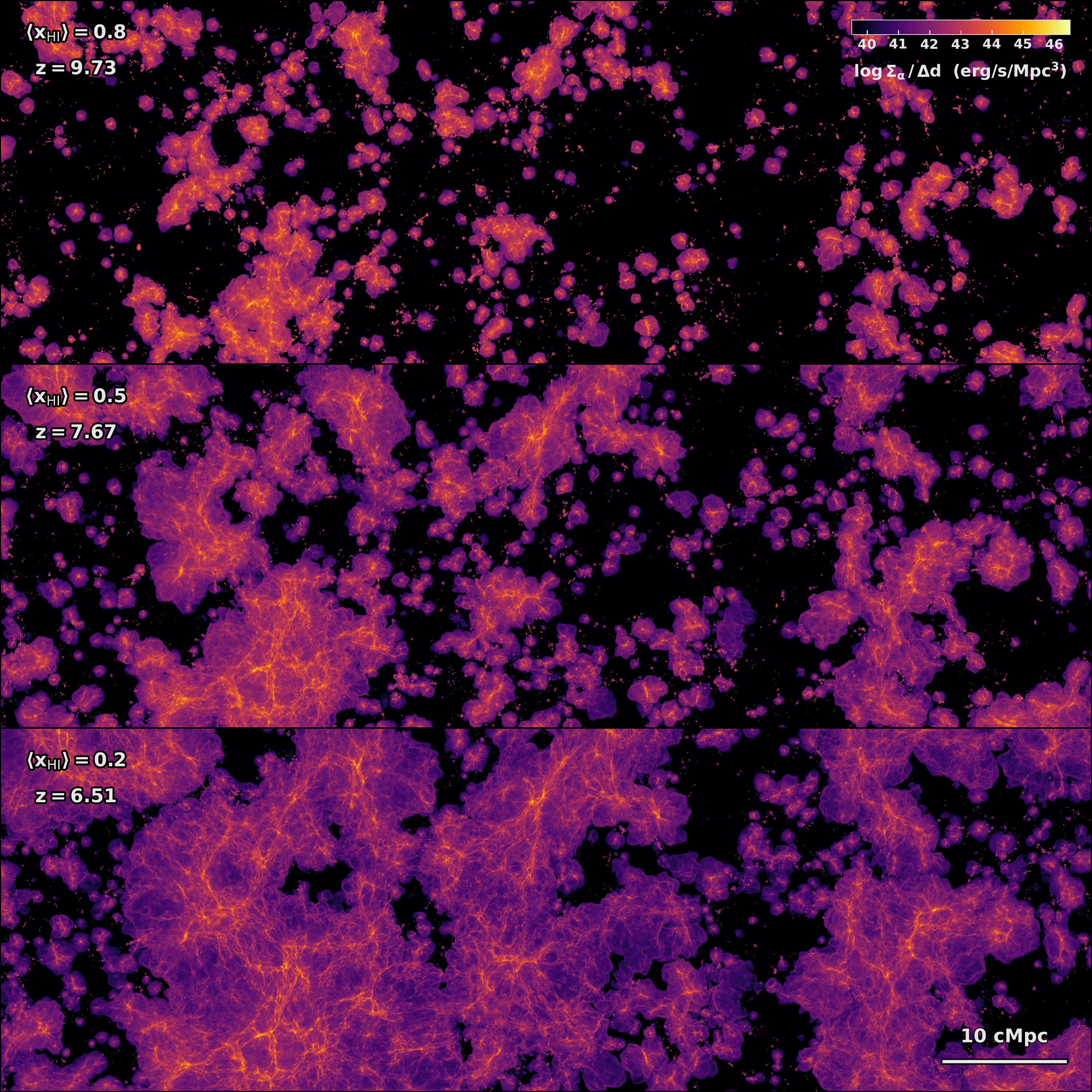}
  \caption{Intrinsic Ly$\alpha$ surface brightness covering the same $90 \times 30 \times 3\,\text{cMpc}^3$ subvolume for snapshots corresponding global neutral hydrogen fractions of $\langle x_\text{\HI} \rangle \approx \{0.8, 0.5, 0.2\}$ or redshifts of $z \approx \{9.73, 7.67, 6.51\}$ from top to bottom, respectively. The images are made with an adaptive quadrature ray-tracing scheme though the Voronoi tessellation to guarantee conservation of the luminosity as given by equations~(\ref{eq:L_rec})--(\ref{eq:L_stars}). Although radiative transfer effects on ISM to IGM scales are not included, the Ly$\alpha$ emissivity is clearly connected to the large-scale cosmic structure and topology of reionization.}
  \label{fig:image}
\end{figure*}

\begin{figure*}
  \centering
  \includegraphics[width=\textwidth]{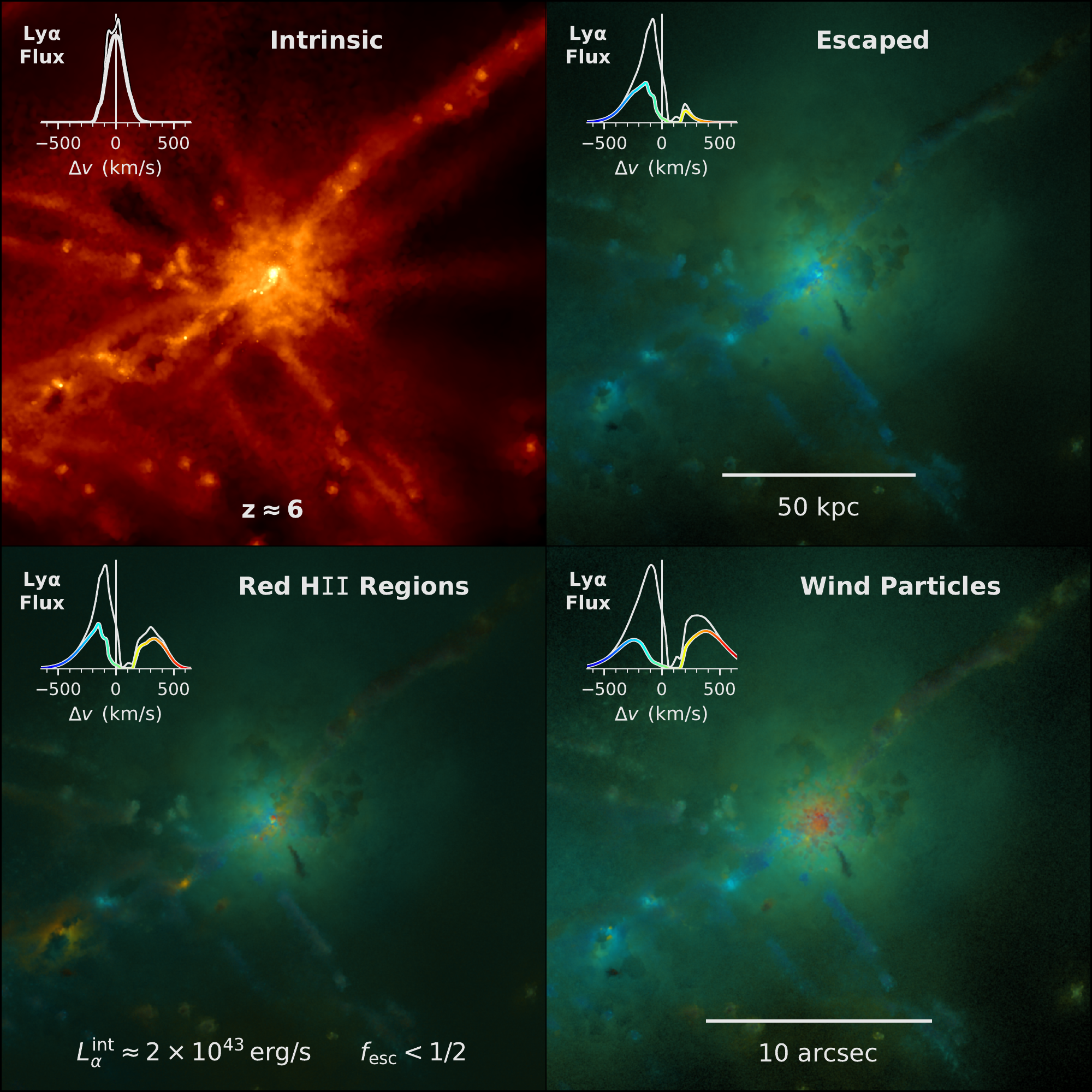}
  \caption{\textit{Upper left-hand panel:} Intrinsic Ly$\alpha$ surface brightness for a galaxy of mass $M_\text{halo} \approx 10^{11}\,\Msun$ at $z = 6$. \textit{Upper right-hand panel:} False colour rendering of the escaped Ly$\alpha$ emission based on synthetic integral field unit (IFU) data generated with the \textsc{colt} Monte Carlo radiative transfer code. The spectroscopic information is blended from blue--green to yellow--red with the image opacity encoding the surface brightness. The rest-frame is defined by the frequency centroid of the intrinsic line profile shown in the left-hand panel, while the emergent spectra in the remaining panels serve as velocity offset colour maps. The bold coloured line profiles are taken from a smaller aperture ($1''$~radius) while the white spectra include (mostly blue peak) photons from a wider area ($\approx 20''$~radius). \textit{Lower left-hand panel:} An alternative emission model that artificially reddens the initial frequency of unresolved \HII\ regions to explore the impact of local feedback-induced outflows on ISM scales. \textit{Lower right-hand panel:} An alternative transport model that incorporates wind particles to explore the role of galactic winds in shaping line profiles on CGM scales.}
  \label{fig:rgb_int}
\end{figure*}

\section{THESAN simulations}
\label{sec:simulations}
In this section we briefly summarize the main features of the \thesan\ simulations, which are introduced in detail in Paper~I \citet{KannanThesan2022}. \thesan\ is a suite of radiation-magnetohydrodynamical simulations run with the moving-mesh hydrodynamics code \textsc{arepo} \citep{Springel2010,Weinberger2020}\footnote{Public code access and documentation available at \href{https://arepo-code.org}{\texttt{www.arepo-code.org}}.}. The code employs a finite-volume method to solve the Euler equations on an unstructured Voronoi tessellation for an accurate treatment of quasi-Lagrangian flows with complex source terms over large dynamic ranges. Gravity calculations utilize a hybrid Tree-PM approach, which splits the force into short- (direct summation) and long-range (particle mesh) contributions computed through an adaptive oct-tree data structure \citep{Barnes1986}. In addition, the \textsc{arepo} implementation includes hierarchical time integration of the resolution elements and randomization of the node centres at each domain decomposition as described in the \textsc{gadget4} paper \citep{Springel2021}.

For self-consistent radiative transfer, we employ the \textsc{arepo-rt} extension described by \citet{Kannan2019}, which solves the first two moments of the RT equation assuming the M1 closure relation \citep{Levermore1984}. This scheme reaches second-order accuracy by replacing the piecewise constant approximation of the \citet{Godunov1959} scheme with a slope-limited piecewise linear spatial extrapolation utilizing a local least-squares fit for gradient estimates \citep{Pakmor2016}. A first-order time extrapolation based on half time-steps is employed to obtain the primitive variables on both sides of the interface \citep{vanLeer1979}. For computational efficiency, we choose to only model the ionizing part of the radiation spectrum and discretize photons into energy bins defined by the following thresholds: $[13.6, 24.6, 54.4, \infty)$\,eV. Each resolution element tracks the comoving photon number density and flux for each bin. To partially compensate the loss of resolution in the frequency sampling, we assume the radiation within each bin follows the spectrum of a $2$\,Myr old, quarter-solar metallicity stellar population with amplitude given by the local photon number density. This choice is physically motivated as young stars contribute most of the ionizing budget. Still, the effective photon properties are relatively insensitive to age and metallicity. The stellar spectra employ the Binary Population and Spectral Synthesis models \citep[BPASS version 2.2.1;][]{BPASS2017}, assuming a Chabrier IMF \citep{Chabrier2003}. The bin average values of the \HI, \HeI, and \HeII\ photoionization cross-sections ($\sigma$), energy injected into the gas per interacting photon ($\mathcal{E}$), and mean energy per photon ($e$) are reported in Table~1 of Paper~I. The RT equations are coupled to a non-equilibrium solver that accurately computes the ionization state of hydrogen and helium, as well as the temperature change due to photoheating, atomic, metal and Compton cooling. Finally, we employ a reduced speed of light (RSLA) approximation with an effective value of $\tilde{c} = 0.2\,c$ \citep[e.g.][]{Gnedin2016}. We emphasize that although \citet{Ocvirk2019} argue for higher values, in Appendix~A of Paper~I we demonstrate that in the context of our model this value is large enough to accurately capture the propagation of ionization fronts and post-reionization gas properties.

The \thesan\ simulations were designed to simultaneously capture the assembly of primeval galaxies and their impact on reionization as realistically as possible. For this reason, we employ the state-of-the-art IllustrisTNG galaxy formation model, which updates the previous Illustris model \citep{Genel2014,Vogelsberger2014,Vogelsberger2014b} to include subresolution physics tuned to reproduce a wide range of galaxy properties that are consistent with available observations down to low redshift \citep{Marinacci2018,Naiman2018,Nelson2018,Pillepich2018b,Springel2018}. This choice ensures that, although the \thesan\ simulations are only evolved to $z\simeq 5.5$, the physical model can be trusted throughout the history of the Universe \citep[e.g. for especially relevant high-$z$ galaxy predictions see][]{Shen2020,Shen2022,Vogelsberger2020b}. Additionally, the only new free parameter is the stellar escape fraction $f_\text{esc}^\text{ion}$\footnote{Employing a constant value for the ionizing escape fraction at the level of the birth cloud is intended to approximately capture subresolution internal neutral hydrogen and dust self-absorption. In reality, the local escape fraction transitions from zero early on to order unity after feedback disperses the high-density star-forming gas \citep[e.g.][]{Kimm2019}. The variation in time- and rate-averaged values is still an open question, and likely depends on complex environmental properties requiring improved resolution and ISM modelling.}, which we calibrate to match constraints for the global reionization history\footnote{For reference, the global reionization history is included in Fig.~\ref{fig:T_IGM_z}.} (0.37 for the flagship simulation). Furthermore, we also include a state-of-the-art dust model developed by \citet{McKinnon2017}.

The full \thesan\ simulation set and parameters  are catalogued in Table~2 of Paper~I. All runs follow the evolution of a cubic patch of the universe with linear comoving size $L_\text{box} = 95.5\,\text{cMpc}$. The initial conditions employ a method in which the initial Fourier mode amplitudes are fixed to the ensemble average power spectrum to suppress variance \citep{AnguloPontzen2016}. Throughout this paper, we employ a \citet{Planck2015_cosmo} cosmology with simulation constants of $h = 0.6774$, $\Omega_0 = 0.3089$, and $\Omega_\text{b} = 0.0486$, where all the symbols have the usual meaning.

The flagship \mbox{\textsc{thesan-1}} simulation is designed to resolve atomic cooling haloes with virial temperatures of $T_\text{vir} \gtrsim 10^4$\,K and masses of $M_\text{halo} \gtrsim 10^8\,h^{-1}\Msun$ \citep[see e.g.][]{BrommYoshida2011}. Thus, the total number of dark matter and (initial) gas particles is $\text{N}_\text{particles} = 2100^3$ each with mass resolutions of $m_\text{DM} = 3.1 \times 10^6\,\Msun$ and $m_\text{gas} = 5.8 \times 10^5\,\Msun$, respectively. The gravitational softening length for the star and dark matter particles is set to $2.2\,\text{ckpc}$, while the gas cells employ adaptive softening according to the cell radius. Gas cells are (de-)refined to ensure masses remain within a factor of two from the target mass, thus the minimum cell radius at $z = 5.5$ is $\sim 10$ pc for a dynamic resolution range of six orders of magnitude.

To quantify the resolution in the IGM we define the effective radius of each cell to be $r_\text{cell} \equiv (3 V_\text{cell} / 4 \pi)^{1/3}$, where $V_\text{cell}$ denotes the cell volume. We calculate the average cell radius as $\langle r_\text{cell} \rangle_V \equiv \sum r_\text{cell} V_\text{cell} / \sum V_\text{cell} \approx 5.28\,(2.82)\,\text{kpc}$ at $z = 5.5\,(10)$, noting that due to the Lagrangian nature of the code the spatial resolution is significantly better for higher density gas near galaxies and IGM structures where most absorption occurs. In particular, \citet{RahmatiSchaye2018} showed that the main \HI\ sinks of ionizing radiation, namely Lyman-limit systems with expected sizes of 1--10\,kpc, are well resolved in their reference runs that have 3.5 times coarser baryonic mass resolution than the \mbox{\textsc{thesan-1}} simulation used in this study (see also Paper~II). Thus, \thesan\ provides a state-of-the-art framework for connecting resolved galaxy and IGM properties throughout the EoR, ideal for this study of Ly$\alpha$ transmission and other topics relevant to the high-redshift Universe. However, there is also the question of achieving adequate spatial resolution throughout the extended CGM of galaxies. Although it is beyond the current state-of-the-art to uniformly require $\lesssim 1\,\text{kpc}$ resolution for such large-volume simulations, it may be worth making steps in this direction through various optimization trade offs to study the impact of CGM resolution on reionization, especially as this has been found to increase the covering fraction of Lyman-limit systems around isolated Milky Way-mass galaxies \citep{vandeVoort2019}. In this paper, we focus exclusively on the main \mbox{\textsc{thesan-1}} simulation, deferring comparisons with the other simulations to a future study.

\section{Galaxy emission catalogues}
\label{sec:galaxies}
For all snapshots we produce Ly$\alpha$ catalogues directly mirroring the friends-of-friends (FoF) halo catalogues and \textsc{subfind} subhalo catalogues. These post-processing files provide supplemental data for each group and subhalo, corresponding to identifications of dark matter haloes and galaxies, respectively. There is a single \verb"Lya_*" HDF5 file for each snapshot containing the following groups: Header, Group, Subhalo, Inner, Outer, and Total. For convenience, the header contains information about the simulation and the others contain global sums of various Ly$\alpha$ luminosities over all groups, subhaloes, inner/outer ``fuzz'' of unbound particles, and the entire simulation box. More importantly, we provide local sums for each individual group or subhalo, summarized in Table~\ref{tab:catalog} and explained below.

\begin{table}
  \centering
  \caption{Brief description of fields in the Ly$\alpha$ group and subhalo catalogues.}
  \label{tab:catalog}
  \begin{tabular}{ccc}
  \hline
  Field 
  & Units & Description \\
  \hline
  $L_\alpha$ & $\text{erg\,s}^{-1}$ & Total Ly$\alpha$ luminosity ($L_\alpha = L_\alpha^\text{rec} + L_\alpha^\text{col} + L_\alpha^\text{stars}$) \\
  $L_\alpha^\text{rec}$ & $\text{erg\,s}^{-1}$ & Ly$\alpha$ luminosity from resolved recombination \\
  $L_\alpha^\text{col}$ & $\text{erg\,s}^{-1}$ & Ly$\alpha$ luminosity from collisional excitation \\
  $L_\alpha^\text{stars}$ & $\text{erg\,s}^{-1}$ & Ly$\alpha$ luminosity from unresolved \HII\ regions \\
  $L_{\lambda,1216}$ & \!\!\!$\text{erg\,s}^{-1}\text{\AA}^{-1}$\!\!\!\!\! & Stellar continuum spectral luminosity at 1216\,\AA \\
  $L_{\lambda,1500}$ & \!\!\!$\text{erg\,s}^{-1}\text{\AA}^{-1}$\!\!\!\!\! & Stellar continuum spectral luminosity at 1500\,\AA \\
  $L_{\lambda,2500}$ & \!\!\!$\text{erg\,s}^{-1}\text{\AA}^{-1}$\!\!\!\!\! & Stellar continuum spectral luminosity at 2500\,\AA \\
  $L_\text{ion}^\text{AGN}$ & $\text{erg\,s}^{-1}$ & Ionizing luminosity from active galactic nuclei \\
  $\bmath{r}_\alpha$ & kpc & Centre of Ly$\alpha$ luminosity position in the box \\
  $\bmath{v}_\alpha$ & $\text{km\,s}^{-1}$ & Centre of Ly$\alpha$ luminosity peculiar velocity \\
  $\sigma_\alpha$ & $\text{km\,s}^{-1}$ & Centre of Ly$\alpha$ luminosity 1D velocity dispersion \\
  \hline
  \end{tabular}
\end{table}

\subsection{Ly\texorpdfstring{$\balpha$}{α} production}
The total Ly$\alpha$ luminosity $L_\alpha$ gives the intrinsic emission from recombinations, collisional excitation, and local stars. We calculate the resolved luminosity due to radiative recombination as
\begin{equation} \label{eq:L_rec}
  L_\alpha^\text{rec} = h \nu_\alpha \int P_\text{B}(T) \alpha_\text{B}(T)\,n_e n_p\,\text{d}V \, ,
\end{equation}
where $h \nu_\alpha = 10.2$\,eV, the Ly$\alpha$ conversion probability per recombination event is $P_\text{B} \approx 0.68$ \citep{Cantalupo2008}, the case B recombination coefficient is $\alpha_\text{B}$ \citep{HuiGnedin1997}, and the number densities $n_e$ and $n_p$ are for free electrons and protons, respectively \citep{Dijkstra2019}. In addition, we calculate the resolved contribution of radiative de-excitation of collisional excitation of neutral hydrogen by free electrons as
\begin{equation} \label{eq:L_col}
  L_\alpha^\text{col} = h \nu_\alpha \int q_{1s2p}(T)\,n_e n_\text{\HI}\,\text{d}V \, ,
\end{equation}
where the temperature-dependent rate coefficient $q_{1s2p}$ is taken from \citet{Scholz1991}. Due to the uncertainties surrounding the effective equation of state (EoS) for cold gas above the density threshold $n_\text{H} \approx 0.13\,\text{cm}^{-3}$, we isolate recombinations and collisional excitation emission from non-EoS cells as identified by the SFR being identically zero, noting that these components each contribute at the $\sim10\%$ level. The non-equilibrium radiation hydrodynamics solver provides accurate thermal and ionization states for such gas. However, we note that the ordering of the black hole energy injection routine leads to hot gas above the EoS that is artificially neutral for half a time-step. Therefore, for each cell we compare the neutral hydrogen fraction state from the simulation output to the maximum expected value assuming collisional ionization equilibrium (CIE), updating the ionization states if $x_\text{\HI} > x_\text{\HI,CIE}$. We note that we employ iteration so the rate coefficients reflect the correct temperatures. We find this pre-conditioning step robustly eliminates unphysical collisional excitation emission that will easily be avoided in future \thesan\ simulations by adjusting the ordering of the cooling physics. Of course, there will be additional Ly$\alpha$ cooling emission in the unresolved EoS cells, therefore our collisional excitation luminosities are conservative values. Furthermore, due to the RHD coupling of the thermochemistry a fraction of the ionizing budget is lost to the EoS cells. To account for this we also track the EoS recombination emission assuming these cells become ionized by local sources during radiation subcycles. We show later that this approach provides good agreement with the expectation for the global production of Ly$\alpha$ photons assuming an escape fraction of zero.

We account for unresolved \HII\ regions in the vicinity of stellar populations by converting self-absorbed ionizing photons to sources of Ly$\alpha$ emission at the subgrid level as
\begin{equation} \label{eq:L_stars}
  L_\alpha^\text{stars} = 0.68 h \nu_\alpha (1 - f_\text{esc}^\text{ion}) \dot{N}_\text{ion} \, .
\end{equation}
Here, the factor $0.68$ is the fiducial conversion probability assuming a temperature for emitting gas of $10^4\,\text{K}$, $f_\text{esc}^\text{ion}$ denotes the escape fraction of ionizing photons, and $\dot{N}_\text{ion}$ is the emission rate of ionizing photons from stars. The latter is a complex function of age and metallicity taken from the BPASS models (v2.2.1), which include binary stellar poplulations accounting for mass transfer, common envelope phases, binary mergers, and quasi-homogeneous evolution at low metallicities \citep{BPASS2017}. Thus, our Ly$\alpha$ emission catalogues correspond to the same model as the intrinsic sources of reionization within the simulations. Overall, the relative contribution of resolved and unresolved sources provides some intuition about the uncertainties in our Ly$\alpha$ emission modelling.

To better understand the intrinsic production of Ly$\alpha$ photons in the \thesan\ simulations in Fig.~\ref{fig:image} we show the Ly$\alpha$ surface brightness for a $90 \times 30 \times 3\,\text{cMpc}^3$ sub-volume for snapshots corresponding global neutral hydrogen fractions of $\langle x_\text{\HI} \rangle \approx \{0.8, 0.5, 0.2\}$ or redshifts of $z \approx \{9.73, 7.67, 6.51\}$, respectively. We employ an adaptive quadrature ray-tracing scheme though the Voronoi tessellation to ensure conservation of the luminosity as given by equations~(\ref{eq:L_rec}--\ref{eq:L_stars}). Specifically, convergence is achieved via the iterative refinement algorithm described in Appendix~A of \citet{Yang2020}. Although radiative transfer effects on ISM to IGM scales are not included, the Ly$\alpha$ emissivity is clearly connected to the large-scale structure (LSS) and topology of reionization. This visualization also emphasizes the strong impact of cosmic variance on Ly$\alpha$ studies based on small ($\lesssim 30\,\text{cMpc}$) reionization simulations as they may lead to biased statistics for certain quantities such as clustering and IGM transmission \citep[for a discussion of implications for observing the post-reionization cosmic web in Ly$\alpha$ emission see][]{Witstok2021}. The \thesan\ resolution and box size are sufficient to provide representative galaxy populations for Ly$\alpha$ emission and correctly capture EoR fluctuations \citep{Gnedin2014b,Iliev2014}.

In Fig.~\ref{fig:rgb_int} we demonstrate that \thesan\ is fully capable of capturing IGM scale effects, but detailed LAE modelling is strongly affected by ISM scale sourcing and radiation transport through the circumgalactic medium (CGM). In the left-hand panel, we illustrate the intrinsic Ly$\alpha$ emissivity for a galaxy of mass $M_\text{halo} \approx 10^{11}\,\Msun$ at $z = 6$. In the remaining panels we exhibit false colour renderings of the escaped Ly$\alpha$ emission based on synthetic integral field unit (IFU) data generated with the Cosmic Ly$\alpha$ Transfer code \citep[\textsc{colt};][which describe the physics and code implementation in detail]{Smith2015,Smith2019,Smith2021}\footnote{For public code access and documentation see \href{https://colt.readthedocs.io}{\texttt{colt.readthedocs.io}}.}. The Monte Carlo radiative transfer calculations employ $10^8$ photon packets sourced according to equations~(\ref{eq:L_rec})--(\ref{eq:L_stars}) with photons drawn within cell volumes for resolved recombinations ($32\%$) and collisional excitation ($8\%$) or from star particle locations for unresolved \HII\ regions ($60\%$). Ray integrations are performed using the native Voronoi cells with the neutral hydrogen and dust densities, gas temperatures, and bulk velocities taken directly from the simulation. We assume a dust opacity of $\kappa_\text{d} = 5.8 \times 10^4\,\text{cm}^2\,\text{g}^{-1}$ of dust, scattering albedo of $A = 0.325$, and asymmetry parameter of $\langle \cos\theta \rangle = 0.676$, based on the Milky Way dust model from \citet{Weingartner2001}. All images and spectra are for the same line-of-sight, although we have checked that all positive and negative coordinate direction observations result in similar qualitative conclusions. The spectroscopic data are blended in colour space with the image opacity encoding the surface brightness. The rest-frame is defined with respect to the frequency centroid of the intrinsic H$\alpha$ line profile inset in the first panel. The emergent flux is dominated by a blue peak shaped by the cosmological environment, which reflects the insufficient resolution and inconsistencies with the density, temperature, velocity, and ionization state regulated by the effective EoS model, but also the wide integration aperture ($\approx 20\arcsec$~radius). To isolate this last effect, we show that in comparison line profiles taken from a smaller aperture ($1''$~radius) have reduced blue peaks.

In the lower panels, we explore two alternative scenarios designed to produce spectra with enhanced red peaks in better agreement with observations. First, we artificially redden the initial frequency of unresolved \HII\ regions (i.e. photons from star particles are injected as a red peak) to model local feedback-induced outflows on subgrid ISM scales. We note that the resolved recombination and collisional excitation contributions ($\sim$ half) remain unchanged (including their frequency initialization). Lastly, we incorporate hydrodynamically decoupled wind particles (i.e. as additional gas cells) to emulate a clumpy outflow on resolved scales tracing galactic winds out to the CGM. Multiphase winds help regulate star formation and transport metals out of galaxies, but we reduce the speeds by a factor of four to better track the cold neutral components. Encouragingly, in each case the red peak is enhanced reflecting the physically motivated improvements, but further adjustments and calibrations may be required for comparison to LAE surveys. These may include dust rescaling, subgrid clumping, bulk velocities informed by winds, or modifications to EoS cell properties, each affecting the robustness of predicted spectra \citep[see also][]{Li2020,Byrohl2021,Gronke2021}. We are pursuing more accurate Ly$\alpha$ radiative transfer studies from zoom-in simulations of \thesan\ galaxies, which will set anchor points on these uncertainties. In fact, our preliminary results reveal that red peak dominated spectra naturally arise in the context of realistic multiphase ISM and CGM environments shaped by feedback in these simulations. For now, the Ly$\alpha$ line profile results in Fig.~\ref{fig:rgb_int} provide important insights and warnings to the community regarding galaxy scale Ly$\alpha$ radiative transfer calculations from large-volume reionization simulations. Specifically, it remains difficult to produce red peak dominated spectra and ultimately the solution must rely on improved galaxy formation models rather than empirical corrections. As IGM transmission can be highly sensitive to the emergent spectra, it is non-trivial to robustly connect back to the intrinsic Ly$\alpha$ emission. Therefore, the remainder of this paper focuses on emission and transmission, allowing dedicated follow-up explorations to more fully address radiative transfer uncertainties. We return to this example after presenting our main IGM transmission findings in Section~\ref{sec:covering_fractions}.

\begin{figure}
  \centering
  \includegraphics[width=\columnwidth]{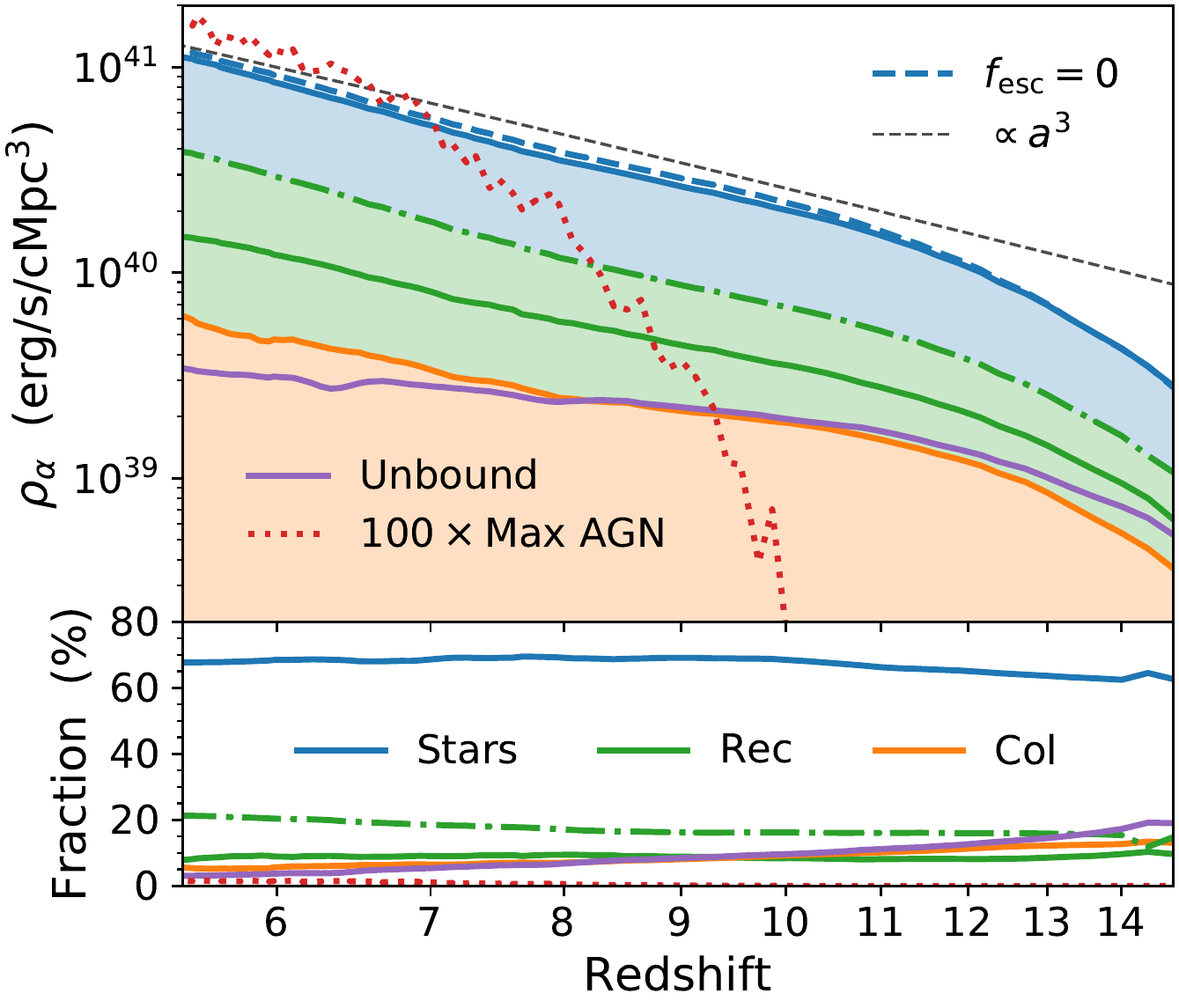}
  \caption{\textit{Top:} Redshift evolution of the global Ly$\alpha$ intrinsic luminosity density including cumulative contributions from resolved collisional excitation (orange) and recombination (green; the dash-dotted curve includes EoS cells) emission and unresolved \HII\ regions (blue). The total emission is well described by a power law of $\rho_\alpha = 8.3 \times 10^{40}\,[(z+1)/7]^{-3}\,\text{erg\,s}^{-1}\,\text{cMpc}^{-3}$ over the range $z \in (5.5, 11)$, offset for visualization purposes. For comparison we show that AGN (red dotted) contribute at the per cent level if ionizing photons are efficiently converted to Ly$\alpha$ photons according to $\rho_\alpha^\text{AGN} \approx 0.68 h\nu_\alpha \dot{N}_\text{ion}^\text{AGN}$. \textit{Bottom:} The fraction of luminosity from each channel illustrating that local stars dominate the emission budget. For clarity, we include separate contributions from recombinations, collisional excitation, and sources not bound to any subhalo (purple).}
  \label{fig:rho_alpha_time}
\end{figure}

\begin{figure}
  \centering
  \includegraphics[width=\columnwidth]{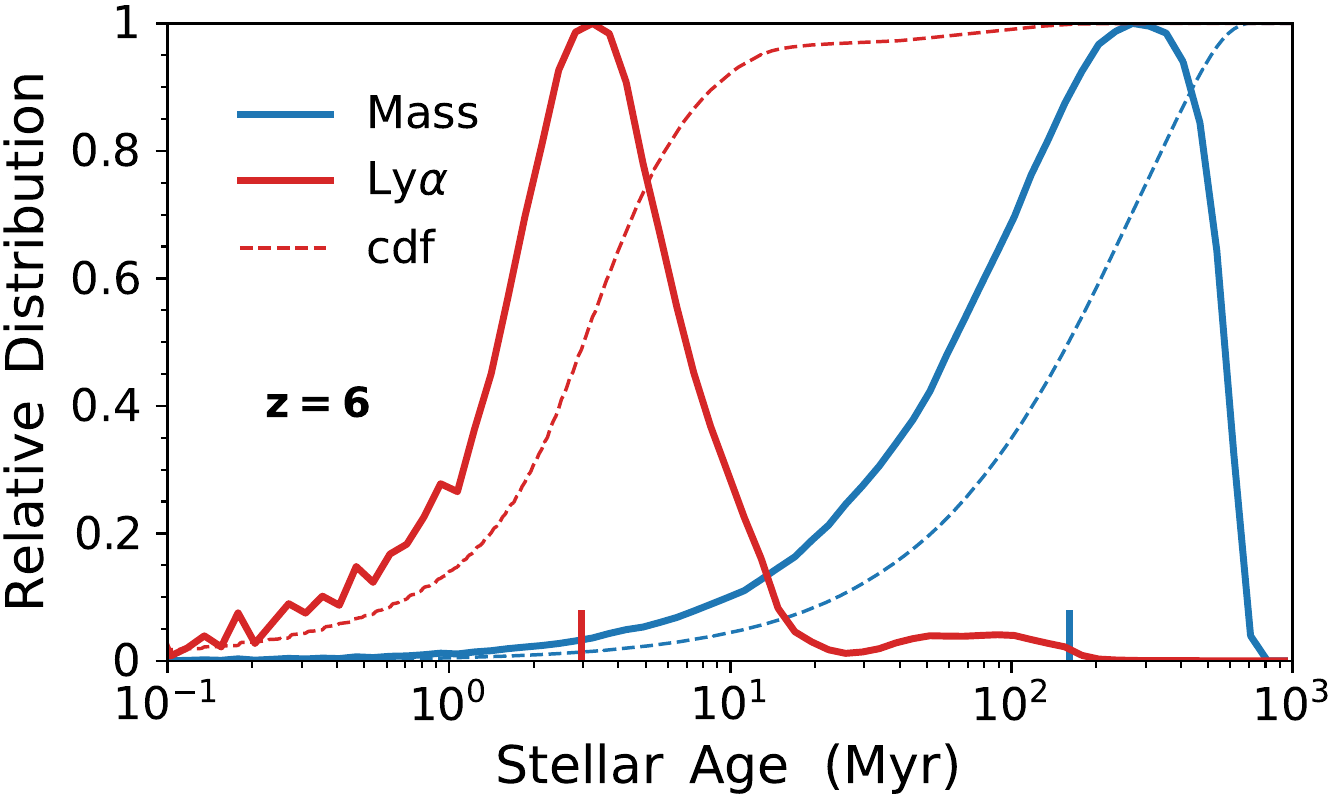}
  \caption{Relative distribution of ages contributing to the intrinsic ionizing radiation that is locally reprocessed into Ly$\alpha$ photons, with cumulative distribution functions shown as dashed curves and vertical markers denoting medians. The emission traces the youngest stellar populations with a median age of roughly $3$\,Myr, while the mass-weighted median age is $160$\,Myr at $z = 6$. Thus, the star formation history acts as the primary driver of ionizing sources rather than the stellar mass.}
  \label{fig:age_hist}
\end{figure}

In Fig.~\ref{fig:rho_alpha_time}, we show the redshift evolution of the global Ly$\alpha$ intrinsic luminosity density including contributions from unresolved \HII\ regions and resolved collisional excitation and recombination emission. The total emission is well described by a power law of $\rho_\alpha = 8.3 \times 10^{40}\,[(z+1)/7]^{-3}\,\text{erg\,s}^{-1}\,\text{cMpc}^{-3}$ over the range $z \in (5.5, 11)$. Local stars (blue) dominate the emission budget followed by recombination (green) and cooling radiation (orange), thus radiative transfer effects are expected to result in drastic reprocessing of the observed emission beyond what we are able to capture here. Interestingly, the fraction from sources not bound to any subhalo (purple), i.e. the inner and outer ``fuzz'', declines with time to a few per cent by $z \lesssim 6$. Likewise, AGN only contribute at the per cent level even if ionizing photons are efficiently converted to Ly$\alpha$ photons according to $\rho_\alpha^\text{AGN} \approx 0.68 h\nu_\alpha \dot{N}_\text{ion}^\text{AGN}$ (red dotted; see Section~\ref{sec:additional_Lya_fields}), which may be boosted by a factor of a few due to harder ionizing spectra inducing multiple ionizations \citep{Raiter2010}. In fact, the total emission budget agrees with the expectation of a resolution agnostic model ($f_\text{esc} = 0$) assuming every ionizing photon eventually produces Ly$\alpha$ emission (blue dashed).

\begin{figure}
  \centering
  \includegraphics[width=\columnwidth]{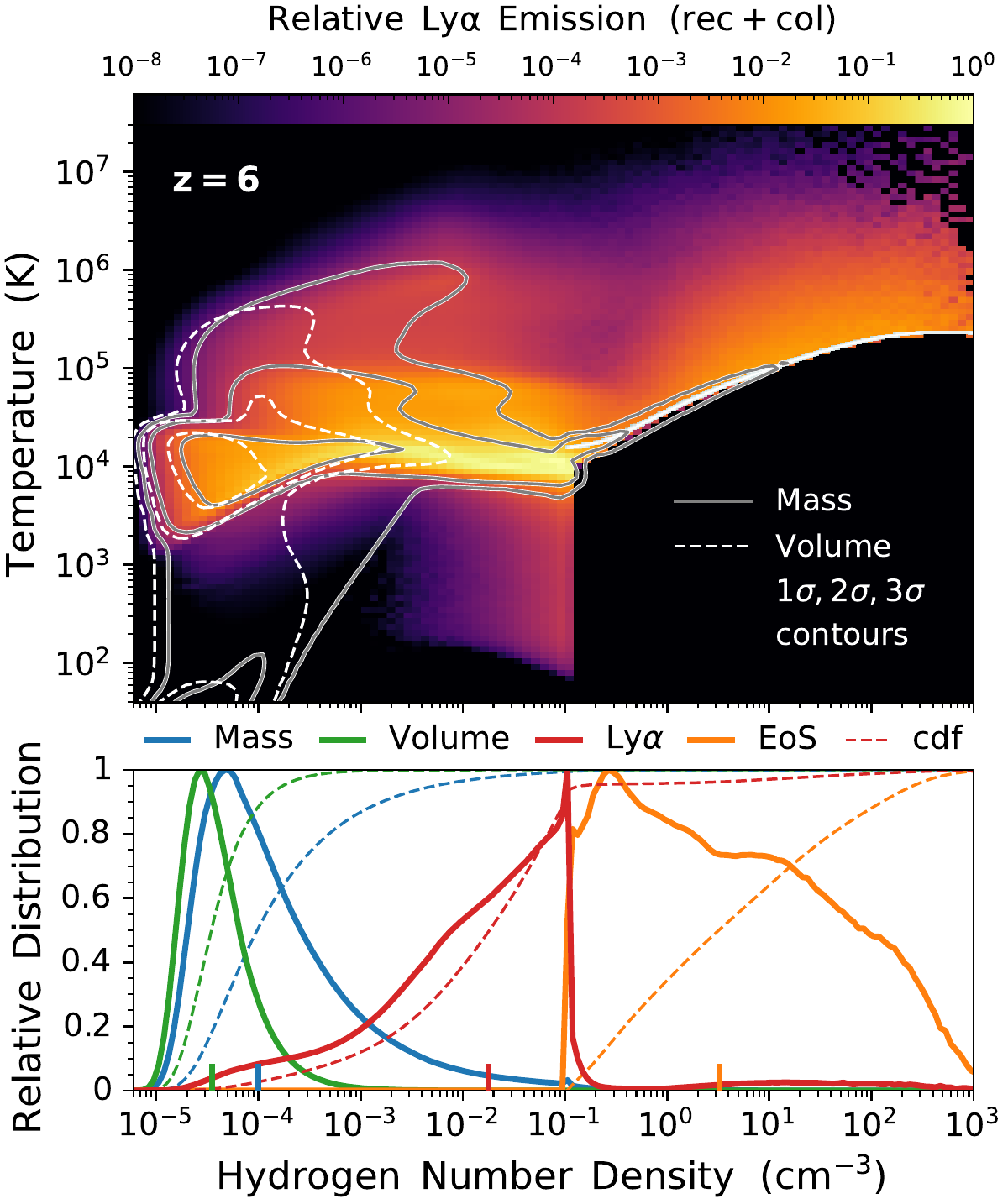}
  \caption{\textit{Top:} Relative Ly$\alpha$ intrinsic luminosity originating from resolved recombination and collisional excitation emission in the gas hydrogen number density and temperature ($n_\text{H}$--$T$) phase plane at $z = 6$. \textit{Bottom:} Distributions showing that Ly$\alpha$ photons are mainly produced by gas around the effective equation of state threshold of $\sim 0.13\,\text{cm}^{-3}$, dominated by ionized gas heated to $\sim 10^4\,\text{K}$. The various curves are for mass (blue), volume (green), and Ly$\alpha$ emission separated into resolved (red) and EoS (orange) components. The dashed curves are cumulative distribution functions, with vertical markers denoting the median volume, mass, and Ly$\alpha$ luminosity-weighted number densities at $3.5 \times 10^{-5}$, $9.9 \times 10^{-5}$, $0.018$, and $3.2\,\text{cm}^{-3}$, respectively.}
  \label{fig:nT_phase}
\end{figure}

To explore the sources of Ly$\alpha$ emission further, in Fig.~\ref{fig:age_hist} we provide the age distribution of stars emitting ionizing radiation that is locally reprocessed into Ly$\alpha$ photons at $z = 6$. The emission traces the youngest stellar populations with a median age of roughly $3$\,Myr, while the mass-weighted median age is $160$\,Myr at $z = 6$, emphasizing the important role of the star formation history beyond the stellar mass alone. Similarly, in Fig.~\ref{fig:nT_phase} we illustrate the relative Ly$\alpha$ intrinsic luminosity originating from resolved recombination and collisional excitation emission as functions of hydrogen number density and temperature ($n_\text{H}$--$T$) at $z = 6$. Ly$\alpha$ photons are mainly produced by gas around the effective equation of state threshold of $\sim 0.13\,\text{cm}^{-3}$. The majority of the emission ($> 90\%$) is from photoheated gas slightly above $10^4\,\text{K}$. Overall, the properties of Ly$\alpha$ production are in line with our expectations considering the limitations of the galaxy formation model on ISM scales.

\begin{figure*}
  \centering
  \includegraphics[width=.507\textwidth]{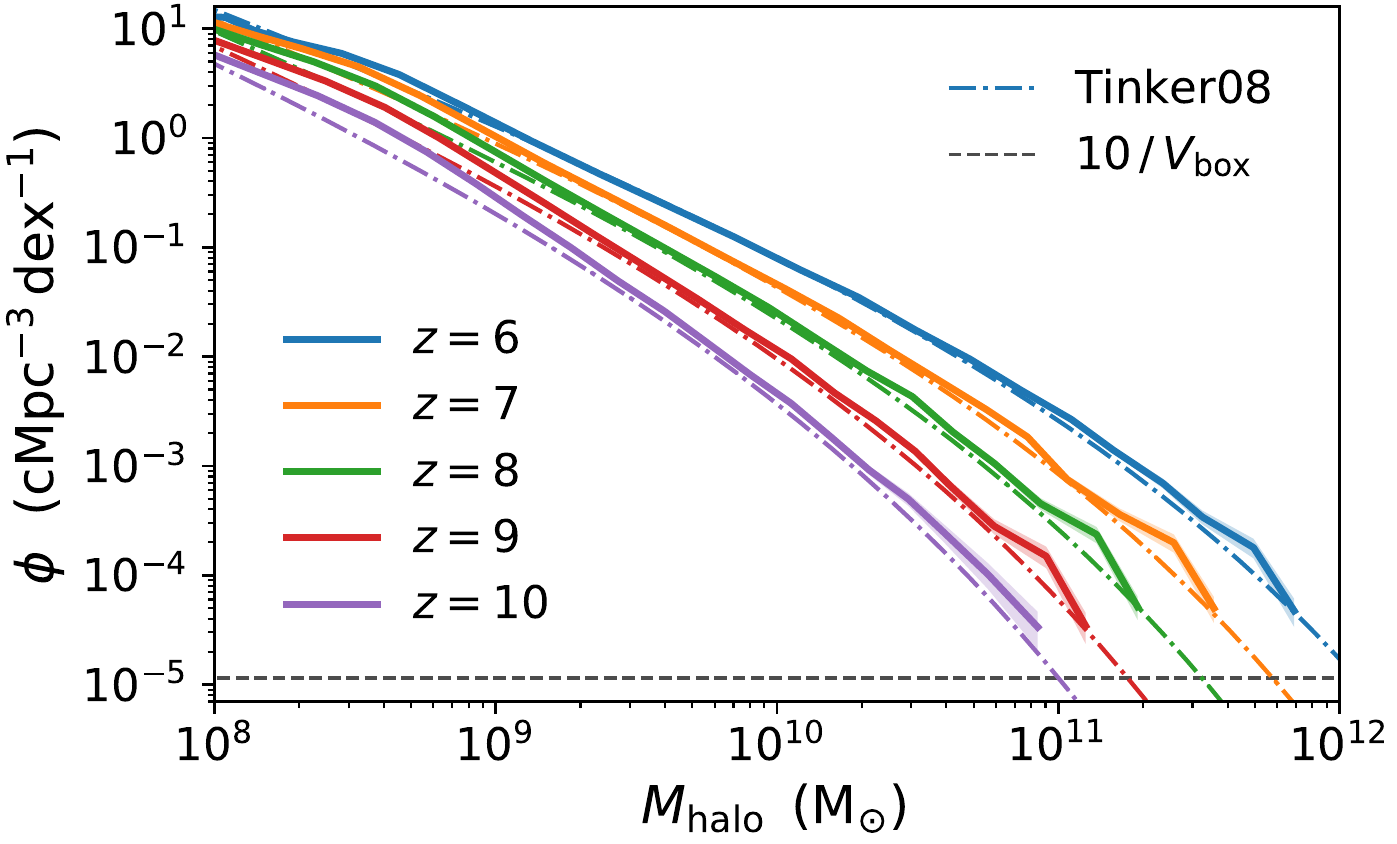}
  \includegraphics[width=.4885\textwidth]{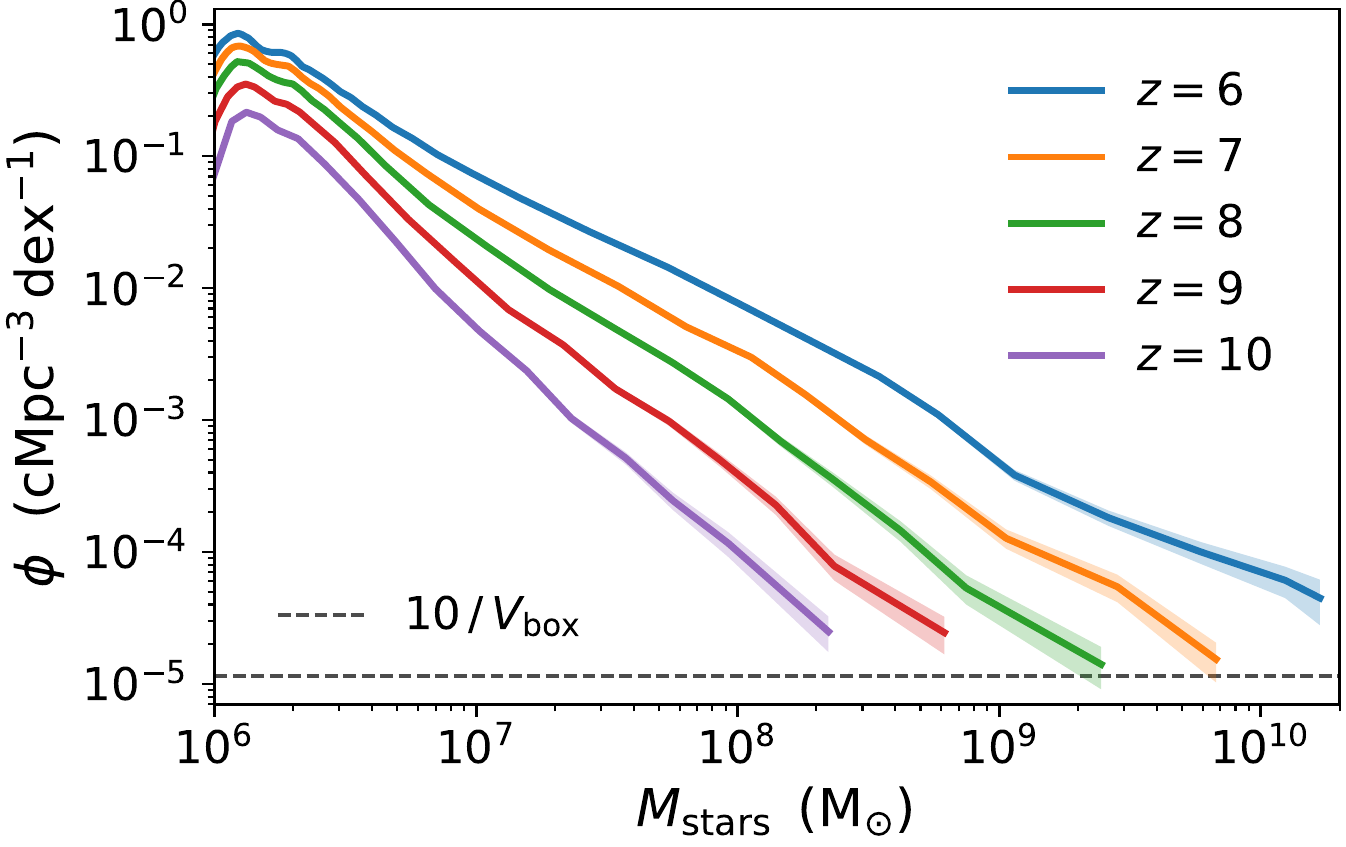}
  \caption{\textit{Left-hand panel:} Galaxy halo mass functions for integer redshifts over the range $z \in [6,10]$ showing the evolution of resolved structures ($M_\text{halo} \gtrsim 10^8\,\Msun$). \textit{Right-hand panel:} Galaxy stellar mass functions down to the clustering resolution of the simulation ($M_\text{stars} \gtrsim 10^6\,\Msun$). The shaded regions represent the Poisson error ($\propto \sqrt{N}$) on the number counts in each bin. For reference we include a grey dashed line representing 10 objects within the simulation box. For comparison, in the left-hand panel we also include the analytic halo mass function model from \citet{Tinker2008}, which was calibrated at higher masses and lower redshifts, specifically they optimized the agreement for halo masses of $M_\text{halo} \sim 10^{11-15}\,\Msun\,h^{-1}$ at $z \sim 0$.}
  \label{fig:phi_Mhalo_Mstar}
\end{figure*}

\begin{figure*}
  \centering
  \includegraphics[width=.475\textwidth]{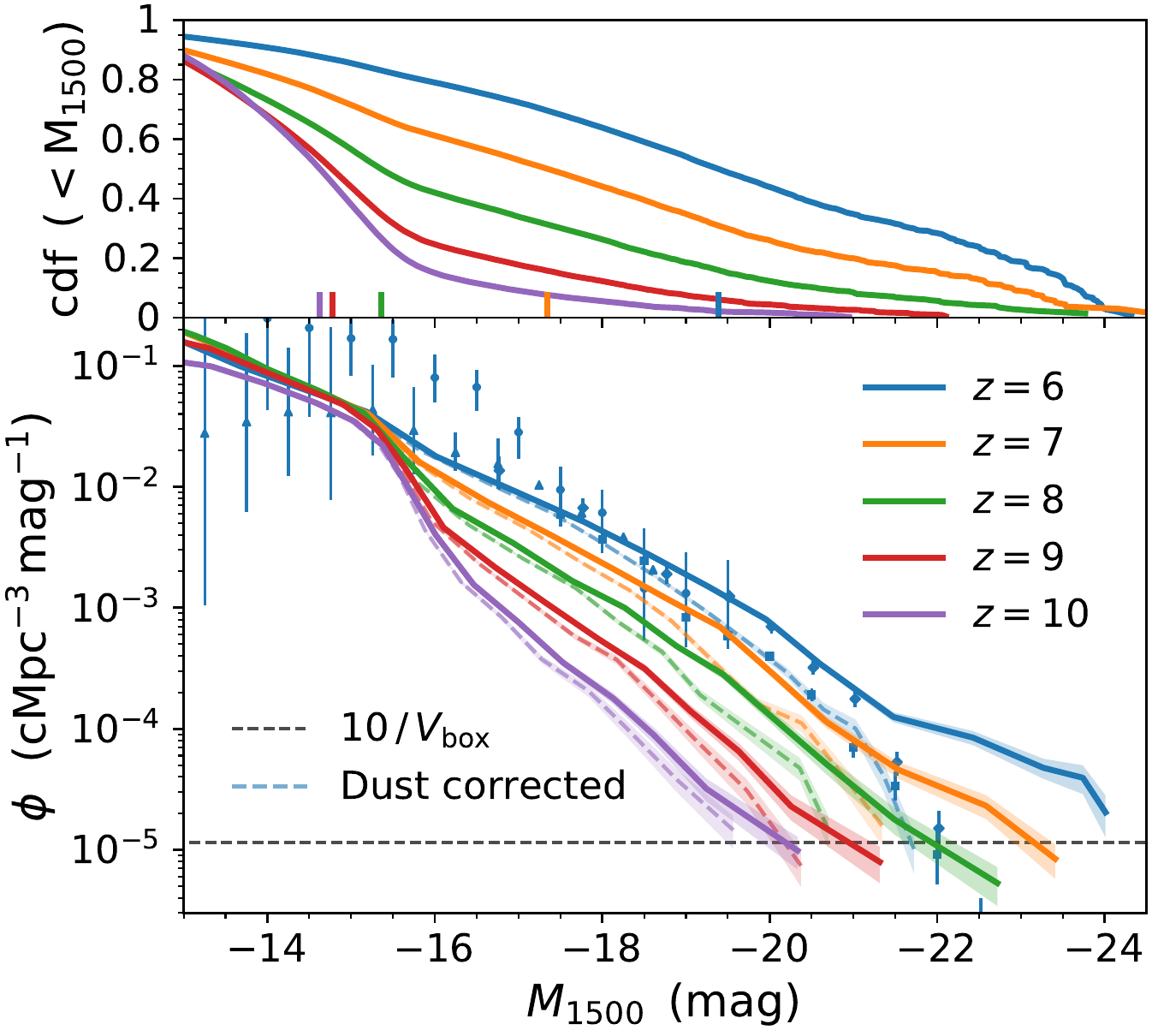}
  \quad
  \includegraphics[width=.492\textwidth]{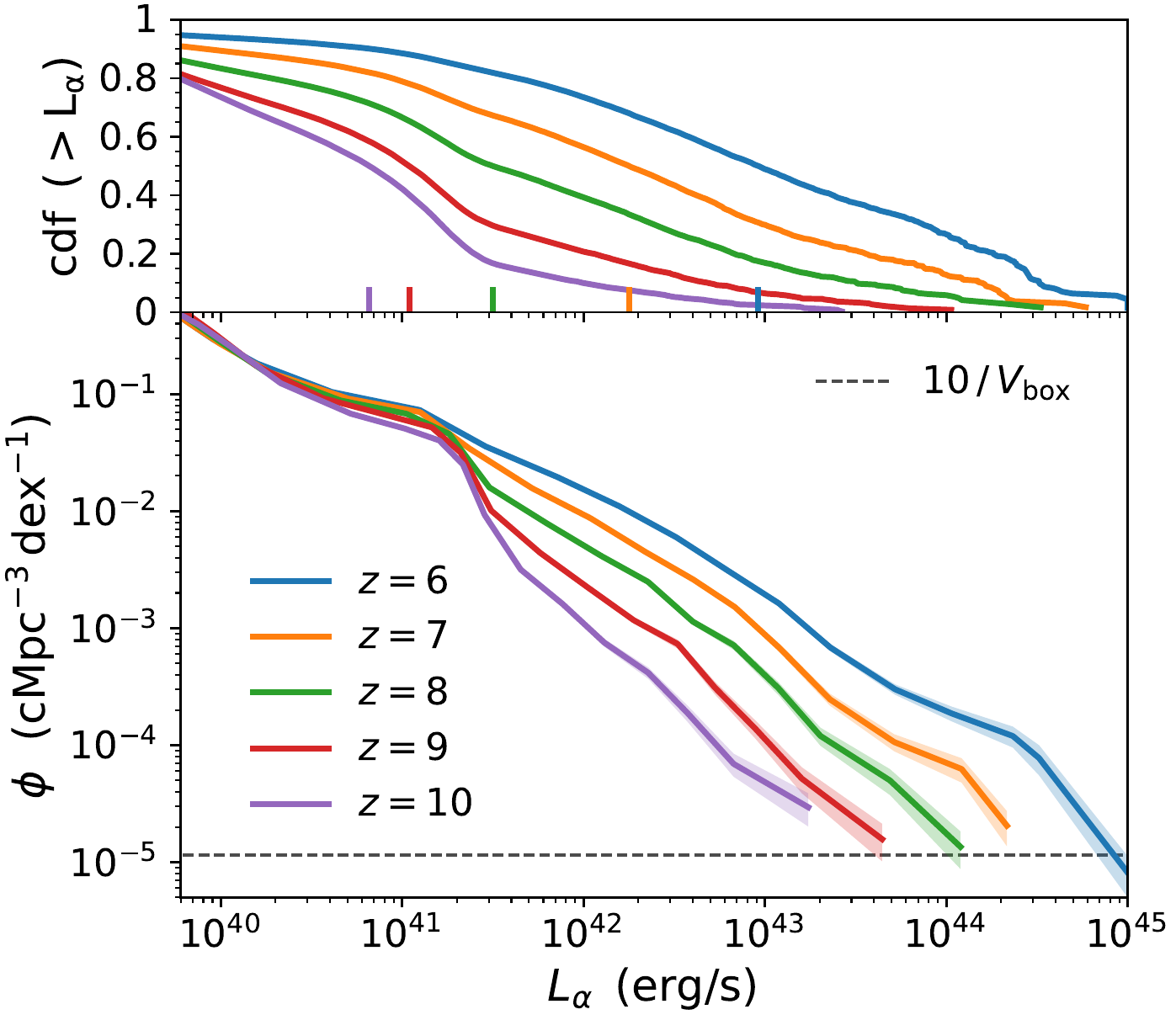}
  \caption{\textit{Left-hand panel:} Galaxy UV (rest-frame 1500\,\AA) luminosity functions for integer redshifts over the range $z \in [6,10]$ showing the emergence of bright galaxies and moderately steep faint-end slopes near the resolution limit $(M_{1500} \gtrsim -15)$. The simulated results match observational estimates from \citet[][diamonds]{Bouwens2015}, \citet[][squares]{Finkelstein2015}, \citet[][circles]{Livermore2017} and \citet[][triangles]{Atek2018} over a wide range of magnitudes after applying the empirical relation presented in \citet{Gnedin2014} to account for dust attenuation, which strongly affects luminous sources. The dust opacity is scaled according to the redshift-dependent dust-to-metal ratio given in \citet{Vogelsberger2020}. \textit{Right-hand panel:} Galaxy intrinsic Ly$\alpha$ luminosity functions showing smooth evolution mirroring the UV luminosity functions but with additional recombination and collisional excitation emission. To give a sense of population convergence, in the upper panels we also show the normalized cumulative luminosity from haloes above a given brightness threshold, including vertical markers for the medians. As in Fig.~\ref{fig:phi_Mhalo_Mstar} the shaded regions represent the Poisson error ($\propto \sqrt{N}$) and the horizontal dashed lines represent 10 objects within the simulation box.}
  \label{fig:phi_M1500_Lya}
\end{figure*}

\subsection{Additional Ly\texorpdfstring{$\balpha$}{α}-centric fields}
\label{sec:additional_Lya_fields}
We also provide the stellar continuum spectral luminosities $L_{\lambda,\text{cont}}$ at $\lambda_\text{cont} = \{1216, 1500, 2500\}$\,\AA, derived by taking the logarithmic average of the BPASS SEDs over windows of $\{50, 20, 20\}$\,\AA\ around the reference wavelengths. The larger window around $\lambda_\alpha = 1215.67$\,\AA\ reduces the sensitivity to Ly$\alpha$ absorption features in the BPASS spectra. For convenience, the absolute magnitude as if viewing the stars from a distance of 10\,pc at these wavelengths is $M_\text{cont} = -2.5 \log[(\lambda_\text{cont}/\text{\AA})^2 L_{\lambda,\text{cont}}/(\text{erg/s/\AA})] + 97.78683$, which is used throughout this study. Likewise, the rest-frame equivalent width characterizing the strength of the Ly$\alpha$ line relative to the continuum flux is given by $\text{EW}_{\alpha,0} \approx L_\alpha / L_{\lambda,1216}$. When the local continuum is not detected in observations, the equivalent width may be derived by extrapolating continuum values based on the UV slope assuming a power-law $L_\lambda \propto \lambda^\beta$ \citep{Hashimoto2017}, such that $\beta = \log(L_{\lambda,2500}/L_{\lambda,1500}) / \log(2500/1500)$ and the estimated continuum around the Ly$\alpha$ line is $L_{\lambda,1216}^\text{est} = L_{\lambda,1500}\,(1215.67 / 1500)^\beta$. The relative difference of these two approaches is explored in Appendix~\ref{appendix:UV_slope} where we find that extrapolations systematically underpredict intrinsic Ly$\alpha$ equivalent widths by approximately 30 per cent.

Similarly, the ionizing luminosity from active galactic nuclei (AGN) $L_\text{ion}^\text{AGN}$ gives the escaping emission from supermassive black holes. We note that the AGN spectral energy distribution (SED) uses the \citet{Lusso2015} parametrization with $35.5$ per cent of the bolometric AGN luminosity at energies above $13.6$\,eV. After converting this quantity to c.g.s.\ units the equivalent number of ionizing photons is determined by dividing by the average energy per photon, i.e. $5.29 \times 10^{-11}\,\text{erg} \approx 33\,\text{eV}$. In this paper we do not explore the properties of AGN bright galaxies beyond confirming that the global contribution is subdominant (see Fig.~\ref{fig:rho_alpha_time}). However, defining the fraction of ionizing photons originating from AGN as $f_\text{AGN} \equiv \dot{N}_\text{ion}^\text{AGN} / (\dot{N}_\text{ion}^\text{AGN} + \dot{N}_\text{ion}^\text{stars})$, we find at $z = \{5.5, 6, 7, 8, 9\}$ there are $\{23, 6, 4, 0, 1\}$ galaxies with $f_\text{AGN} \geq 0.5$ and $\{170, 104, 41, 15, 7\}$ with $f_\text{AGN} \geq 0.1$. Thus, by the end of the simulation rare but very luminous quasars can be dominant for a handful of galaxies, which becomes even more relevant for larger boxes and is worth pursuing in detail in future investigations.

Finally, we also calculate the centre of Ly$\alpha$ luminosity position $\bmath{r}_\alpha$ within the periodic box, peculiar velocity $\bmath{v}_\alpha$, and 1D velocity dispersion $\sigma_\alpha$ for each group and subhalo. For notational convenience we define the Ly$\alpha$ luminosity-weighted average of a quantity as $\langle f \rangle_\alpha = \sum_i f_i L_{\alpha,i} / \sum_i L_{\alpha,i}$, where the sum is over all Ly$\alpha$ sources including gas and stars. Thus, the centre of Ly$\alpha$ luminosity quantities are respectively $\bmath{r}_\alpha \equiv \langle \bmath{r} \rangle_\alpha$, $\bmath{v}_\alpha \equiv \langle \bmath{v} \rangle_\alpha$, and $\sigma_\alpha^2 \equiv (\langle \bmath{v} \bmath{\cdot} \bmath{v} \rangle_\alpha - \langle \bmath{v} \rangle_\alpha \bmath{\cdot} \langle \bmath{v} \rangle_\alpha) / 3$, where the individual $\bmath{r}$ and $\bmath{v}$ are centre of mass positions and peculiar velocities. We note that these quantities are weighted by the intrinsic Ly$\alpha$ emission so are not directly observable due to radiative transfer effects. However, they are still useful for connecting Ly$\alpha$ emission to other galaxy properties, or to account for spatial and velocity offsets and line broadening as the $n_\text{H}^2$ dependence for emission can lead to biases here (discussed in Section~\ref{sec:correlations}).

\begin{figure*}
  \centering
  \includegraphics[width=.48\textwidth]{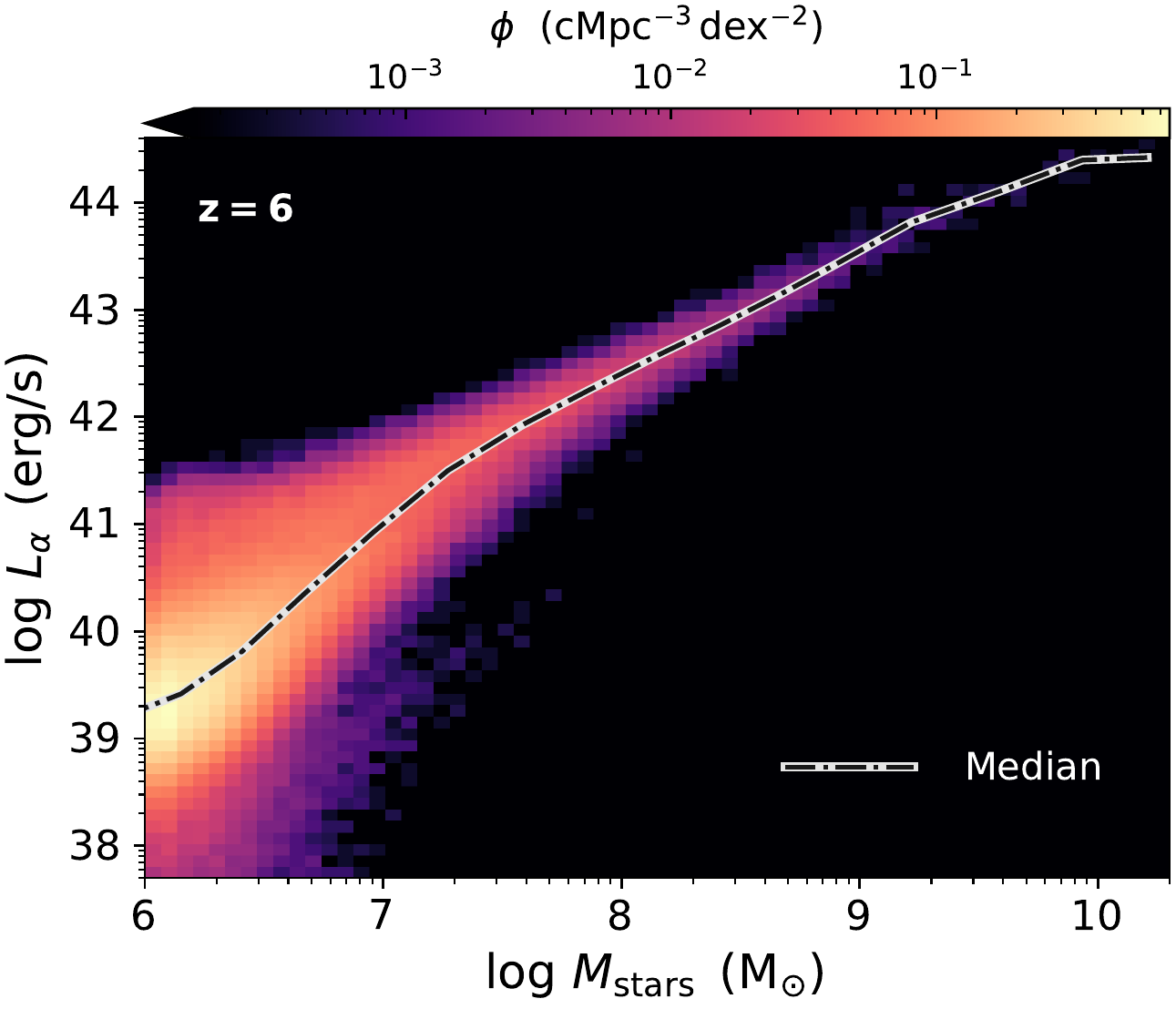}
  \quad
  \includegraphics[width=.485\textwidth]{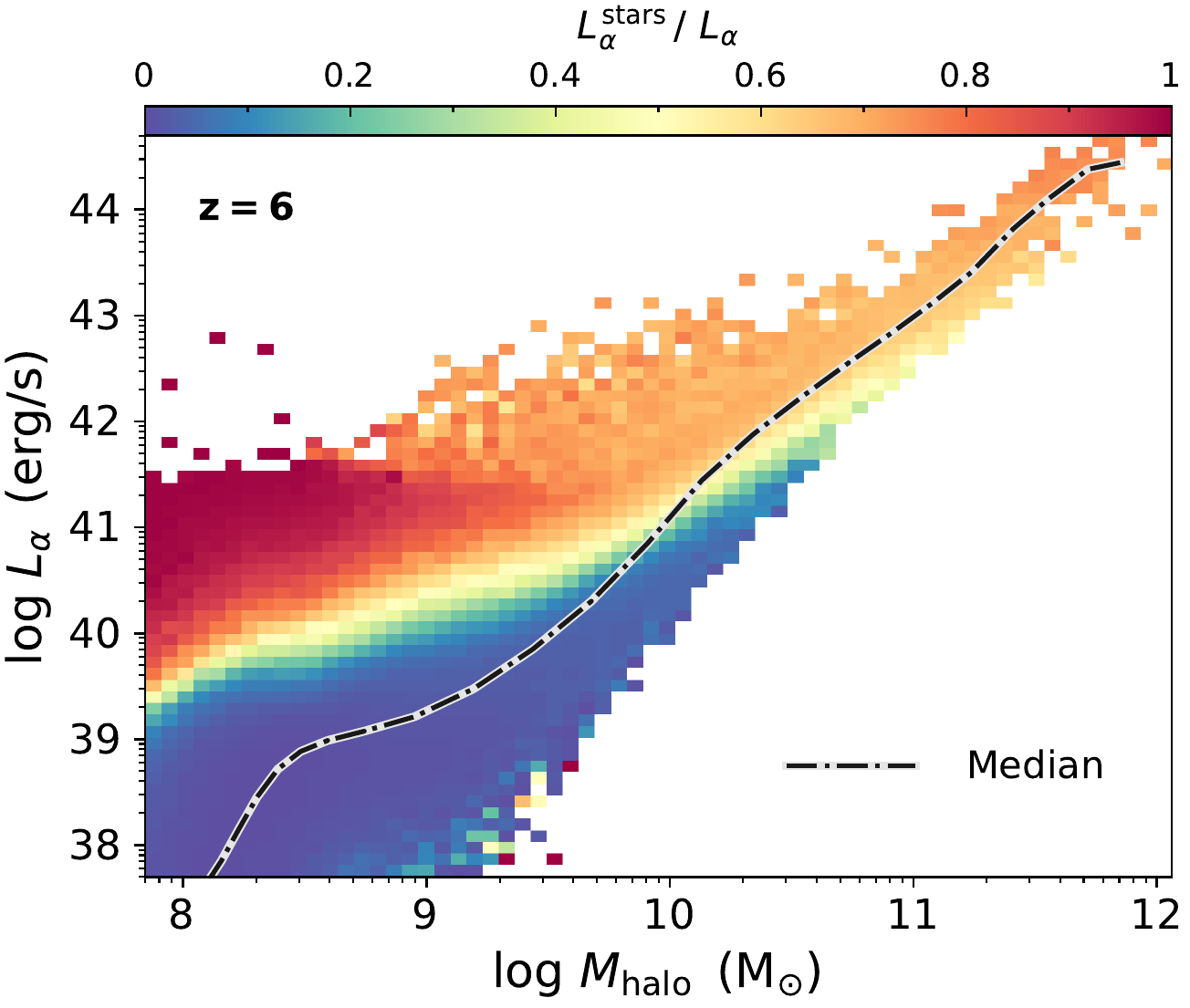}
  \caption{\textit{Left-hand panel:} Total intrinsic Ly$\alpha$ luminosity $L_\alpha$ as a function of galaxy stellar mass, $M_\text{stars}$, at $z = 6$. The colour axis shows the number density of haloes, which provides a sense for the scatter around the median (dash--dotted curve) caused by the wide range of star formation histories. \textit{Right-hand panel:} Total intrinsic Ly$\alpha$ luminosity $L_\alpha$ and galaxy halo mass, $M_\text{halo}$, at $z = 6$. The colour gradient in the fraction of emission from unresolved \HII\ regions ($\sum L_\alpha^\text{stars} / \sum L_\alpha$) within each bin) mirrors the physical diversity of star formation across different haloes. The median reveals where most of the haloes reside within the relation.}
  \label{fig:Mstar_Mhalo_Lya}
\end{figure*}

\subsection{Luminosity functions}
We now consider the evolution of the occurrence and luminosities of galaxies throughout the EoR. In Fig.~\ref{fig:phi_Mhalo_Mstar}, we show galaxy halo and stellar mass functions over the redshift range $z \in [6,10]$ covering four orders of magnitude in resolved structure formation ($M_\text{halo} \gtrsim 10^8\,\Msun$) down to the clustering resolution of the simulation ($M_\text{stars} \gtrsim 10^6\,\Msun$). For reference, we include a horizontal dashed line representing a volume limit of 10 objects within the simulation box. Similarly, in Fig.~\ref{fig:phi_M1500_Lya} we provide galaxy far UV (rest-frame 1500\,\AA) and Ly$\alpha$ intrinsic luminosity functions for the same redshift range. The evolution is smooth and relatively steep due to the continual formation of young stars as galaxies assemble, merge, and are fed by streams of cold gas accretion. The distributions for $L_\alpha$ closely follow that of UV magnitude $M_{1500}$ with the caveat that galaxies can be brighter in Ly$\alpha$ due to the recombination and collisional excitation emission. To provide a sense of population convergence, e.g. above $M_{1500} \approx -15.5$ or $L_\alpha \approx 3 \times 10^{41}\,\text{erg\,s}^{-1}$, we also show the normalized cumulative luminosity from haloes above a given brightness threshold. The details of these luminosity functions may depend on modelling choices and resolution but the shape is robust as it follows the star formation density evolution, which IllustrisTNG is designed to match for currently available observational data at lower redshifts.

We note that the \citet{Springel2003} model employs stochastic sampling to determine when new star particles are created in the simulation, with comparable mass resolutions for star and gas particles. This prescription is correct in the global sense and favourable for avoiding numerical artefacts, e.g. with gravitational force calculations being more robust against artificial mass segregation due to different resolutions for stars and gas \citep{Ludlow2019}. However, the age discretization leads to characteristically bursty star formation histories (SFHs) in marginally resolved haloes \citep{Iyer2020}. To extend the reliability of halo-by-halo predictions for galaxies at the faint-end of the luminosity function we employ the following SFH smoothing procedure. We first calculate the total mass of young stars ($< 5\,\text{Myr}$) in each subhalo and combine this with the instantaneous SFR to define a duration over which we expect these stars to have formed: $\Delta t_\text{SFH} = M_\text{stars}(<5\,\text{Myr}) / \text{SFR}$. In the event that this duration is longer than 5\,Myr we reassign the young stars as having all been formed with a constant SFR with the expected ages. This is done with 100 equal age bins in the interval $[0, \Delta t_\text{SFH}]$, retaining the mass and metallicity distribution of each stellar population. This prescription helps us to smooth out an artificial bump around $M_{1500} \sim -15$ or $L_\alpha \sim 10^{41}\,\text{erg\,s}^{-1}$, corresponding to newly spawned $\sim 5 \times 10^5\,\Msun$ star particles in low mass haloes, while the rest of the luminosities are essentially unaltered. Unless stated otherwise, we employ this SFH smoothing method for all quantities derived from star luminosities when required for individual subhaloes, including Ly$\alpha$-centric fields.

\begin{figure*}
  \centering
  \includegraphics[width=.497\textwidth]{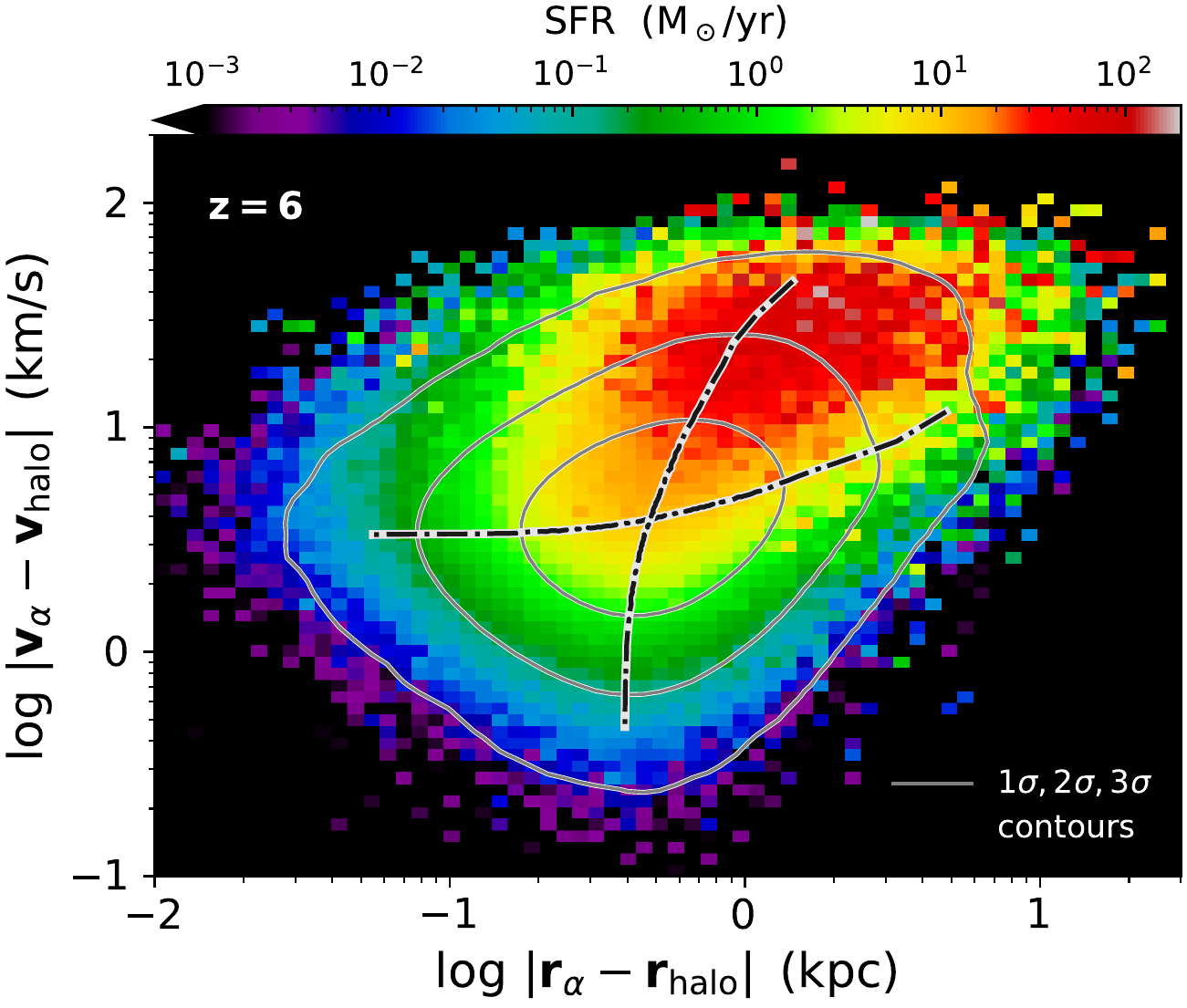}
  \quad
  \includegraphics[width=.4823\textwidth]{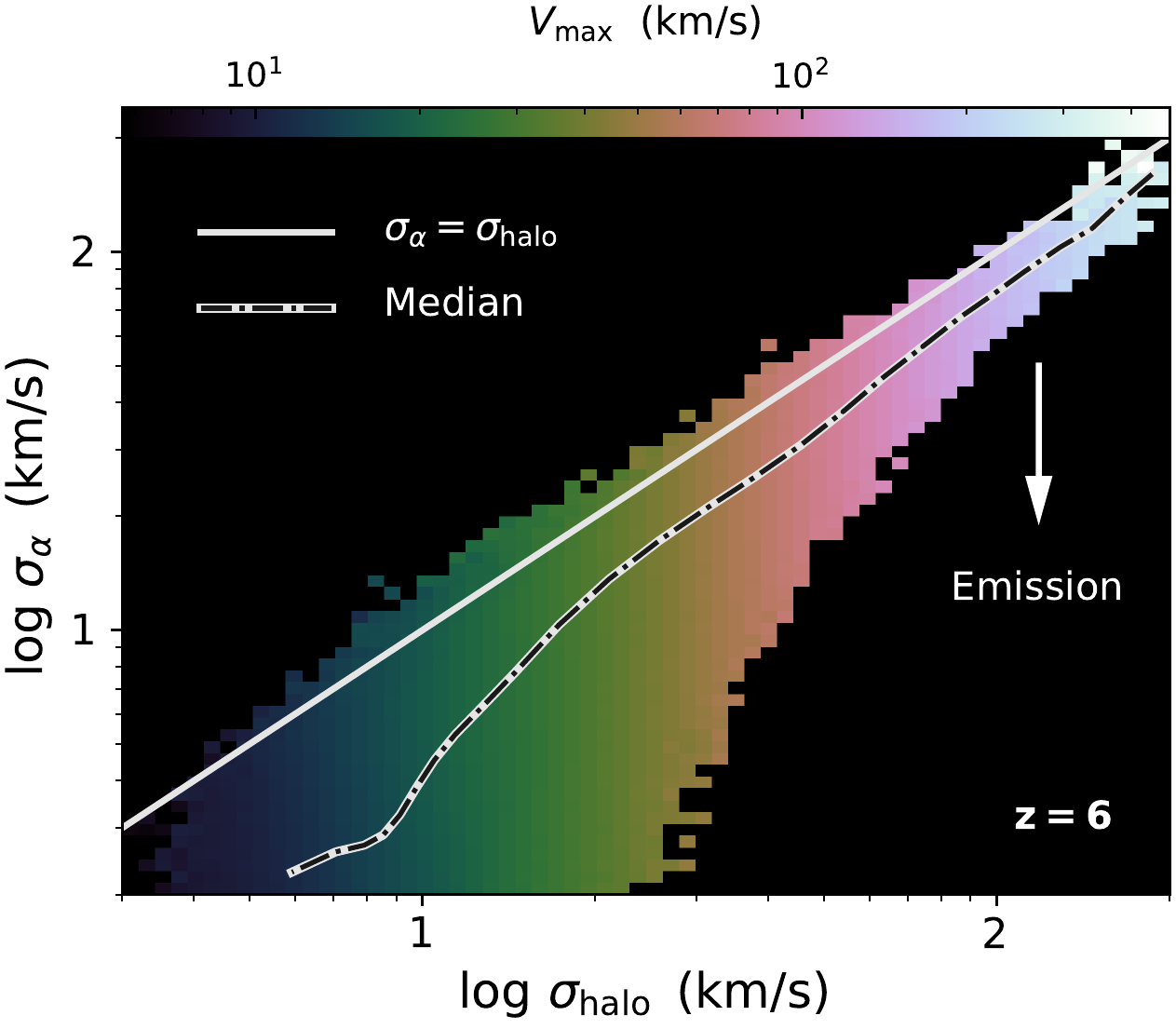}
  \caption{\textit{Left-hand panel:} Position and velocity offsets between the intrinsic Ly$\alpha$ luminosity and mass centroids for each halo at $z = 6$. Galaxies with higher SFRs (denoted by colour) can differ by up to $\sim 10\,\text{kpc}$ and $\sim 100\,\text{km\,s}^{-1}$, affecting the systemic position $\bmath{r}_\alpha$ and velocity $\bmath{v}_\alpha$ for IGM transmission. We show halo number density contours and medians of position and velocity offset bins. \textit{Right-hand panel:} 1D velocity dispersions with intrinsic Ly$\alpha$ luminosity ($\sigma_\alpha$) and mass ($\sigma_\text{halo}$) weights for each halo. For reference, we include a line of equality, median halo counts, and bin-averaged colours showing the maximum of the rotation curve $V_\text{max}$. The intrinsic Ly$\alpha$ line widths before resonant scattering are up to $\approx 2$ times narrower than would be inferred by $\sigma_\text{halo}$ or $V_\text{max}$.}
  \label{fig:pos_vel_sigma}
\end{figure*}

Finally, before proceeding further we emphasize that the results presented in this paper are all without accounting for galaxy-scale Ly$\alpha$ radiative transfer effects. At the bright end, the observed luminosity function is expected to differ by up to two orders of magnitude \citep[comparing intrinsic values to][]{Taylor2020,Taylor2021} highlighting the crucial role of a proper RT treatment \citep{Laursen2019,Garel2021}. A comparison of luminosity functions in the literature reveals that such predictions remain highly uncertain, even from studies that carry out careful radiative transfer calculations. Given this, it would be surprising to see agreement between simulation groups given subtle but important differences in resolution, convergence, algorithm implementations, environmental effects, cosmic variance, and star formation, ISM, or dust modelling choices.

\subsection{Galaxy correlations}
\label{sec:correlations}
We finish this section by exploring various Ly$\alpha$ correlations across galaxy populations. In Fig.~\ref{fig:Mstar_Mhalo_Lya} we show the tight power-law relationship between the total intrinsic Ly$\alpha$ luminosity $L_\alpha$ and galaxy stellar mass $M_\text{stars}$ at $z = 6$. The colour axis illustrates the number density of haloes, which provides a sense for the variation around the median within stellar mass bins (shown by the dash--dotted curve). In particular, we find a relatively large scatter in low mass haloes mostly as a result of the wide range of star formation histories. To investigate this further, we also show the relationship between the Ly$\alpha$ luminosity and galaxy halo mass $M_\text{halo}$. There is a slightly larger scatter mirroring the physical diversity of star formation across different haloes. However, the colours provide information about the fraction of Ly$\alpha$ luminosity originating from unresolved \HII\ regions, i.e. $\sum L_\alpha^\text{stars} / \sum L_\alpha$ within each bin. In this case, the median curve is important to show where most haloes reside. At the low mass end there is a clear gradient from haloes almost entirely dominated by stars due to a recent starburst (red) to haloes with very little recent star formation such that recombination and cooling emission powers the luminosity (blue). At the high mass end the ratio matches the global value with minor deviations (see Fig.~\ref{fig:rho_alpha_time}). This picture is consistent with star formation duty cycles and suppression prior to transitioning into a steady and sustained growth mode for higher mass haloes.

In examining the additional Ly$\alpha$-centric fields we focus on the most relevant quantities for the study of IGM transmissivity in the next section. In Fig.~\ref{fig:pos_vel_sigma} we illustrate the position and velocity offsets between the intrinsic Ly$\alpha$ luminosity centroids ($\bmath{r}_\alpha$, $\bmath{v}_\alpha$) and equivalent centre-of-mass quantities ($\bmath{r}_\text{halo}$, $\bmath{v}_\text{halo}$) for each halo at $z = 6$. While there is only a weak correlation and the majority of haloes are in reasonable agreement, we find that galaxies with higher SFRs can differ by up to $\sim 10\,\text{kpc}$ and $\sim 100\,\text{km\,s}^{-1}$ as can be seen by the gradient towards the red average colour ($\gtrsim 10\,\Msun/\text{yr}$). As this can have an impact on IGM transmission calculations, in this study we prefer to set the systemic position $\bmath{r}_\alpha$ and velocity $\bmath{v}_\alpha$ based on these values from the Ly$\alpha$ catalogue. To provide further information about the statistics of galaxy populations, the contours show the halo number density and the dash--dotted curves are medians of position and velocity offset bins, respectively. In the right-hand panel of Fig.~\ref{fig:pos_vel_sigma} we compare the 1D velocity dispersions calculated with intrinsic Ly$\alpha$ luminosity ($\sigma_\alpha$) and mass ($\sigma_\text{halo}$) weights for each halo. For reference, we include a line denoting equality ($\sigma_\alpha = \sigma_\text{halo}$), median halo counts, and bin-averaged colours showing the maximum value of the spherically averaged rotation curve $V_\text{max}$. In general, we find the intrinsic Ly$\alpha$ line widths before resonant scattering and other RT effects can be up to $\approx 2$ times narrower than would be inferred by either $\sigma_\text{halo}$ or $V_\text{max}$. Of course, this is the expected behavior for emission processes due to both the clustered nature of starbursts and the selection bias towards high density gas for two-body ($\propto \rho^2$) recombination and collisional excitation emission.

\section{Transmission curves from galaxies}
\label{sec:IGM}
The strong absorption of photons near the Ly$\alpha$ line provides a powerful probe of the structure and evolution of reionization. We thus explore various connections between galaxies and the IGM via Ly$\alpha$ transmission statistics from the flagship \thesan\ simulation.

\subsection{Transmission catalogue procedures}
\begin{figure}
  \centering
  \includegraphics[width=\columnwidth]{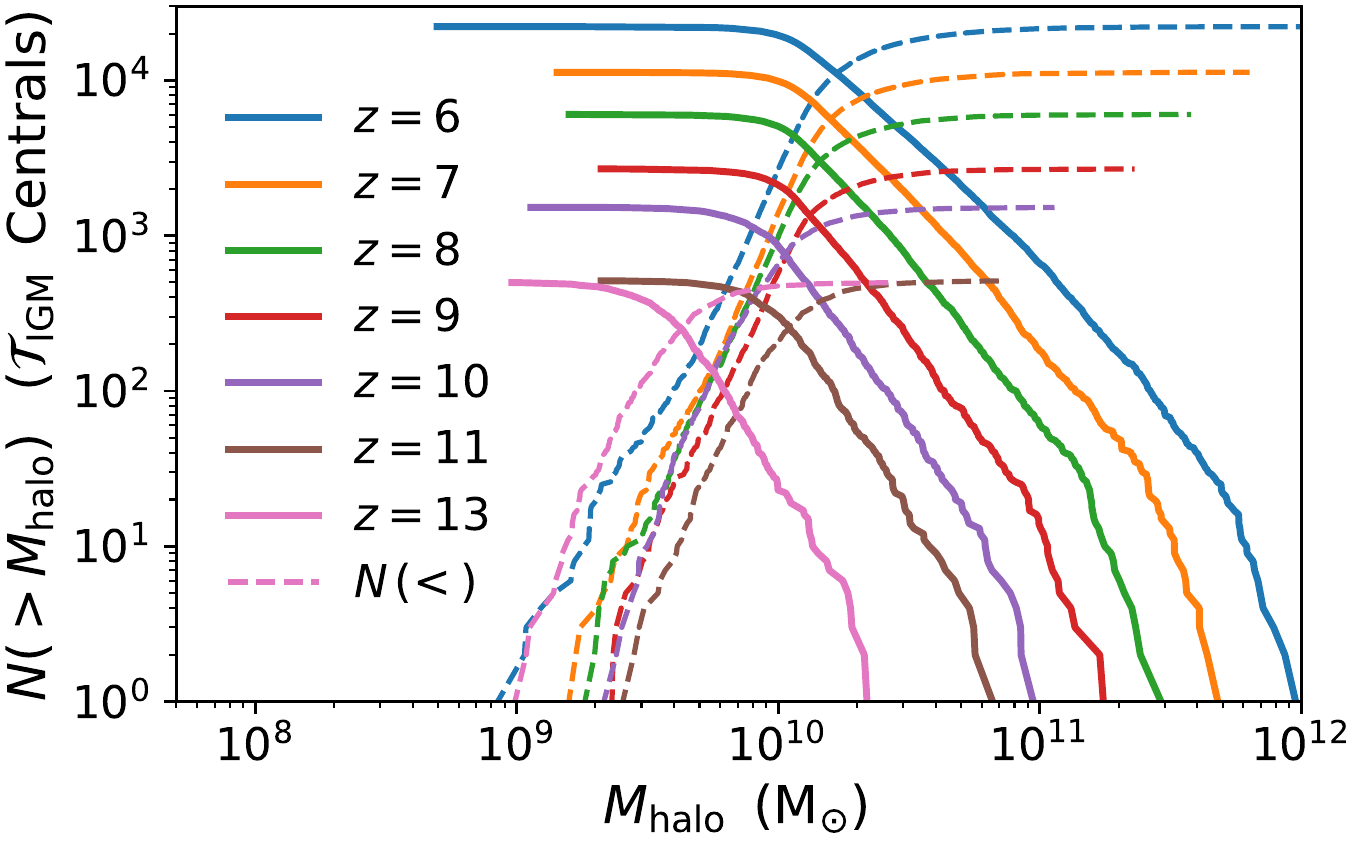}
  \caption{Cumulative number of central galaxies above (solid curves) and below (dashed curves) a given halo mass $M_\text{halo}$ that are included in the initial IGM transmission catalogues for each redshift $z = \{6,7,8,9,10,11,13\}$.}
  \label{fig:N_IGM_catalog}
\end{figure}

In this work we provide a catalogue of Ly$\alpha$ IGM transmission curves that are designed to be as robust as possible for reuse with future studies. While we follow many of the procedures suggested by previous and concurrent authors, including \citet{Laursen2011}, \citet{Jensen2014}, \citet{Byrohl2020}, \citet{Gronke2021}, \citet{Garel2021}, and \citet{Park2021}, we briefly describe our parameter choices in case they differ. In particular, for each selected galaxy we extract 768 radially outward rays corresponding to equal area healpix directions of the unit sphere. To cut back on the amount of data storage and computation we focus on the central galaxies of each group. We expect satellites to generally be fainter than the central and have similar cosmological scale IGM transmission with the exception of minor localized sightline and velocity offset effects. Our catalogue includes all centrals with at least 32 star particles, but also includes less resolved centrals if they are among the 90\% most massive centrals. Thus, our initial sample covers a large range of environments and galaxy histories for 44\,700 galaxies and $3.43 \times 10^7$ ray extractions at $z = \{6,7,8,9,10,11,13\}$ -- see Fig.~\ref{fig:N_IGM_catalog} for cumulative number counts with halo mass. In fact, if satellites inherit transmission properties from the central halo then the catalogue can be viewed as nearly complete for all galaxies with resolved star formation histories.

As we focus on central galaxies, we start the rays at initial distances of $R_\text{vir}$ for the entire group, defined as the radius within which the mean density becomes 200 times the cosmic value ($R_{200}$). Our choice is a compromise between sufficient proximity to capture the local imprint of the CGM and distance to ensure most photons will not scatter back into the line-of-sight, and in Appendix~\ref{appendix:initial_radius} we show that the resulting median statistics are unaffected by doubling this value. We take the systemic location and rest-frame from the subhalo catalogue described in Section~\ref{sec:galaxies}, which accounts for the intrinsic Ly$\alpha$ luminosity averaged position and velocity including all emission sources, i.e. recombinations, collisional excitation, and stars. Although the Ly$\alpha$ spectra will be altered in the process of escaping galaxies, we believe these spatial and spectral anchors are slightly more accurate than using the halo position and velocity directly. We select a broad wavelength range of $\Delta v \in [-2000, 2000]\,\text{km\,s}^{-1}$ sampled at a high spectral resolution of $5\,\text{km\,s}^{-1}$ or a resolving power of $R \approx 60\,000$ for convergence and to cover a variety of future use cases. To ensure the bluemost frequencies are redshifted well into the red wing we perform the integrations out to a distance of $4000\,\text{km\,s}^{-1} / H(z) \approx 40\,\text{cMpc}\,[(1+z) / 7]^{-1/2}$.

\subsection{Ray extractions}
Radiative transfer on Voronoi meshes has recently been adopted in the Ly$\alpha$ community \citep{SmithDCBH2017,Byrohl2020,Byrohl2021,Camps2021,Smith2021}. However, while there are many approaches to integration for optically thin radiative transfer, we have developed a low memory method for exact ray-tracing through the native Voronoi unstructured mesh data (implemented as an additional module within \textsc{colt}). This was done to avoid constructing the tessellation for the entire simulation box each time a new set of rays is desired, especially as this may not be feasible on local in-house computing facilities. The idea is to select particles from a cylindrical region around the ray, construct a localized tessellation from these points, and output 1D data objects based on ray-tracing through the smaller tessellation. This process produces identical results compared to ray-tracing through the full grid as long as each intersecting cell is included in the selection. Thus, we first calculate the maximum cell volume $V_\text{max}$ over the entire simulation and set the cylinder impact radius as $r_\text{max} = \eta V_\text{max}^{1/3}$, where $\eta$ is a factor accounting for the geometry. Based on convergence tests we use $\eta = 0.75$, slightly larger than the value for spherical cells of $(3 / 4\pi)^{1/3} \approx 0.62$. In most cases cells are significantly smaller than this value so an adaptive impact parameter would likely be more efficient but we did not experiment with such an approach.

The extraction process is simplified by recentring the box so that the origin is the same for all rays under consideration, and we assume this is true in the equations that follow. We note that periodic boundary conditions are implemented by checking all possible tilings and duplicating particle data as many times as is necessary, although the rays for our Ly$\alpha$ transmission study are smaller than half the box size so this is not necessary here. The distance from each point $\bmath{r} = (x, y, z)$ to the line defined by the unit vector direction $\bmath{n} = (n_x, n_y, n_z)$ passing through the origin is $b = \| \bmath{r} - (\bmath{r} \bmath{\cdot} \bmath{n}) \bmath{n} \|$. The squared distance along the ray is then $d^2 = r^2 - b^2$, where the distance from the point to the origin is $r = \| \bmath{r} \|$ and the sign of $d$ is the same as that of $\bmath{r} \bmath{\cdot} \bmath{n}$. Thus a point is within the cylindrical region if $b < r_\text{max}$ and $d \in (r_0 - r_\text{max}, r_0 + l + r_\text{max})$ where $r_0$ and $l$ denote the starting radial offset and length of the ray, respectively. For convenience with ray-tracing after extraction we rotate all points to align with the $z$-axis and shift the start of the ray to the origin. The new points are located at $\bmath{r}' = (x', y', z') = (x - n_x C_z, y - n_y C_z, \bmath{r} \bmath{\cdot} \bmath{n} - r_0)$, where the rotation constant is $C_z = (\bmath{r} \bmath{\cdot} \bmath{n} + z) / (1 + n_z)$. Finally, after selection and ray-tracing the raw particle data is written as a compact file containing the ray properties, origins, directions, length segments, and raw data for each cell in the traversed order.

\subsection{Integration with continuous Hubble flow}
For frequency-dependent optically thin Ly$\alpha$ radiative transfer it is convenient to convert to the dimension-less frequency
\begin{equation}
  x \equiv \frac{\nu - \nu_0}{\Delta \nu_\text{D}} \, ,
\end{equation}
where $\nu_0 = 2.466 \times 10^{15}$\,Hz denotes the frequency at line centre and $\Delta \nu_\text{D} \equiv (v_\text{th}/c)\nu_0$ the Doppler width of the profile. The frequency dependence of the absorption coefficient is given by the Voigt profile $\phi_\text{Voigt}$. For convenience we define the Hjerting--Voigt function $H(x) = \sqrt{\pi} \Delta \nu_\text{D} \phi_\text{Voigt}(\nu)$ as the dimension-less convolution of Lorentzian and Maxwellian distributions,\footnote{We employ the standard notation of $H$ for both the Hubble parameter and the line profile, but the meaning can be understood from the context.}
\begin{align} \label{eq:H}
  H(x) &= \frac{a}{\pi} \int_{-\infty}^\infty \frac{e^{-y^2}\text{d}y}{a^2+(y-x)^2} \approx
    \begin{cases}
      e^{-x^2} & \quad \text{`core'} \\
      {\displaystyle \frac{a}{\sqrt{\pi} x^2} } & \quad \text{`wing'}
    \end{cases} \\
    &= \text{Re}\left( e^{(a - i x)^2} \text{erfc}(a - i x) \right) \approx e^{-x^2} + \frac{2 a}{\sqrt{\upi}} \big( 2 x F(x) - 1 \big) \, . \notag
\end{align}
Here the `damping parameter', $a \equiv \Delta \nu_L /2 \Delta \nu_D \approx 4.7 \times 10^{-4}\,T_4^{-1/2}$, describes the relative broadening compared to the natural line width $\Delta \nu_\text{L} = 9.936 \times 10^7$\,Hz. The final approximation is the first order expansion in $a$, and the (complex) complementary error function is related to the area under a Gaussian by $\text{erfc}(z) \equiv 1 - 2 \int_0^z e^{-y^2}\text{d}y / \sqrt{\pi}$ and the Dawson integral is $F(x) \equiv \int_0^x e^{y^2 - x^2}\text{d}y$.

We now introduce a new method for calculating the traversed optical depth based on continuous Doppler shifting due to velocity gradients encountered during propagation. For the specific case of cosmological Hubble flow the expansion induces a constant redshift per unit distance such that the change in velocity is $\Delta v = H(z) \Delta \ell$. Thus, photons experience continuous Doppler shifting that can be modelled by a position-dependent frequency as
\begin{equation}
  \frac{\Delta\lambda}{\lambda} = \frac{\Delta v}{c} \quad \Rightarrow \quad x' = x - \frac{H(z)}{v_\text{th}} \Delta\ell \, ,
\end{equation}
where we have used the relation between Doppler frequency and velocity: $x = -\Delta v / v_\text{th}$. The resulting optical depth in this case is
\begin{align} \label{eq:tau_Hubble}
  \tau &= k_0 \int_0^\ell H\left( x - \mathcal{K} \ell' \right)\,\text{d}\ell' \\
  &\approx \frac{\sqrt{\pi} k_0}{2 \mathcal{K}} \big[ \text{erf}(x) - \text{erf}(x - \mathcal{K} \ell) \big] + \frac{2 a k_0}{\sqrt{\pi} \mathcal{K}} \big[ F(x - \mathcal{K} \ell) - F(x) \big] \, , \notag
\end{align}
where $\mathcal{K} \equiv H(z) / v_\text{th}$ and the absorption coefficient at line centre is $k_0 \equiv n_\text{\HI} \sigma_0$ with cross-section of $\sigma_0 = f_{12} \sqrt{\pi} e^2 / (m_e c \Delta \nu_\text{D})$ and oscillator strength of $f_{12} = 0.4162$. The final expression employs the first-order expansion from equation~(\ref{eq:H}), which is sufficiently accurate for the Ly$\alpha$ line although it is possible to include higher order terms in $a$ if desired. We note that for numerical stability if $\mathcal{K} \ell \ll 1$, corresponding to $\ell \ll 18.4\,\text{kpc}\,T_4^{1/2} [(1+z) / 7]^{-3/2}$, then it is suitable to use the static approximation for the optical depth: $\tau = k_0 H(x) \ell$. Finally, this scheme can also be used to incorporate exact Doppler shifting for other scenarios such as underresolved galactic winds by incorporating the local LOS velocity gradient relative to the comoving frame of the gas. A similar gridless Monte Carlo radiative transfer scheme has also been employed to accurately integrate through arbitrary density gradients \citep{LaoSmith2020}.

\begin{figure}
  \centering
  \includegraphics[width=\columnwidth]{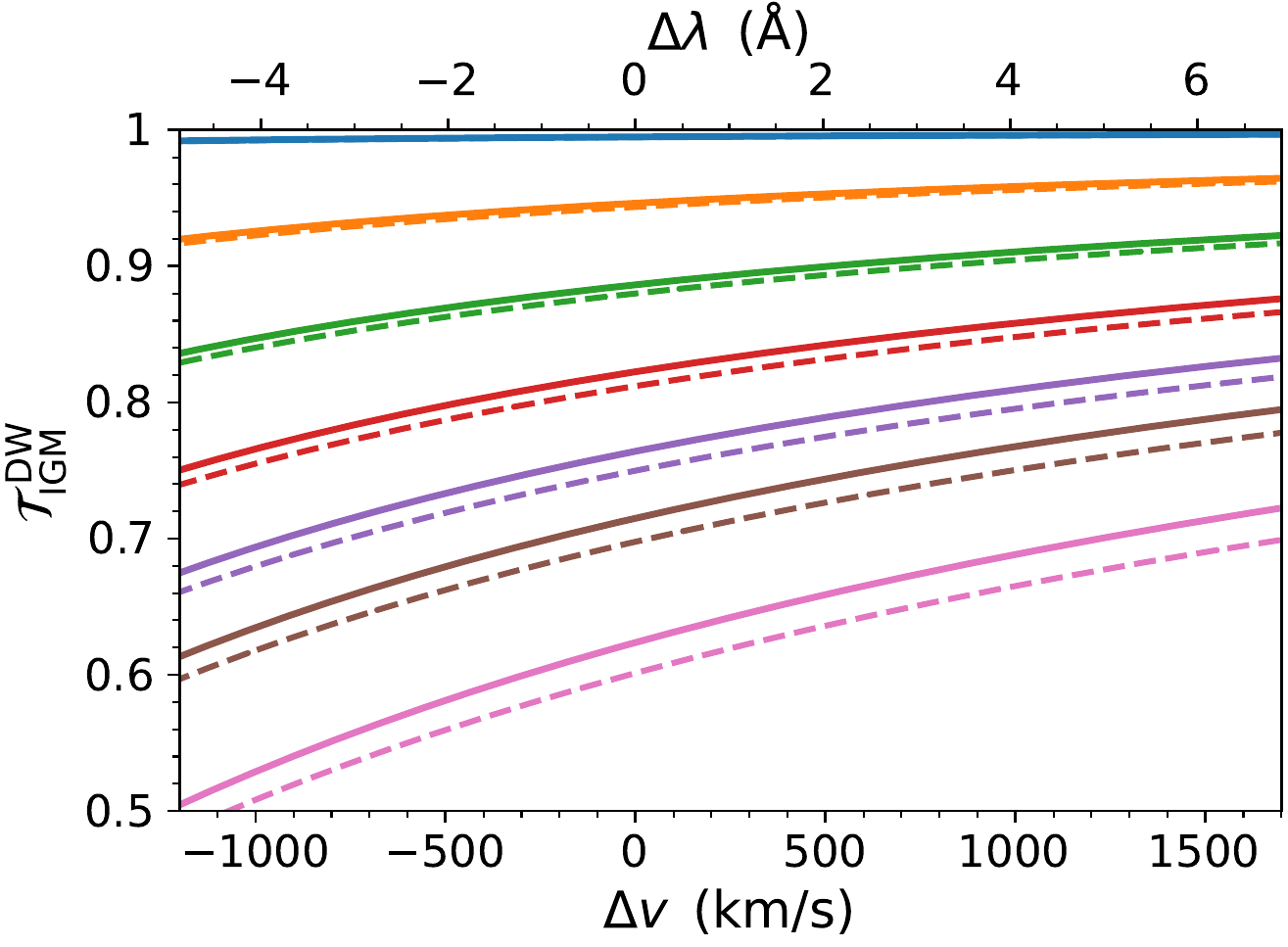}
  \includegraphics[width=\columnwidth]{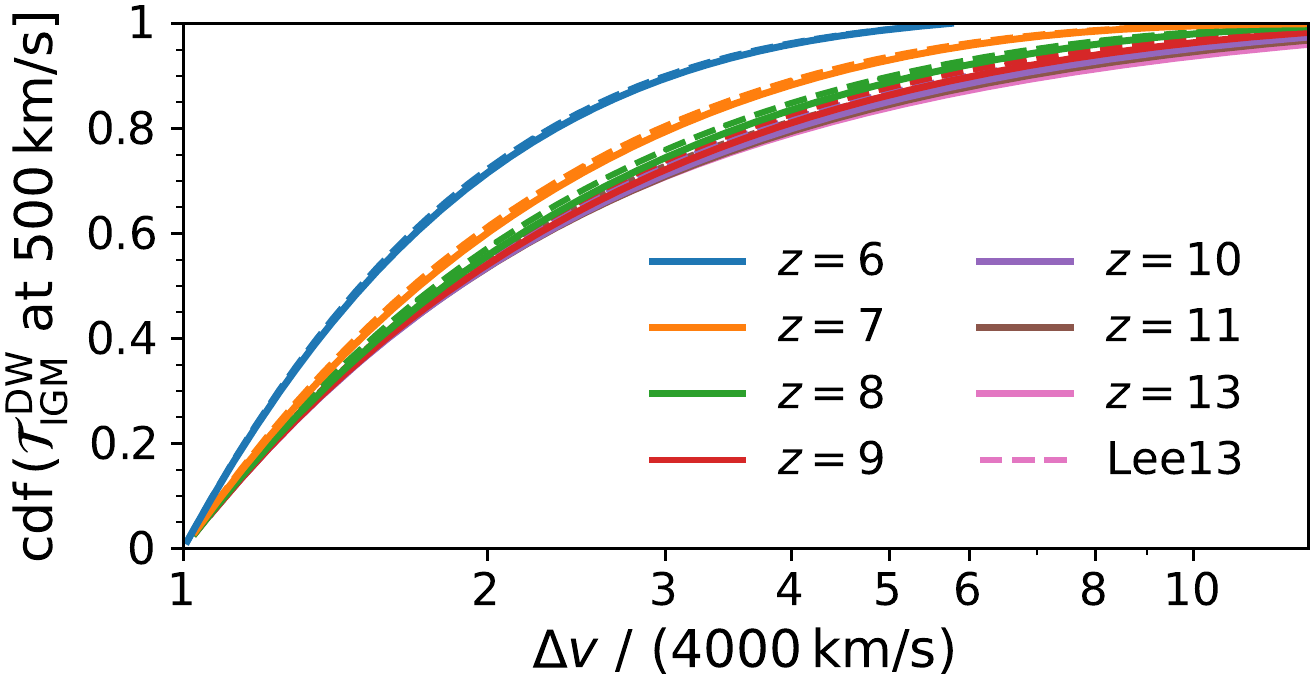}
  \caption{\textit{Top:} Damping-wing IGM transmission $\mathcal{T}_\text{IGM}^\text{DW} = \exp(-\tau_\text{DW})$ beyond the local rays of $\Delta v_s = 4000\,\text{km\,s}^{-1}$ as a function of velocity offset $\Delta v$ and rest-frame wavelength offset $\Delta \lambda$. This large-scale contribution cannot be ignored at pre-reionization redshifts. The solid curves assume a standard Lorentzian profile while the dashed curves include quantum-mechanical corrections to the Voigt profile \citep{Lee2013}. \textit{Bottom:} Cumulative distribution functions of the damping-wing absorption as a function of traversed velocity offset. Full convergence requires light-cone distances of up to $\sim 10$ times longer than the detailed local calculations considered in this study. }
  \label{fig:DW_absorption}
\end{figure}

As we have already extracted the ray segments into small data arrays the IGM absorption approximation results in a series of independent radiative transfer calculations for each input frequency and segment. Specifically, we adopt notation for the $i^\text{th}$ frequency index and $j^\text{th}$ path segment such that for a given initial velocity offset $\Delta v_i$ (i.e.\ reference frequency), unit direction $\bmath{n}$, LOS peculiar velocity relative to the systemic velocity $v_j \equiv \bmath{n} \bmath{\cdot} (\bmath{v}_{\text{pec},j} - \bmath{v}_\text{sys})$, and segment starting distance $r_j$ (all in physical units), the Doppler frequency becomes $x_{i,j} = -(\Delta v_i + v_j + H(z) r_j) / v_{\text{th},j}$. The optical depth $\Delta\tau_{i,j}$ contributed by each segment is the result of evaluating equation~(\ref{eq:tau_Hubble}) with this frequency over the path length $\Delta \ell_j$. In practice, the total optical depth $\tau(\Delta v_i) = \sum_j \Delta\tau_{i,j}$ defines the frequency-dependent transmission function for each ray:
\begin{equation}
  \mathcal{T}(\Delta v) \equiv \exp\big[ -\tau(\Delta v) \big] \, ,
\end{equation}
which describes the fraction of flux not attenuated by the IGM after escaping the halo. Thus, we have a statistical framework for Ly$\alpha$ transmission directly connected to the galaxy catalogues.

\begin{figure}
  \centering
  \includegraphics[width=\columnwidth]{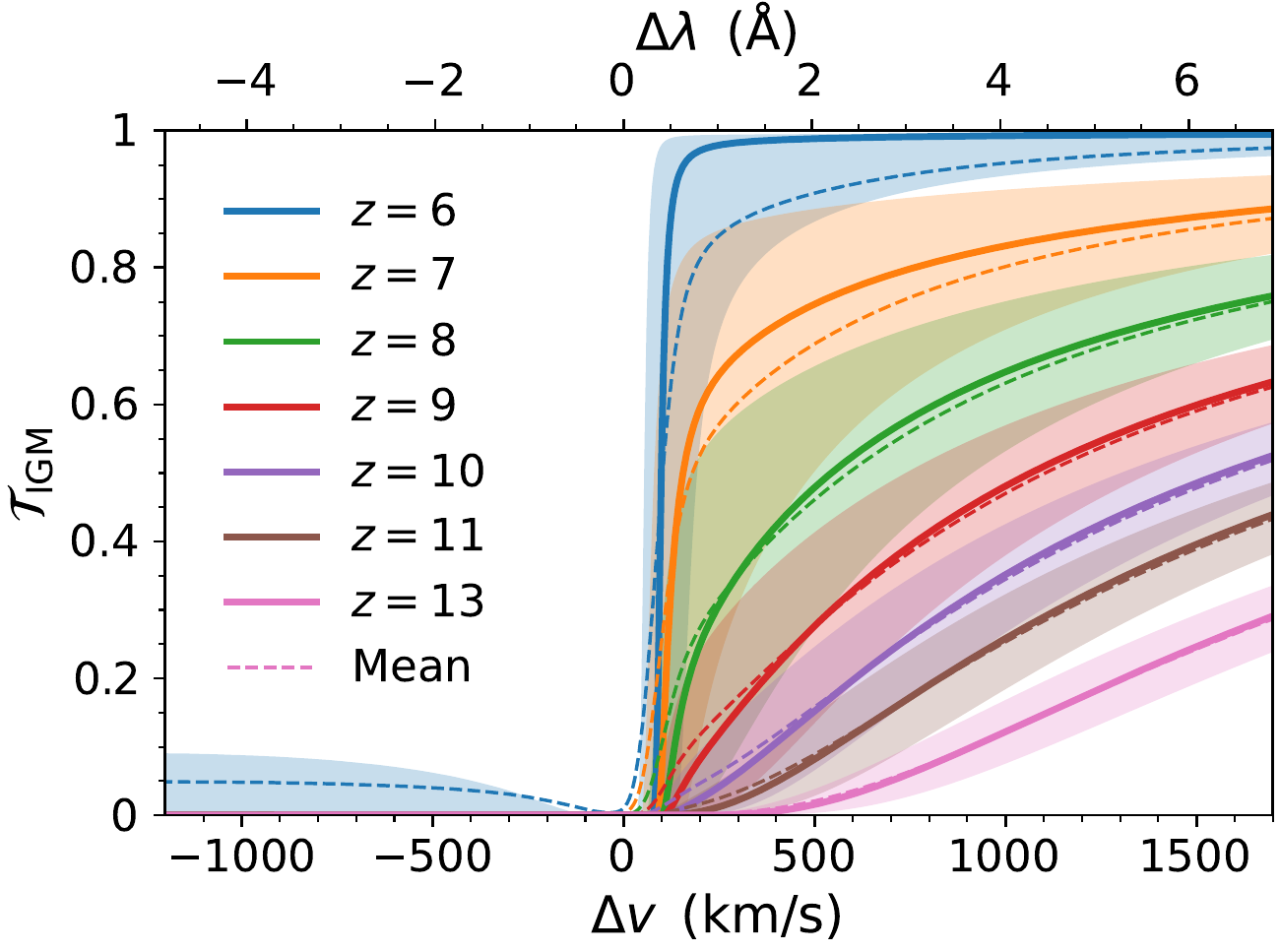}
  \caption{IGM transmission $\mathcal{T}_\text{IGM}$ as a function of velocity offset $\Delta v$ and rest-frame wavelength offset $\Delta \lambda$ around the Ly$\alpha$ line at each redshift. The solid (dashed) curves show the catalogue median (mean) statistics and shaded regions give the $1\sigma$ confidence levels. With equal weight to all galaxies and sightlines, this view is biased towards lower mass that dominate the catalogues by halo count. However, this overview nicely illustrates blue peak suppression and red damping-wing absorption throughout the EoR.}
  \label{fig:T_IGM_catalog}
\end{figure}

\subsection{Damping-wing absorption}
\label{appendix:DW_absorption}
Similar to \citet{Park2021}, we choose to incorporate the distant IGM beyond the local rays in a statistical sense based on the global reionization history. A more accurate treatment requires considering the time-evolution of ionized bubbles through $\gtrsim 200\,\text{cMpc}$ light cones, which we will explore in a future study. We emphasize that this additional damping-wing absorption has a very minor impact after the EoR but is increasingly important at higher redshifts \citep{Miralda-Escude1998}. We choose to frame the integration in terms of the source redshift $z_s$, initial velocity offset $\Delta v$, and global volume-averaged neutral hydrogen fraction $\langle x_\text{\HI} \rangle$. The cosmological neutral hydrogen number density is then
\begin{equation}
  \langle n_\text{\HI} \rangle = \frac{\Omega_b X}{m_\text{H}} \frac{3 H_0^2}{8 \pi G} \langle x_\text{\HI} \rangle (1 + z)^3 \, ,
\end{equation}
where the hydrogen mass fraction is $X = 0.76$ and Hubble constant is $H_0 = 100\,h\,\text{km/s/Mpc}$. The wing cross-section depends only on the photon frequency, which in terms of velocity offsets becomes
\begin{equation}
  \sigma(\Delta v) \approx \frac{a \sigma_0}{\sqrt{\pi} x^2} = \frac{f_{12} c e^2 \Delta \nu_\text{L}}{2 m_e \nu_0^2 \Delta v^2} \, .
\end{equation}
However, cosmological redshifting continuously changes the frequency such that if we ignore peculiar motions then the relation $\nu(z) = \nu(z_s) (1 + z) / (1 + z_s)$ translates to offsets of
\begin{equation}
  \frac{\Delta v(z)}{c} = \frac{\nu_0 - \nu(z)}{\nu_0} = \frac{z_s - z + (1 + z) \Delta v/c}{1 + z_s} \, .
\end{equation}
The effective redshift corresponding to the end of the local rays is $z_{s,\text{eff}} = z_s - (1 + z_s) \Delta v_s / c$, where the ray length is characterized by the local velocity offset, which in our case is $\Delta v_s = 4000\,\text{km\,s}^{-1}$. Thus, the traversed optical depth in physical units is
\begin{equation} \label{eq:tau_DW}
  \tau_\text{DW} = \int_0^{z_{s,\text{eff}}} \langle n_\text{\HI} \rangle \sigma\left[\Delta v(z)\right] \frac{c\,\text{d}z}{(1+z) H(z)} \, .
\end{equation}
If the reionization history is a step function then exact analytical expressions can be found for both pure Lorentzian wings and with the common modification of an additional $(\nu / \nu_0)^4$ dependence appropriate for Rayleigh scattering \citep{Miralda-Escude1998}. However, we employ a numerical integration over the high redshift resolution ($\Delta z \lesssim 0.005$) reionization history directly from the simulations. Furthermore, we adopt the complete first-order quantum-mechanical correction to the Voigt profile presented by \citet{Lee2013},
\begin{equation} \label{eq:Lee}
  \sigma_\text{Lee}(\Delta v) \approx \sigma(\Delta v)\,\left( 1 + 1.792\,\Delta v / c \right) \, ,
\end{equation}
which strengthens the red wing due to positive interference of scattering from all other levels.

\begin{figure}
  \centering
  \includegraphics[width=\columnwidth]{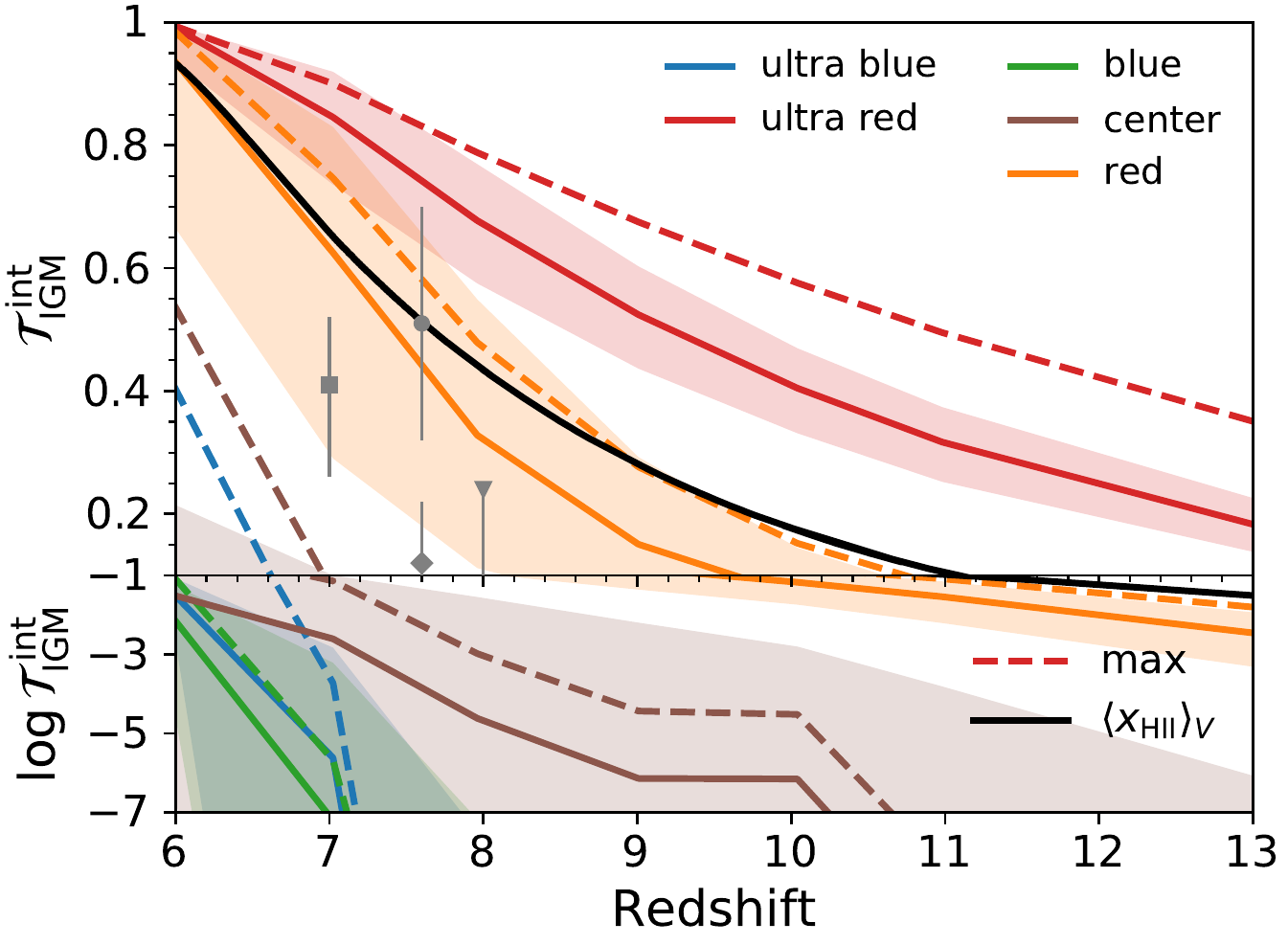}
  \caption{Evolution of the integrated IGM transmission $\mathcal{T}_\text{IGM}^\text{\,int}$ over five broad wavelength ranges from (ultra) blue to red (see the text for details). The median blue peak suppression remains strong until the universe is fully ionized, while red peaks are sensitive probes of the global reionization history (black curve). We also show the median maximum transmission spike in each spectral window yielding more optimistic prospects (dashed curves). For reference, we also include observational constraints from the detection of Ly$\alpha$ emission in Lyman break selected galaxies, which may exhibit properties of both the red and central bands (\citealp{Mason2018z7} -- square; \citealp{Mason2019} -- triangle; \citealp{Hoag2019} -- diamond; \citealp{Jung2020} -- circle).}
  \label{fig:T_IGM_z}
\end{figure}

\begin{figure*}
  \centering
  \includegraphics[width=\textwidth]{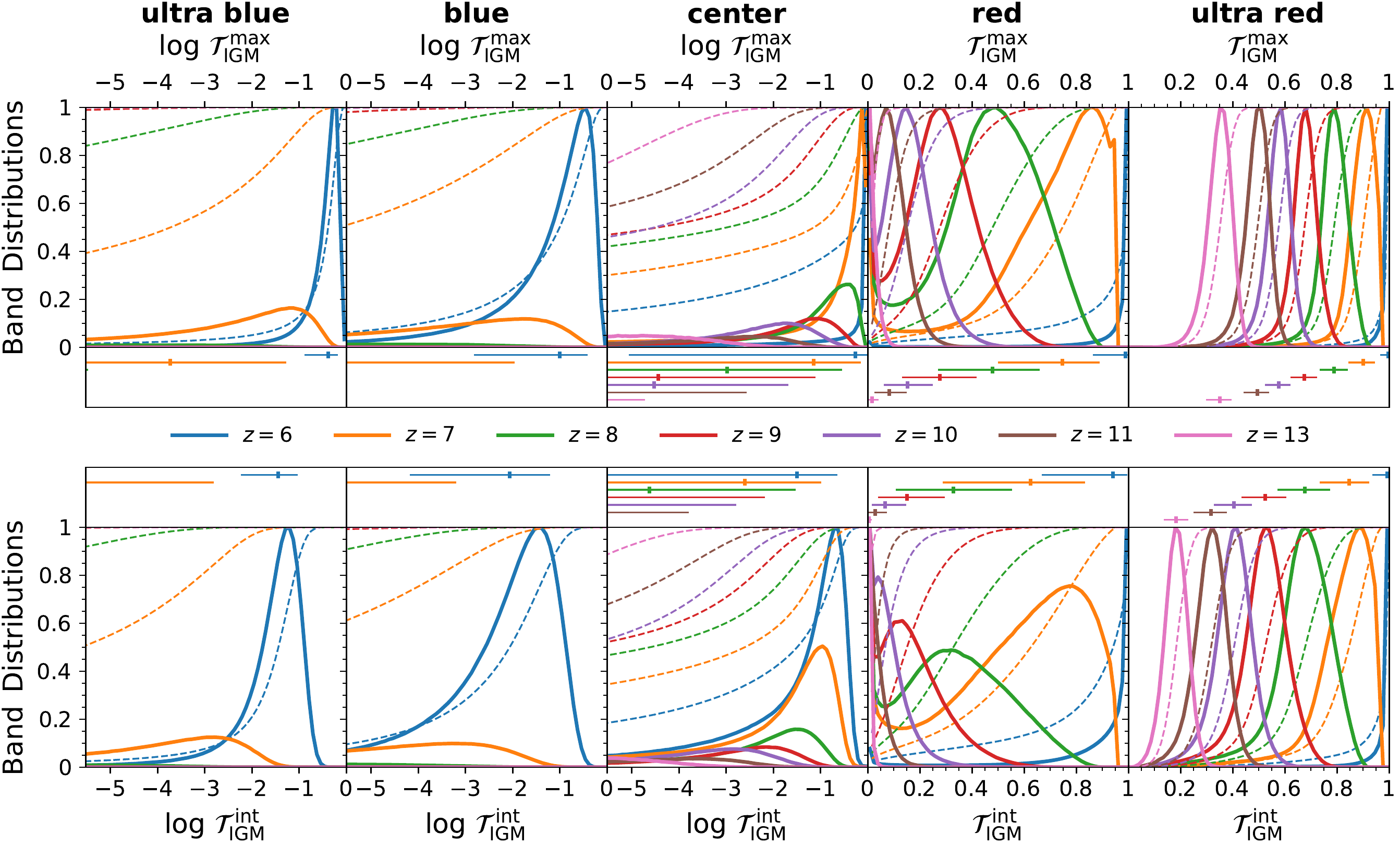}
  \caption{Relative probability distributions for a given integrated ($\mathcal{T}_\text{IGM}^\text{\,int}$, bottom panels) and maximum ($\mathcal{T}_\text{IGM}^\text{\,max}$, top panels) band transmission at each redshift (different colour curves) considering all haloes. For reference we also show the cumulative distribution functions as dashed curves and the median and $1\sigma$ summary statistics in the middle panels. See the text for further discussion, but this perspective reveals a complex landscape of broad, skewed, or bimodal distributions.}
  \label{fig:T_IGM_bands_both}
\end{figure*}

In Fig.~\ref{fig:DW_absorption} we show this transmission $\mathcal{T}_\text{IGM}^\text{DW} = \exp(-\tau_\text{DW})$ beyond the local rays as a function of velocity offset $\Delta v$ and rest-frame wavelength offset $\Delta \lambda$. In the idealized model given by equation~(\ref{eq:tau_DW}) the statistical absorption depends mainly on the redshift, reionization history, and local ray lengths. In detail, we expect a range of values due to variations in the neutral hydrogen density of individual light cones. We also warn that our assumption of a homogeneous Universe in the damping-wing calculation will result in too much absorption on average. While we plan to investigate this further in a future study, we expect our conservative estimates will be most uncertain around the midpoint of reionization. This is because the clumping factor in the IGM remains close to unity beforehand ($\mathcal{C}_{100} \approx 1$; see fig.~17 in Paper~I) while the damping-wing transmission saturates to unity afterwards ($\mathcal{T}_\text{IGM}^\text{DW} \approx 1$). Still, this large-scale contribution cannot be ignored at high redshifts and is therefore included in all of the analysis in this paper, including the quantum-mechanical correction in equation~(\ref{eq:Lee}). To explore where this extra opacity comes from we also plot the corresponding cumulative distribution functions against the traversed velocity offset. About half of the large-scale damping-wing scattering optical depth originates within an additional $\approx 3000\,\text{km\,s}^{-1}$. However, we emphasize that achieving per cent level accuracy for the EoR damping-wing opacity requires a light-cone procedure in simulations with distances of up to $\sim 10$ times longer than the detailed local calculations considered carried out in this study.

\subsection{Redshift dependence}
\label{sec:trans_z}
We first consider the redshift evolution of global catalogue statistics, in which we give equal weight to all galaxies and sightlines. In Fig.~\ref{fig:T_IGM_catalog} we show the IGM transmission $\mathcal{T}_\text{IGM}$ including the local and damping-wing contributions as a function of velocity offset $\Delta v$ and rest-frame wavelength offset $\Delta \lambda$ around the Ly$\alpha$ line at redshifts of $z = \{6,7,8,9,10,11,13\}$. The solid (dashed) curves show the full median (mean) statistics while the shaded regions give the $1\sigma$ confidence levels. We emphasize that this view is biased towards low mass haloes that dominate the number counts. Therefore, our focus is primarily to provide a reference comparison for similar works. Still, the full frequency dependence clearly and intuitively illustrates the blue peak suppression and red damping-wing absorption throughout the EoR. In the following subsections we explore the rich diversity of transmission properties in greater detail.

\begin{figure*}
  \centering
  \includegraphics[width=\textwidth]{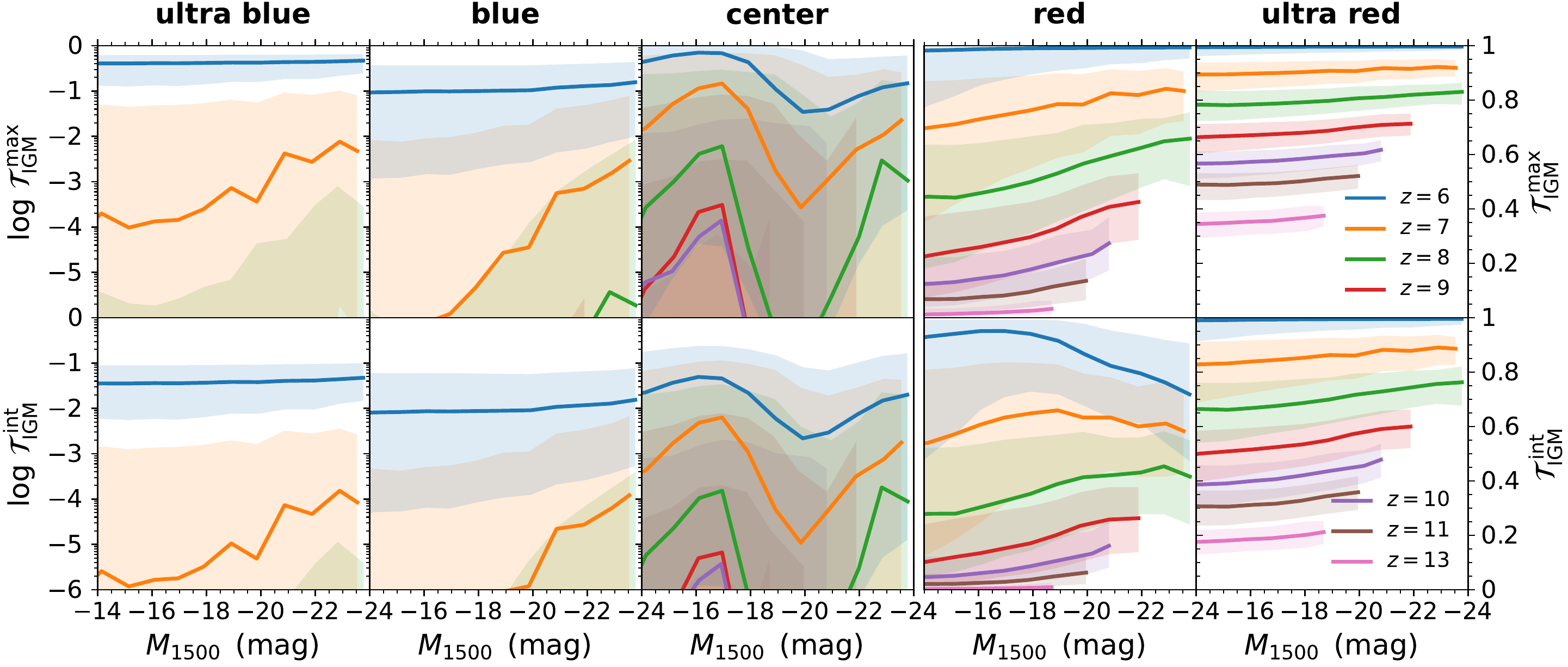}
  \caption{Band integrated ($\mathcal{T}_\text{IGM}^\text{\,int}$, bottom panels) and maximum ($\mathcal{T}_\text{IGM}^\text{\,max}$, top panels) IGM transmission median and $1\sigma$ statistics as a function of UV magnitude $M_{1500}$ for each redshift. The highly suppressed blue bands are shown with a log scale axis, while the red bands employ a linear axis. We find a clear trend of high transmission for UV bright galaxies, which is especially evident for $\mathcal{T}_\text{IGM}^\text{\,max}$. However, there is significant absorption that encroaches into the red band for $\mathcal{T}_\text{IGM}^\text{\,int}$ due to the complex distribution of high-velocity neutral gas structure around these galaxies.}
  \label{fig:T_IGM_M1500}
\end{figure*}

For a complementary perspective of the rapid change in transmissivity throughout the EoR we consider band averages. In particular, we focus on five velocity offset $\Delta v$ spectral windows that are relevant for Ly$\alpha$ related science. The `ultra blue' band of $(-2000, -500)\,\text{km\,s}^{-1}$ represents a baseline for absorption where most of the flux is free from the immediate proximity of the host halo before redshifting into resonance. The `blue' band of $(-500, -100)\,\text{km\,s}^{-1}$ is sensitive to both the local halo overdensity and the ionized bubble statistics. The `centre' band of $(-100, 100)\,\text{km\,s}^{-1}$ is meant to give wiggle room for systemic and peculiar velocities that blend the blue and red bands. The `red' band of $(100, 500)\,\text{km\,s}^{-1}$ is most relevant for LAE surveys as this corresponds to the location of observed red peaks at low and high redshifts \citep[e.g.][]{Ouchi2020}. The `ultra red' band of $(500, 2000)\,\text{km\,s}^{-1}$ is also mostly detached from proximity effects and thus reflects damping-wing absorption in the context of LSS and bubble size distributions.

To summarize the band properties we consider the following metrics taken over each wavelength range: first, the integrated or mean transmission defined as $\mathcal{T}_\text{IGM}^\text{\,int} \equiv \int \mathcal{T}_\text{IGM}\,\text{d}\Delta v / \int \text{d}\Delta v$, and secondly, the maximum transmission defined as $\mathcal{T}_\text{IGM}^\text{\,max} \equiv \max(\mathcal{T}_\text{IGM})$. In Fig.~\ref{fig:T_IGM_z} we show the redshift evolution of the integrated and maximum IGM transmission over each band. Although neither of these statistics are directly observable, as transmission also depends on the input Ly$\alpha$ spectra emerging from the galaxy, we expect these roughly trace what would be inferred by isolating IGM effects, especially for the non-saturated red and ultra red bands. We find that the median blue peak suppression remains strong until the universe is fully ionized, with slightly higher transmission for the ultra blue band compared to the blue band due to the reduced proximity effects. Overall, the maximum transmission statistics yield systematically more optimistic prospects for LAE detectability, especially if these transmission spikes coincide with emergent Ly$\alpha$ peaks (shown by the dashed curves). We find it encouraging that the red band seems to provide a sensitive probe of the global reionization history (shown by the black curve). The ultra red band does the same for sources in the damping wing such as faint quasars or gamma-ray burst afterglows found in futuristic high-redshift surveys \citep[e.g.][]{Lidz2021}.

A more equitable view of the catalogue is given by considering the distributions of band values across all haloes. In Fig.~\ref{fig:T_IGM_bands_both} we quantify the relative probability for a given integrated ($\mathcal{T}_\text{IGM}^\text{\,int}$, bottom panels) and maximum ($\mathcal{T}_\text{IGM}^\text{\,max}$, top panels) band transmission at each redshift. For reference we also show the cumulative distribution functions as dashed curves and the median and $1\sigma$ summary statistics in the middle panels. It is extremely unlikely to have non-negligible transmission spikes in the blue bands at $z \gtrsim 8$, although this is certainly allowed (int) or even common (max) at $z \lesssim 6$. When also considering the central band in rare cases it is plausible to witness multiple-peaked or otherwise complex spectral line profiles \citep[e.g. see the discussions by][]{Byrohl2020,Mason2020,Gronke2021,Park2021}. This also stresses the need for systemic tracers beyond Ly$\alpha$ to distinguish IGM signatures and pinpoint the origins of various spectral features. Perhaps more significant is the broad distribution in the red band. The transmission is relatively high after the midpoint of reionization $z \lesssim 7.67$, and roughly follows the global ionized fraction (see Fig.~\ref{fig:T_IGM_z}). However, as will be shown later the exponential sensitivity on optical depth ($\mathcal{T}_\text{IGM} = e^{-\tau_\text{IGM}}$) allows the same galaxy to have both large and small values, i.e. the bimodality is not due to halo mass or environment alone. In fact, there will always be sightlines with very low transmission due to filaments and other self-shielding structures common at high-$z$. The location and broad nature of the higher transmission peak is redshift dependent, which is an important consideration when using LAEs as probes of reionization. We emphasize that the sharp cutoffs below $\mathcal{T}_\text{IGM} \approx 1$ are not physical but are due to the global treatment of the long-range damping-wing absorption. For example, at $z = 7$ the red band values cannot exceed $\mathcal{T}_\text{IGM} \approx 0.95$ as every sightline on average experiences at least $5\%$ absorption via cosmological integration throughout the remainder of the EoR. Finally, we note that the ultra red band is less susceptible to resonant scattering and halo proximity effects and is therefore a cleaner probe of the global state of the IGM if this can be robustly measured with deep spectroscopy for large numbers of high-$z$ galaxies.

\subsection{Dependence on UV magnitude}
We now explore the dependence on UV magnitude (without any dust correction), which serves as an observational indicator for young stellar populations. In fact, UV bright galaxies are expected to have high ionizing photon budgets and therefore promote inside out reionization of their local bubbles. On the other hand, these same galaxies may give rise to relatively low and sightline-dependent Lyman continuum escape fractions. High dust contents, crowded environments, and cosmic streams of infalling gas introduce additional complexity that can offset the large bubble advantage for IGM transmission. In Fig.~\ref{fig:T_IGM_M1500} we explore the band integrated ($\mathcal{T}_\text{IGM}^\text{\,int}$, bottom panels) and maximum ($\mathcal{T}_\text{IGM}^\text{\,max}$, top panels) IGM transmission as a function of UV magnitude $M_{1500}$ for each redshift based on the median and $1\sigma$ statistics. We find that at $z \gtrsim 7$ the highly suppressed blue bands exhibit less transmission for fainter galaxies, although this seems to wash out by $z = 6$ at the tail end of reionization. However, the most interesting aspect of the UV dependence is in the red band, where the maximum statistic $\mathcal{T}_\text{IGM}^\text{\,max}$ clearly shows increasing transmissivity for brighter galaxies at all redshifts. This means that fainter galaxies are more likely to be universally suppressed across the entire red spectral window, or conversely that brighter galaxies have an advantage for transmission somewhere within $\le 500\,\text{km\,s}^{-1}$. We note that the effect is present but less dramatic for the ultra red band.

\begin{figure}
  \centering
  \includegraphics[width=\columnwidth]{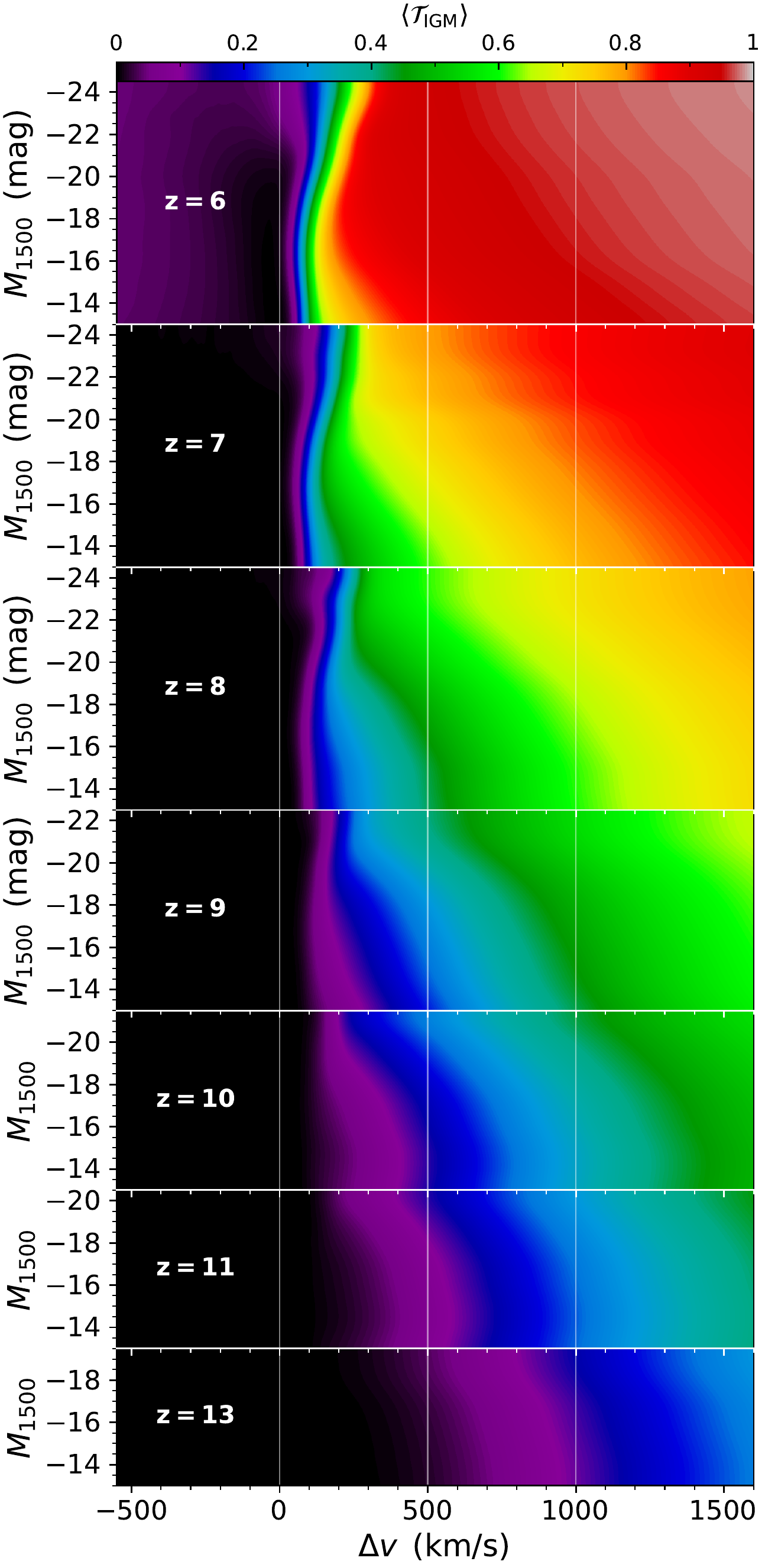}
  \caption{Average IGM transmission $\langle \mathcal{T}_\text{IGM} \rangle$ as a function of velocity offset $\Delta v$ and UV magnitude $M_{1500}$. This perspective illustrates the non-trivial dependence on both of these quantities. Brighter galaxies provide a clear advantage for red wing photons. At the same time, the sharp transition frequency can reach higher velocity offsets, undercutting IGM transparency for Ly$\alpha$ photons escaping near line centre in these environments.}
  \label{fig:M1500_Dv}
\end{figure}

Somewhat counter-intuitively, the integrated red band transmission $\mathcal{T}_\text{IGM}^\text{\,int}$ seems to level off or even dip for the brightest haloes. This can be understood in terms of a step function model in the frequency dependence with an increasing fraction of galaxies and sightlines being cut off above the lower edge of $100\,\text{km\,s}^{-1}$. The basic argument is that if a photon originates on the blue side of line centre then given the high Gunn--Peterson absorption it will be eliminated as soon as it is Hubble redshifted into resonance. However, if the peculiar gas velocity is infalling near the source then even photons redwards of the systemic line centre may be viewed as blue in the comoving frame of the gas. Thus, the frequency range likely to encounter a resonance point is extended to approximately the circular velocity, which is often used to estimate gravitational infall velocities \citep[for further discussion see e.g.][]{Santos2004,Dijkstra2007}:
\begin{equation}
  V_c = \sqrt{\frac{G M_\text{halo}}{R_\text{vir}}} \approx 144\,\text{km\,s}^{-1}\,\left(\frac{M_\text{halo}}{10^{11}\,\Msun}\right)^{1/3} \left(\frac{1 + z}{7}\right)^{1/2} \, ,
\end{equation}
assuming the \textit{Planck} cosmological parameters and a standardized overdensity constant of $\Delta_c = 200$. Still, despite this encroaching absorption feature, observational studies are consistent with no evolution in massive galaxies while faint ones exhibit a drop in the LAE population fraction \citep[e.g.][]{Stark2011,Endsley2021}. This is consistent with our results if bright galaxies have larger velocity offsets and broader line widths, as is expected based on lower redshift observations \citep{Yang2016,Verhamme2018} and a number of theoretical arguments in the literature (also recall Fig.~\ref{fig:pos_vel_sigma}).

Unfortunately, the \thesan\ volumes are not large enough to assess the statistical behavior of the brightest galaxies at $z \gtrsim 10$. In the pre-reionized Universe the ensemble average detectability of first galaxy LAEs may be discouragingly low as more and more photons are lost to the diffuse Ly$\alpha$ background \citep{LoebRybicki1999,Visbal2018}. Furthermore, even with the increased sensitivity of next-generation surveys it may be difficult to probe large enough volumes with sufficient depth to detect enough rare bright galaxies to discern between various IGM transmission models, which can differ by an order of magnitude in the predicted emission and transmissivity \citep{Smith2015,Smith2017}. However, the combination of a sharp frequency cutoff and gradual wing opacity is also what makes Ly$\alpha$ (non-)detections a rich probe of EoR physics. In summary, based on extrapolating the current trends there is room for cautious optimism in which fortunate sightlines, bubble sizes, and emergent line profiles facilitate minimal local and damping-wing absorption.

To further investigate this behavior, in Fig.~\ref{fig:M1500_Dv} we show the average IGM transmission $\langle \mathcal{T}_\text{IGM} \rangle$ as a function of velocity offset $\Delta v$ and UV magnitude $M_{1500}$. This image vividly explains the turnover in the red band relation at $M_{1500} \lesssim -20$ seen in Fig.~\ref{fig:T_IGM_M1500}. One of the key features is that brighter galaxies provide a clear advantage for red wing photons at all redshifts, due to residing within larger ionized bubbles. At the same time, the sharp transition frequency tends to curve towards higher velocity offsets with a large spread by $z = 6$. This encroachment into the red band acts to undercut IGM transparency for Ly$\alpha$ photons escaping near line centre in these environments. Another important feature is that (on average) blue photon transmission is permitted by $z = 6$. In fact, lower redshift studies also exhibit the pattern of higher absorption at line centre before transitioning to bluer wavelengths that are less susceptible to halo proximity effects \citep[see e.g.][]{Laursen2011}.

\subsection{Covering fractions}
\label{sec:covering_fractions}
\begin{figure}
  \centering
  \includegraphics[width=\columnwidth]{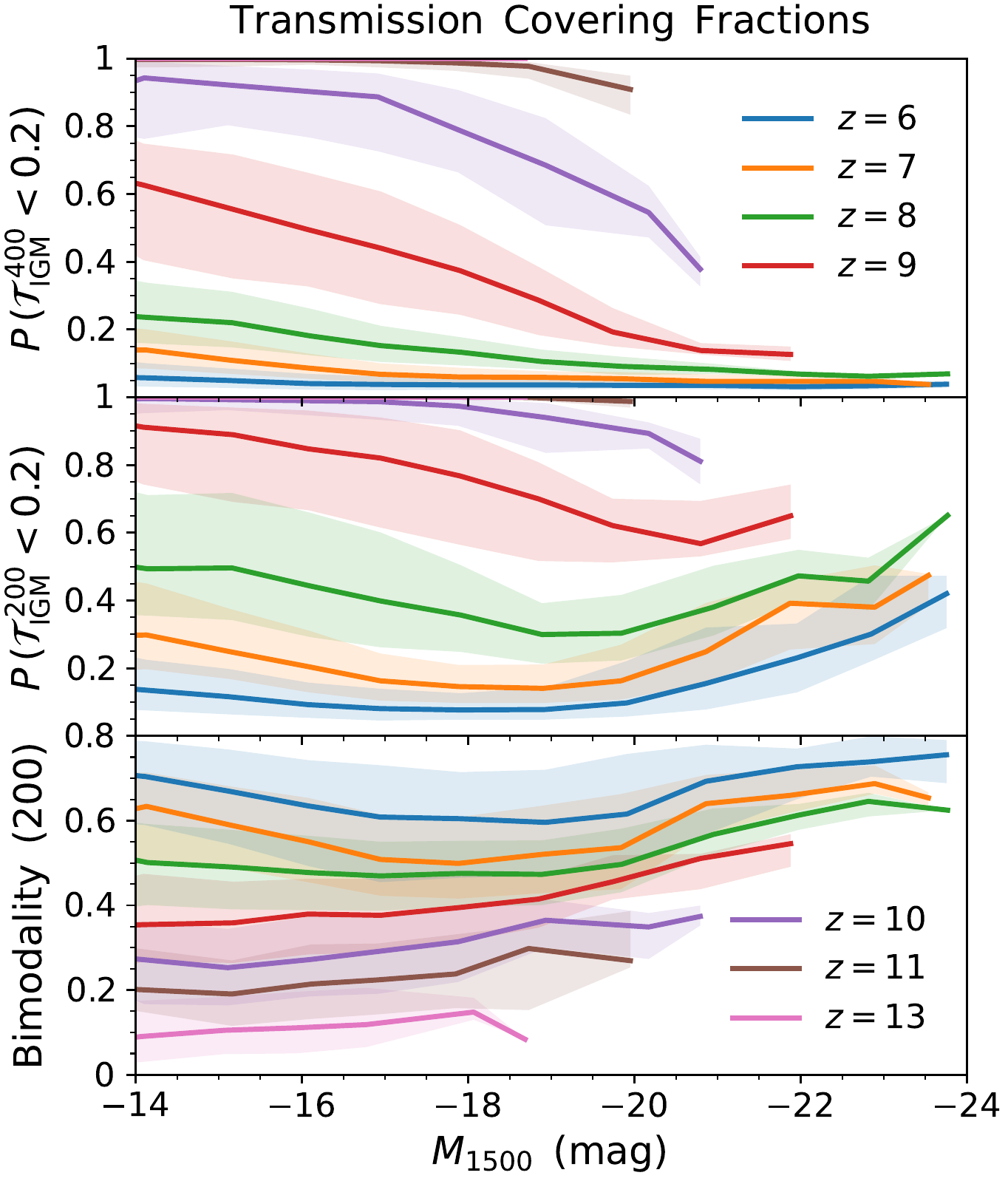}
  \caption{IGM transmission covering fractions defined as the fraction of sightlines around each galaxy with transmission below 20 per cent, i.e. $P(\mathcal{T}_\text{IGM} < 0.2)$. The curves and shaded regions give the median and $1\sigma$ variation as a function of UV magnitude $M_{1500}$ at velocity offsets of $\Delta v = 200\,\text{km\,s}^{-1}$ (top panel) and $\Delta v = 200\,\text{km\,s}^{-1}$ (middle panel). We also show a measure of bimodality to explore the increasing coverings around bright galaxies (bottom panel). See the text for explanations of various trends with redshift and brightness in terms of, e.g., anisotropy and environment.}
  \label{fig:convering_fractions_M1500}
\end{figure}

The concept of covering fractions provides a fundamentally different perspective on the mechanisms responsible for hampering Ly$\alpha$ visibility at the tail end of reionization. Intriguingly, we find that IGM transmission in wavelength ranges where we expect Ly$\alpha$ line emission can be highly anisotropic. In general, emission will be observed as close to line centre as allowed by the halo escape and IGM transmission physics. To quantify this effect, we define the covering fraction of each individual galaxy as the fraction of sightlines with transmission below 20 per cent, i.e. $P(\mathcal{T}_\text{IGM} < 0.2)$. In Fig.~\ref{fig:convering_fractions_M1500} we show the median and $1\sigma$ range of halo covering fractions as a function of UV magnitude $M_{1500}$ (upper panels). For comparison, we provide these statistics at the physically relevant velocity offsets of $\Delta v = 200\,\text{and}\,400\,\text{km\,s}^{-1}$, averaged over wavelength windows of $50\,\text{km\,s}^{-1}$ to match typical spectroscopic instrument capabilities. The velocity offset windows are chosen to capture the impact of IGM transmission on observed red peaks nearer and farther from line centre. The general trend is that fainter galaxies have larger covering fractions, corresponding to more isotropic suppression. Interestingly, there is also an upturn for bright galaxies ($M_{1500} \lesssim -20$) at $200\,\text{km\,s}^{-1}$ caused by anisotropic cold gas accretion around these clustered environments. As expected, this effect washes out by $400\,\text{km\,s}^{-1}$ where the red peak transmission is less affected by the local high velocity infall. Thus, there is a relatively large optimal $M_{1500}$ range for observing transmission. In Table~\ref{tab:covering} we summarize the redshift and frequency dependence of covering fractions by providing $P(\mathcal{T}_\text{IGM} < 0.2)$ including all galaxies with UV brightness $M_{1500} < -19$ for a grid of several observationally relevant velocity offsets, which also demonstrates the relative (in)sensitivity to the chosen velocity offset windows.

We find that these qualitative features are quite robust, although the details depend on the threshold criterion. The value of 20 per cent represents a substantial (but unsaturated) degree of absorption at $z \lesssim 9$. To better understand the impact of redshift evolution we also show a measure of bimodality at $200\,\text{km\,s}^{-1}$ (bottom panel). This calculation is based on a common test employing moments around the mean. Specifically, if the mean is $m_1 \equiv \langle \mathcal{T}_\text{IGM} \rangle$ and unstandardized moments are $m_k \equiv \langle (\mathcal{T}_\text{IGM} - m_1)^k \rangle$, then the standard deviation is $\sigma \equiv \sqrt{m_2}$, skewness is $\gamma \equiv m_3 / \sigma^3$, and kurtosis is $\kappa \equiv m_4 / \sigma^4$, such that the inverted bimodality coefficient is $b^{-1} = 1 / (\kappa - \gamma^2) \in [0, 1]$. We note that we choose to plot $b^{-1}$ as this more intuitively depicts higher bimodality with larger values. Our results demonstrate that brighter galaxies exhibit a more on/off absorption morphology in comparison to less bright galaxies ($M_{1500} \sim -18$). Furthermore, we also find the bimodality is much less significant at higher redshifts ($z \gtrsim 10$) as low opacity sightlines all but disappear. Of course, this result should not be overinterpreted as this does not imply bimodality in optical depths, but rather the presence of viewing angles with $\tau_\text{IGM}^{200}$ greater than and less than unity around the same halo.

\begin{figure*}
  \centering
  \includegraphics[width=\textwidth]{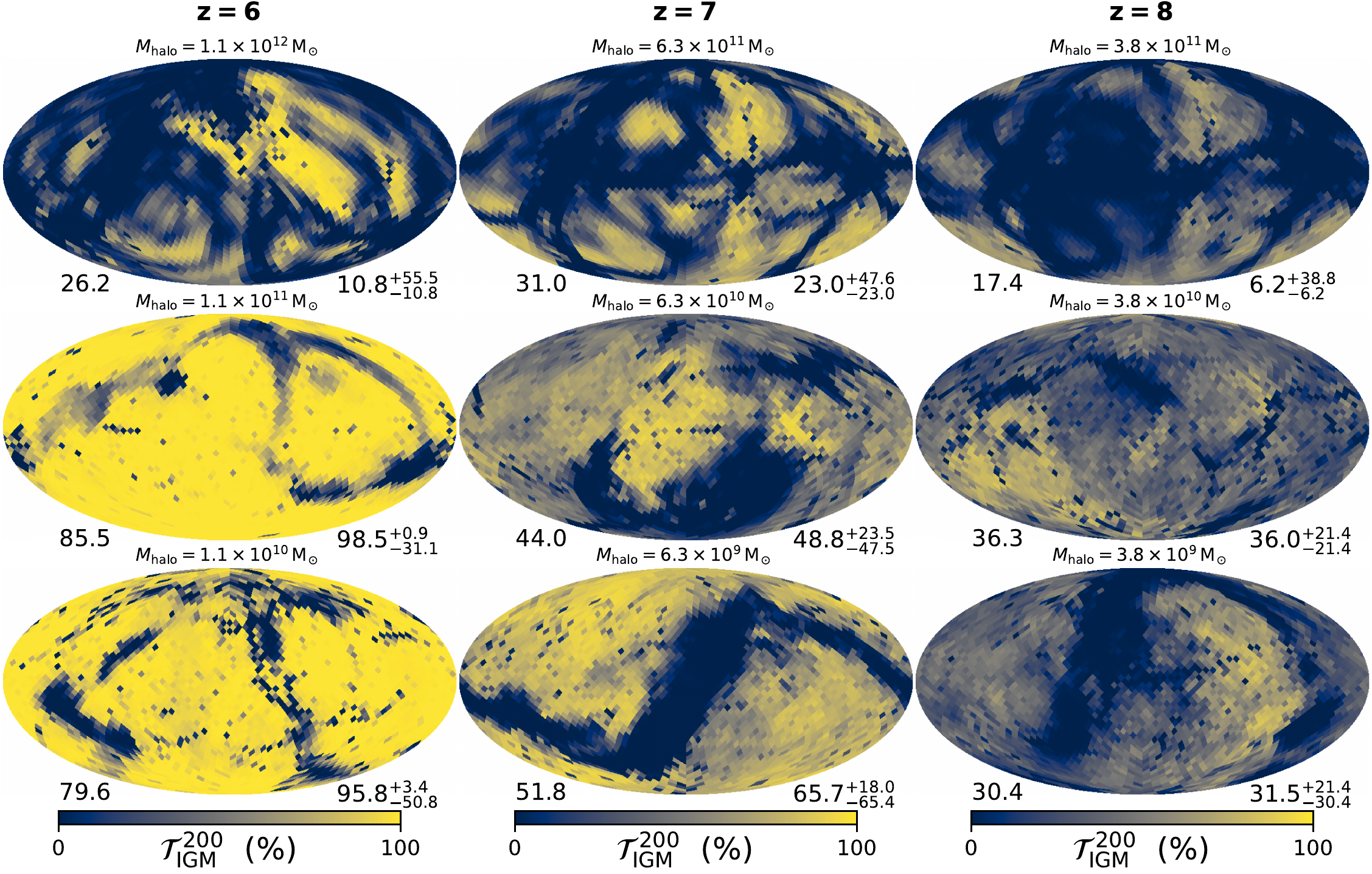}
  \caption{Example angular distributions of the IGM transmission at a velocity offset of $200\,\text{km\,s}^{-1}$, $\mathcal{T}_\text{IGM}^{\,200}$. We provide the mean and median statistics based on 3072 healpix directions of equal solid angle, included as values on the lower left and right of each map, respectively. We select the most massive haloes at $z = \{6, 7, 8\}$ as well as haloes with masses 10 and 100 times smaller. The wide variety of structures include smooth background fluctuations, continent-sized dimming and brightening, and interconnected node and filamentary features from a cosmic web of cold gas streams in projection. The most massive haloes have large covering fractions and dense knots of octopus-like absorption tracks.}
  \label{fig:healpix_200}
\end{figure*}

\begin{figure*}
  \centering
  \includegraphics[width=\textwidth]{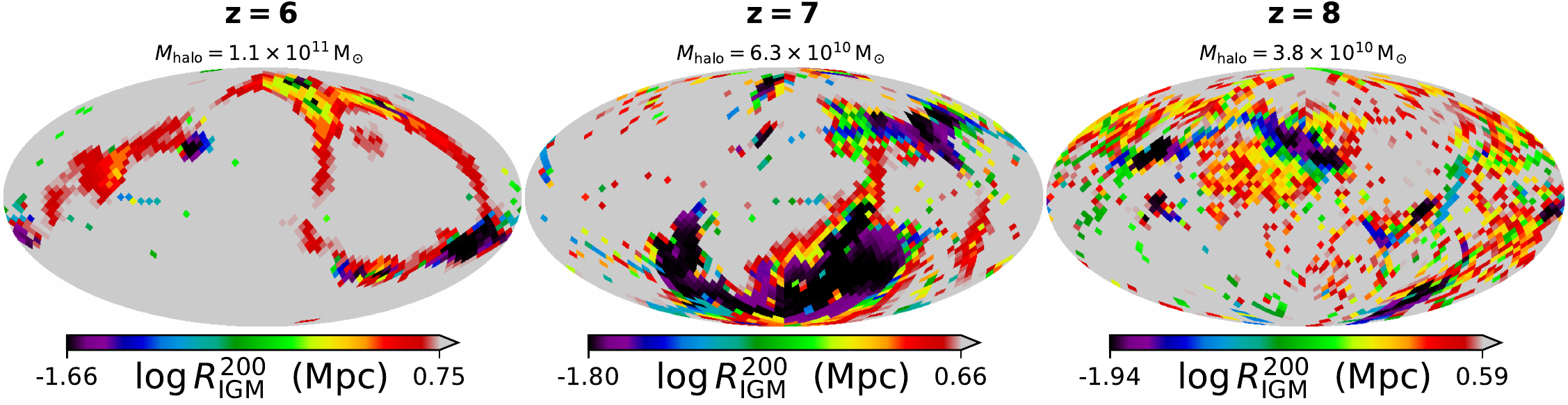}
  \caption{Angular distributions of the distance to the $\tau = 1$ scattering surface $R_\text{IGM}^{200}$ at a velocity offset of $200\,\text{km\,s}^{-1}$ for haloes corresponding to the middle row of Fig.~\ref{fig:healpix_200}. These images highlight that the obstruction morphology can be shaped by nearby gas but pushed over the threshold at cosmological scales ($\gtrsim 1\,\text{Mpc}$). The leftmost galaxy at $z = 6$ is also shown in Fig.~\ref{fig:rgb_int}, providing a visual connection between the angular and spatial representations.}
  \label{fig:healpix_Rtau1_200}
\end{figure*}

Finally, to more clearly visualize the covering fraction results, in Fig.~\ref{fig:healpix_200} we show the angular distributions of the IGM transmission $\mathcal{T}_\text{IGM}^{\,200}$ around a velocity offset of $200\,\text{km\,s}^{-1}$ for a prototypical subset of galaxies. We select the most massive haloes at $z = \{6, 7, 8\}$ as well as haloes with masses 10 and 100 times smaller as listed in the figure. We provide the mean (left-hand side) and median with $1\sigma$ confidence intervals (right-hand side) statistics below each image based on 3072 healpix directions of equal solid angle. The maps reveal a variety of structures including both easily understood and non-trivial dependence on redshift, halo mass, environment, and viewing angle. Broadly speaking the smooth backdrop in each image captures the inhomogeneous reionization process with large patches of dimming and brightening, likely due to moderate to large-scale bubble variations. On the other hand, nearby absorption features highlight shortcomings when separating ISM- and IGM-scale radiative transfer.

Beyond this there are relatively large and interconnected node and filamentary absorption features representing the cosmic web of cold gas streams in projection. It is clear that the environments around the most massive galaxies are significantly different than those of lower masses. In fact, the strong gravitational potential and crowded local volumes give rise to dense knots of octopus-like absorption tracks. Such structures are at the heart of the elevated covering fractions discussed above, and become essentially transparent for red wing photons ($\Delta v \gtrsim 400\,\text{km\,s}^{-1}$). We also observe punctuated extinction from small to moderate solid angle self-shielding clumps, corresponding to small self-shielded clumps and distant damped Lyman-alpha systems in the intervening IGM that persist for red wing photons too \citep[see the discussion by][]{Park2021}. To quantify this further, in Fig.~\ref{fig:healpix_Rtau1_200} we show the distances to the $\tau = 1$ scattering surface $R_\text{IGM}^{200}$ at a velocity offset of $200\,\text{km\,s}^{-1}$ for haloes corresponding to the middle row of Fig.~\ref{fig:healpix_200}. These images highlight that the obstruction morphology can be shaped by nearby neutral gas but pushed over the opacity threshold at cosmological scales ($\gtrsim 1\,\text{Mpc}$).

\begin{figure}
  \centering
  \includegraphics[width=.825\columnwidth]{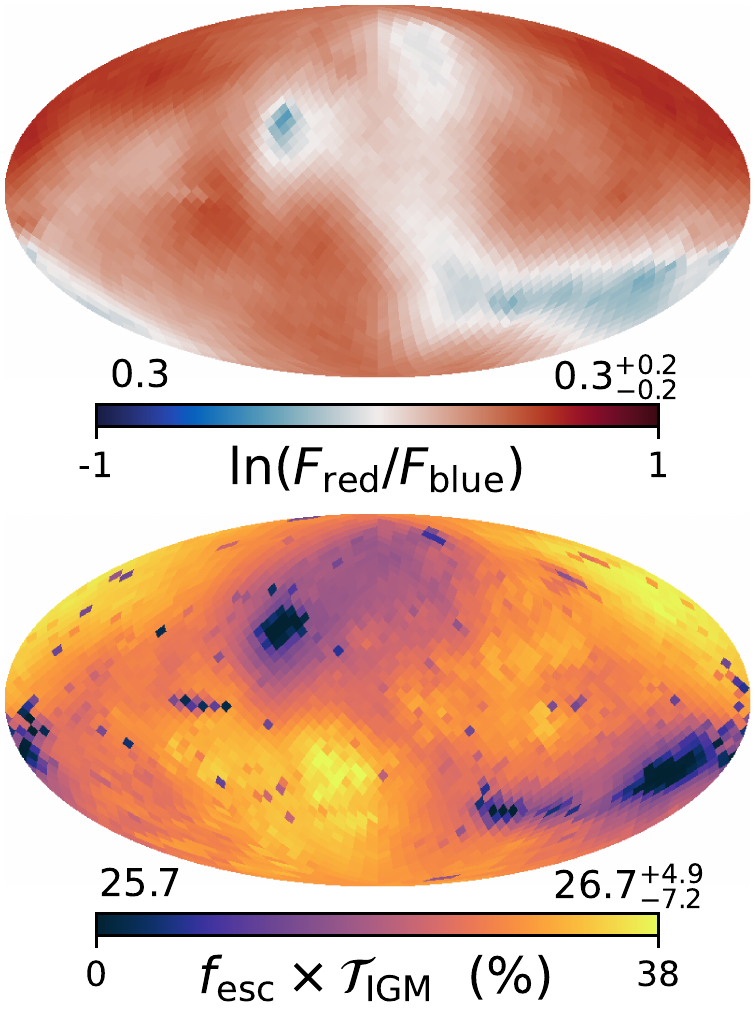}
  \caption{Angular distributions of the red-to-blue flux ratio $F_\text{red} / F_\text{blue}$ (top) and fraction of observed flux (bottom) taken as the product of the dust escape fraction $f_\text{esc}$ and IGM transmission $\mathcal{T}_\text{IGM}$ of the emergent spectra based on Monte Carlo Ly$\alpha$ radiative transfer calculations. This is the same $z = 6$ galaxy ($M_\text{halo} \approx 10^{11}\,\Msun$) as in Figs.~\ref{fig:rgb_int}, \ref{fig:healpix_200}, and \ref{fig:healpix_Rtau1_200} for the wind model, and demonstrates a clear connection between the spectral behaviour and observed flux connecting ISM- and IGM-scale radiative transfer.}
  \label{fig:healpix_esc}
\end{figure}

A proper treatment incorporating radiative transfer effects down to ISM and CGM scales is expected to smooth out some of these features while enhancing others. For example, clumpy media has been shown to produce directional Ly$\alpha$ equivalent width boosting, which vanishes for spatially extended sources once the projected shadow sizes for emitting regions exceed the mean cloud separation distances \citep{GronkeDijkstra2014}. Overall, Ly$\alpha$ resonant scattering tends to reduce the flux anisotropy compared to continuum radiation, but even directional dependence favouring blue or red photons can provide non-trivial variations when accounting for IGM transmission. Similar signatures are found in the viewing angle dependence of high-resolution cosmological Ly$\alpha$ radiative transfer simulations \citep{Behrens2019,Smith2019,Kimock2021,Mitchell2021}. The galaxy and IGM directional effects are certainly correlated to some degree, but it is standard to conceptually include them independently \citep{Laursen2011}. Ultimately, incorporating IGM transmission for individual LAEs is most accurately captured by extending resonant scattering calculations to well beyond the virial radius, especially at increasingly high redshifts.
In fact, nearly saturated IGM transmission covering fractions, i.e. $P(\mathcal{T}_\text{IGM} < 0.2) \approx 1$, are indicative that scattering back into the line of sight is increasingly important, implying that our results are more affected by not taking radiative transfer into account at $z = 10$ compared to $z = 6$. In Fig.~\ref{fig:healpix_esc} we come full circle with the same example $z = 6$ galaxy from Fig.~\ref{fig:rgb_int} by showing the angular distributions of the red-to-blue flux ratio $F_\text{red} / F_\text{blue}$ and the fraction of observed flux taken as the product of the dust escape fraction $f_\text{esc}$ and IGM transmission $\mathcal{T}_\text{IGM}$ of the emergent spectra. As expected, there is a clear correlation in the structure of the spectral behaviour and observed flux as a result of connecting the ISM- and IGM-scale radiative transfer. For simplicity, we only show results for the wind model which produces more realistic enhanced red peaks than the fiducial model. For quantitative comparison the fraction of flux emerging on the red side of line centre is $F_\text{red} / F_\text{tot} \approx 57.4^{+3.9}_{-6.1} (11.3^{+7.5}_{-5.0})$, the dust escape fraction is $f_\text{esc} \approx 46.8^{+5.3}_{-8.6} (58.1^{24.3}_{13.6})$, and the final observed fraction is $f_\text{esc} \times \mathcal{T}_\text{IGM} \approx 26.7^{+4.9}_{-7.2} (7.3^{+3.8}_{-3.4})$ for the wind (fiducial) models, respectively. Overall, this confirms that Ly$\alpha$ radiative transfer effects are important for the observability of high-$z$ LAEs.

\begin{figure}
  \centering
  \includegraphics[width=\columnwidth]{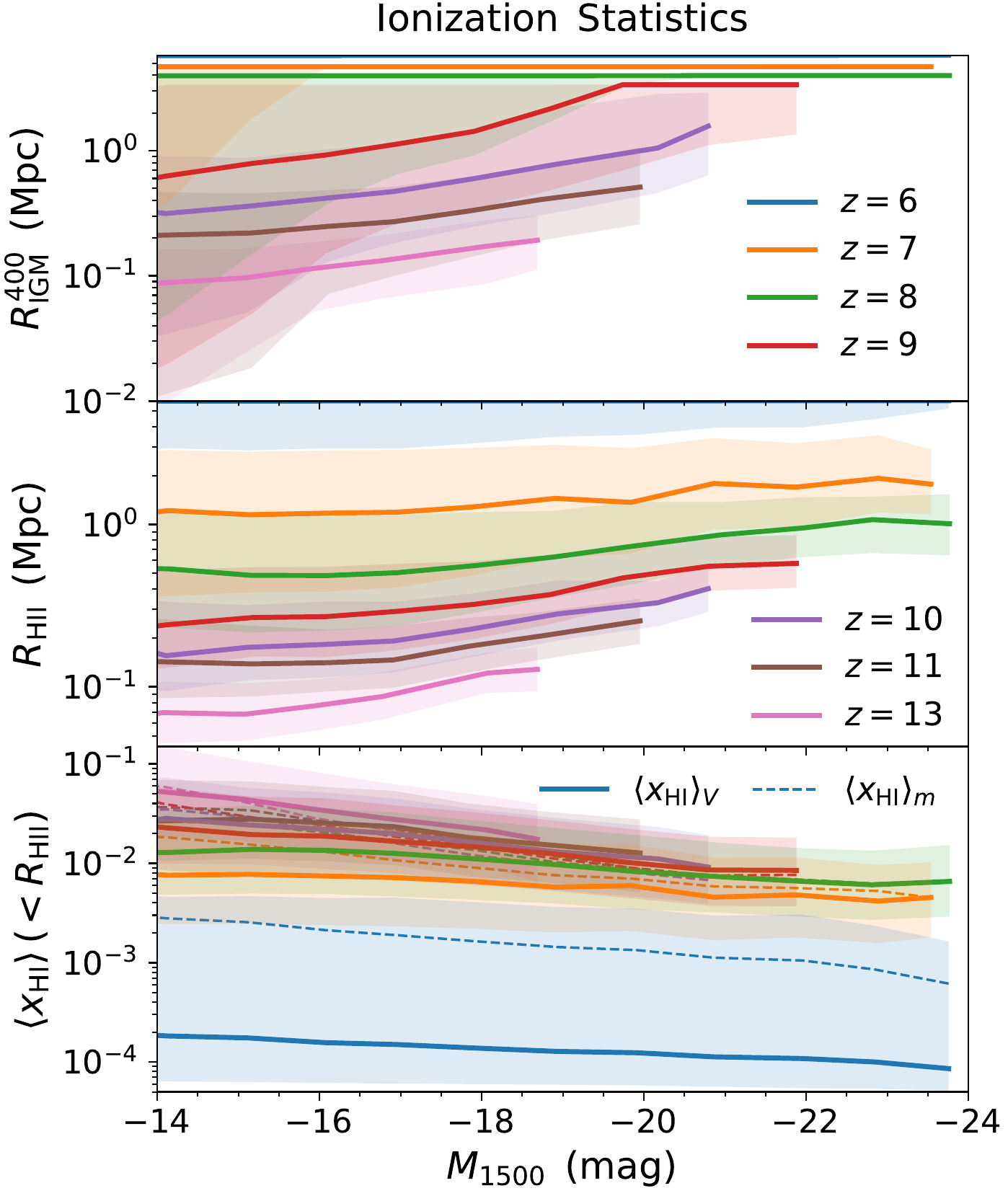}
  \caption{\textit{Top:} Distance to the $\tau = 1$ scattering surface $R_\text{IGM}^{400}$ (physical Mpc) at a velocity offset of $400\,\text{km\,s}^{-1}$ as a function of UV magnitude $M_{1500}$. The curves and shaded regions give the median and $1\sigma$ variation. \textit{Middle:} Size of the local ionized bubble $R_\text{\HII}$ defined as the line-of-sight distance where the neutral fraction exceeds $x_\text{\HI} > 0.9$, ignoring gas within the virial radius. \textit{Bottom:} Average neutral fraction along each ray within the local ionized bubble $\langle x_\text{HI} \rangle (<R_\text{\HII})$, with volume- and mass-weighted averages shown as solid and dashed curves, respectively. Overall, red damping-wing absorption distances generally exceed the local bubble sizes, and that brighter galaxies produce larger ionized bubbles and lower residual neutral fractions.}
  \label{fig:R_HII_M1500}
\end{figure}

\subsection{Ionization statistics}
\label{sec:ionization_statistics}
In Fig.~\ref{fig:R_HII_M1500}, we show several ionization properties for each central galaxy in our catalogue. To better connect the transmission statistics to the large-scale reionization morphology (beyond the example angular distributions), from top to bottom we plot the line-of-sight distance to the $\tau = 1$ scattering surface $R_\text{IGM}^{400}$ at a velocity offset of $400\,\text{km\,s}^{-1}$, the effective size of the local ionized bubble $R_\text{\HII}$, and the neutral fraction within this radius $\langle x_\text{HI} \rangle (<R_\text{\HII})$, all as a function of the intrinsic UV magnitude $M_{1500}$. Specifically, $R_\text{\HII}$ is calculated as the nearest distance along each ray where the neutral fraction exceeds $x_\text{\HI} > 0.9$, i.e. the closest Lyman-limit system apart from gas within the virial radius. As before, the curves and shaded regions give the median and $1\sigma$ variation, which we emphasize includes 768 measurements per halo as incorporating the sightline variance is important for connecting back to observations. For the neutral fraction we show both volume- and mass-weighted averages represented by solid and dashed curves, respectively. Together $R_\text{\HII}$ and $\langle x_\text{HI} \rangle$ reveal clear trends that the line-of-sight environments around brighter galaxies exhibit larger ionized zones and lower residual neutral fractions than fainter ones, in agreement with observed discoveries of very large ionized bubbles \citep[e.g.][]{Jung2020,Endsley2022}. These results also demonstrate that red damping-wing absorption distances $R_\text{IGM}^{400}$ generally exceed the local bubble sizes by up to a factor of a few, although this is no longer the case near line centre where photons experience resonant absorption within ionized gas. This starts to be significant by $200\,\text{km\,s}^{-1}$ around bright, massive haloes with strong infall motions. We also note that at lower redshifts $R_\text{IGM}$ saturates such that the IGM is optically thin as bubbles become large enough for the Hubble flow to efficiently transmit red peaks to the observer. Thus, the visibility of a Ly$\alpha$ emitter is given mainly by the redshift according to the local and global progress of reionization in terms of bubble sizes and residual neutral fractions. Beyond this, the location of absorption can be strongly affected according to the covering fraction of nearby infalling or self-shielding systems.

\begin{table}
  \centering
  \caption{\textit{Top:} Ly$\alpha$ IGM transmission $\mathcal{T}_\text{IGM}$ tabulated as functions of redshift and frequency, averaged over wavelength windows of $50\,\text{km\,s}^{-1}$ to match observations. The summary statistics are calculated including all central galaxies with UV brightness $M_{1500} < -19$ and are given in per cent units with median and asymmetric $1\sigma$ confidence regions. \textit{Bottom:} Ly$\alpha$ transmission covering fractions on the same grid defined as the fraction of sightlines for each galaxy with transmission below 20 per cent, i.e. $P(\mathcal{T}_\text{IGM} < 0.2)$.}
  \label{tab:covering}
  \addtolength{\tabcolsep}{-1.5pt}
  \renewcommand{\arraystretch}{1.1}
  \begin{tabular}{ccccc}
  \hline
  $\mathcal{T}_\text{IGM}$ & $200\,\text{km\,s}^{-1}$ & $300\,\text{km\,s}^{-1}$ & $400\,\text{km\,s}^{-1}$ & $500\,\text{km\,s}^{-1}$ \\
  \hline
  $z = 6$ & $95.2^{+4.1}_{-66.6}$ & $98.7^{+0.8}_{-13.0}$ & $99.1^{+0.4}_{-8.6}$ & $99.2^{+0.3}_{-7.0}$ \vspace{.1cm} \\
  $z = 7$ & $61.5^{+22.4}_{-51.6}$ & $74.1^{+14.1}_{-27.4}$ & $77.2^{+12.1}_{-21.8}$ & $79.2^{+10.8}_{-18.4}$ \vspace{.1cm} \\
  $z = 8$ & $32.8^{+23.7}_{-30.2}$ & $46.2^{+18.4}_{-23.8}$ & $51.5^{+16.1}_{-20.1}$ & $55.5^{+14.3}_{-17.4}$ \vspace{.1cm} \\
  $z = 9$ & $15.0^{+16.4}_{-14.2}$ & $25.5^{+14.6}_{-16.0}$ & $31.2^{+13.3}_{-15.0}$ & $36.0^{+12.0}_{-13.9}$ \vspace{.1cm} \\
  $z = 10$ & $6.5^{+10.3}_{-6.0}$ & $13.1^{+10.7}_{-8.8}$ & $17.9^{+10.3}_{-9.6}$ & $22.4^{+9.8}_{-9.8}$ \vspace{.1cm} \\
  $z = 11$ & $2.4^{+5.6}_{-2.3}$ & $5.8^{+6.9}_{-4.6}$ & $9.4^{+7.4}_{-5.9}$ & $13.1^{+7.5}_{-6.8}$ \\
  \hline
  $P(\mathcal{T}_\text{IGM} < 0.2)$ & $200\,\text{km\,s}^{-1}$ & $300\,\text{km\,s}^{-1}$ & $400\,\text{km\,s}^{-1}$ & $500\,\text{km\,s}^{-1}$ \\
  \hline
  $z = 6$ & $10.1^{+14.8}_{-4.5}$ & $5.1^{+2.5}_{-1.7}$ & $3.5^{+1.6}_{-1.2}$ & $2.6^{+1.2}_{-0.9}$ \vspace{.1cm} \\
  $z = 7$ & $17.0^{+12.7}_{-6.1}$ & $8.1^{+2.6}_{-2.2}$ & $5.6^{+2.0}_{-1.6}$ & $4.0^{+1.4}_{-1.2}$ \vspace{.1cm} \\
  $z = 8$ & $30.9^{+11.7}_{-8.2}$ & $14.1^{+4.0}_{-3.5}$ & $9.2^{+2.7}_{-2.3}$ & $6.5^{+2.1}_{-1.6}$ \vspace{.1cm} \\
  $z = 9$ & $63.4^{+9.7}_{-11.9}$ & $34.4^{+14.2}_{-11.2}$ & $21.6^{+8.3}_{-7.3}$ & $12.8^{+4.3}_{-4.0}$ \vspace{.1cm} \\
  $z = 10$ & $91.5^{+5.5}_{-10.4}$ & $77.9^{+10.0}_{-17.0}$ & $60.7^{+11.4}_{-16.6}$ & $39.3^{+13.7}_{-11.2}$ \vspace{.1cm} \\
  $z = 11$ & $99.4^{+0.5}_{-1.7}$ & $97.6^{+1.5}_{-3.7}$ & $94.2^{+1.8}_{-8.2}$ & $83.4^{+5.1}_{-6.3}$ \\
  \hline
  \end{tabular}
  \addtolength{\tabcolsep}{1.5pt}
  \renewcommand{\arraystretch}{0.9090909090909090909}
\end{table}

\section{Summary and Discussion}
\label{sec:summary}
In this paper, we have constructed catalogues for Ly$\alpha$ emission and IGM transmission for the flagship \thesan\ simulation. In a forthcoming study, we will also present a detailed comparison of the Ly$\alpha$ properties from the remaining suite of lower resolution simulations (presented in Paper~I), including reionization histories affected by halo mass-dependent escape fractions, alternative dark matter, and numerical convergence. In a further study, we will also make predictions for Ly$\alpha$ intensity mapping and other diffuse cosmological radiative transfer statistics. This will utilize the high time cadence on-the-fly redshift-space renderings of Ly$\alpha$ properties to conveniently map out the ensemble radiation field throughout the EoR. We also leave empirical modelling of LAEs to a future study, e.g. constraining idealized galaxy source parameters such as spectral profiles and Ly$\alpha$ escape fractions \citep[as in][]{Jensen2013,Weinberger2019,Gangolli2021}. None the less, our initial explorations presented herein showcase the strengths and weaknesses of pursuing LAE science from state-of-the-art large-volume cosmological reionization simulations. In fact, by augmenting the IllustrisTNG galaxy formation model with fully coupled radiation hydrodynamics and dust physics, \thesan\ already incorporates most of the essential ingredients for realistic and representative simulation-based Ly$\alpha$ surveys. Thus, we welcome collaborations on related science topics or general utilization of Ly$\alpha$-centric catalogues following the upcoming public data release.

One of the main drawbacks is the sub-resolution treatment of the ISM as a two-phase gas \citep{Springel2003}. The predictive power of Ly$\alpha$ radiative transfer calculations based on simulations using the SH03 model without significant modification, and especially in combination with a low escape fraction of ionizing photons from the birth cloud, is reduced. However, we are pursuing a variety of approaches to more closely relate Ly$\alpha$ emission, absorption, and scattering from ISM to IGM scales. Ultimately, the \thesan\ project will include high-resolution zoom-in resimulations of a wide range of galaxies including a multiphase ISM framework \citep{Marinacci2019,Kannan2020} and self-consistent meso-scale reionization environment inherited directly from the flagship simulation. Beyond this, we emphasize that statistical IGM transmission studies also suffer from stitching and resolution effects when attempting to connect to Ly$\alpha$ emission sources \citep[e.g. see][]{Gronke2021}. However, while the emergent spectra are certainly tied to the local ($\lesssim 1\,\text{Mpc}$) radiation transport via environmental and directional biases, the more distant absorption features may be understood as a combination of redshift evolution, UV brightness, and stochastic covering fractions. Different sightlines from the same galaxy may correspond to underdense regions or self-shielded cosmological filaments, with the exponential sensitivity ($\mathcal{T}_\text{IGM} = e^{-\tau_\text{IGM}}$) acting to increase the variation and bimodality in IGM transmissivity. As a result, LAE surveys will require large sample sizes to get a handle on the statistics \citep{Park2021}.

The \thesan\ simulations confirm that large ionized bubbles form around the brightest galaxies at any epoch. Such accelerated reionization regions provide an advantage for Ly$\alpha$ transmission of red peaks, thereby boosting the visibility of LAEs around UV bright galaxies \citep{Mason2018}. However, it is also clear that infall motion plays a significant role in shaping the IGM transmissivity \citep{Santos2004,Dijkstra2007,Sadoun2017,Park2021}. For example, the truncation frequency of the transmission curve is typically set by the maximum infall velocity along the line of sight, which roughly corresponds to the circular velocity of the galaxy $V_c = \sqrt{G M_\text{halo} / R_\text{vir}}$. Of course, ISM and galaxy scale bulk motions set the intrinsic line profile prior to resonant scattering, and the emergent Ly$\alpha$ spectra from the galaxy is modulated by the 3D geometry of the neutral hydrogen and dust distributions \citep{Verhamme2018,Smith2019}. Thus, while infall motions and bubble sizes are crucial for IGM transmission, it is necessary to combine Ly$\alpha$ radiative transfer modelling with IGM reprocessing for realistic luminosity functions \citep[e.g. as in][]{Garel2021}.

Of course, we do not expect a clean separation of ISM and IGM scale Ly$\alpha$ radiative transfer effects into the EoR as these are increasingly coupled for higher redshift galaxies. For example, it is well understood that resonant scattering leads to ubiquitous extended Ly$\alpha$ haloes \citep[e.g.][]{Wisotzki2018}, which only become more exaggerated as the global neutral hydrogen density increases \citep{LoebRybicki1999}. Still, Ly$\alpha$ signatures are shaped from small to large scales with significant correlated physics along the way. For example, feedback can drive outflows through low column density channels that induce redshifting and boost the transmission through ionized windows between clusters of bright galaxies, or vice versa. On the other hand, serendipitous and deleterious circumstances add significant scatter, but sufficient statistics and complementary probes such as non-resonant lines should help distinguish between physics and random processes. Such richness and complexity calls for rigorous theoretical efforts to accurately interpret results from current and upcoming LAE surveys. With this outlook, the \thesan\ project offers a unique framework for studying high-redshift galaxies and the impact of cosmic reionization.

\section*{Acknowledgements}
We thank the referee for constructive comments and suggestions which have improved the quality of this work.
We thank Hyunbae Park, Ewald Puchwein, Sandro Tacchella, Hui Li, and Paul Torrey for insightful discussions related to this work.
AS acknowledges support for Program number \textit{HST}-HF2-51421.001-A provided by NASA through a grant from the Space Telescope Science Institute, which is operated by the Association of Universities for Research in Astronomy, incorporated, under NASA contract NAS5-26555. MV acknowledges support through NASA ATP grants 16-ATP16-0167, 19-ATP19-0019, 19-ATP19-0020, 19-ATP19-0167, and NSF grants AST-1814053, AST-1814259, AST-1909831 and AST-2007355. The authors gratefully acknowledge the Gauss Centre for Supercomputing e.V. (\url{www.gauss-centre.eu}) for funding this project by providing computing time on the GCS Supercomputer SuperMUC-NG at Leibniz Supercomputing Centre (\url{www.lrz.de}). Additional computing resources were provided by the Extreme Science and Engineering Discovery Environment (XSEDE), at Stampede2 through allocation TG-AST200007  and by the NASA High-End Computing (HEC) Program through the NASA Advanced Supercomputing (NAS) Division at Ames Research Center. We are thankful to the community developing and maintaining software packages extensively used in our work, namely: \texttt{matplotlib} \citep{matplotlib}, \texttt{numpy} \citep{numpy} and \texttt{scipy} \citep{scipy}.

\section*{Data Availability}
All simulation data, including snapshots, group and subhalo catalogues (with Ly$\alpha$-centric data), merger trees, and high time cadence Cartesian outputs will be made publicly available in the near future. Data will be distributed via \url{www.thesan-project.com}. Before the public data release, data underlying this article will be shared on reasonable request to the corresponding author(s).

\bibliographystyle{mnras}
\bibliography{biblio}

\appendix

\section{Continuum flux extrapolation}
\label{appendix:UV_slope}
\begin{figure}
  \centering
  \includegraphics[width=\columnwidth]{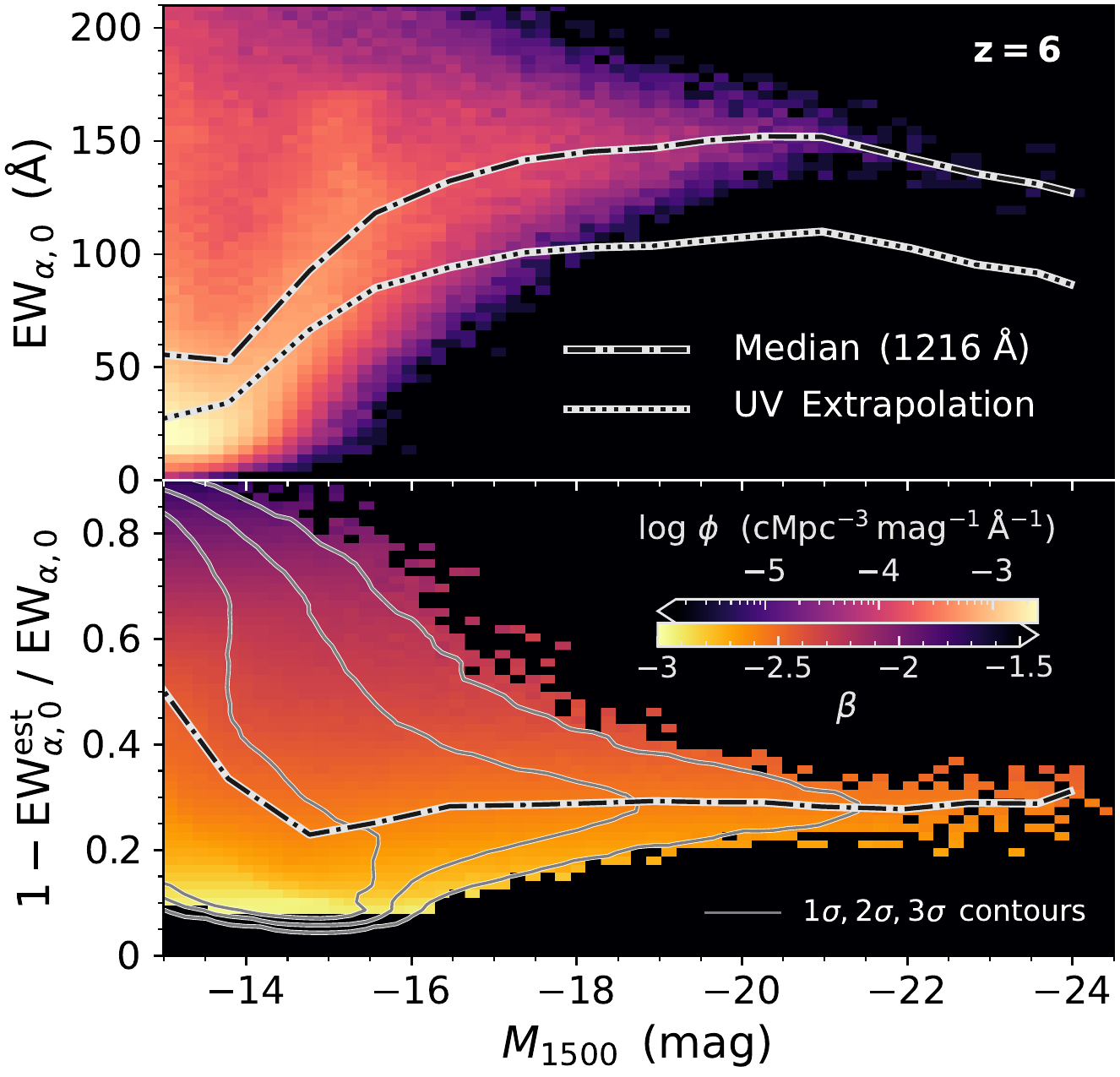}
  \caption{\textit{Upper panel:} Rest-frame Ly$\alpha$ equivalent widths $\text{EW}_{\alpha,0}$ as a function of UV magnitude $M_{1500}$ for each halo at $z = 6$. The colour axis gives the halo number density while the curves show the median for the true and extrapolated values. \textit{Lower panel:} Relative difference in the estimated equivalent widths with the colour denoting the UV slopes $\beta$. Estimates based on extrapolations of the intrinsic UV slope systematically underpredict intrinsic Ly$\alpha$ equivalent widths by approximately 30 per cent.}
  \label{fig:EW}
\end{figure}

We briefly explore the difference between the true and extrapolated Ly$\alpha$ continuum flux levels. This has important consequences when estimating Ly$\alpha$ equivalent widths from both simulated and observed data. Specifically, in Fig.~\ref{fig:EW} we show rest-frame Ly$\alpha$ equivalent widths $\text{EW}_{\alpha,0}$ as a function of UV magnitude $M_{1500}$ for each halo at $z = 6$. We find that estimates based on extrapolations of the intrinsic UV slope systematically underpredict intrinsic Ly$\alpha$ equivalent widths by approximately 30 per cent. We expect a similar degree of uncertainty to remain or be enhanced after dust scattering and absorption. Thus, while a proper radiative transfer treatment is necessary to assess this fully, we caution that extrapolations in general likely incur biases for Ly$\alpha$ equivalent width measurements and predictions.

\section{Initial integration radius}
\label{appendix:initial_radius}
\begin{figure}
  \centering
  \includegraphics[width=\columnwidth]{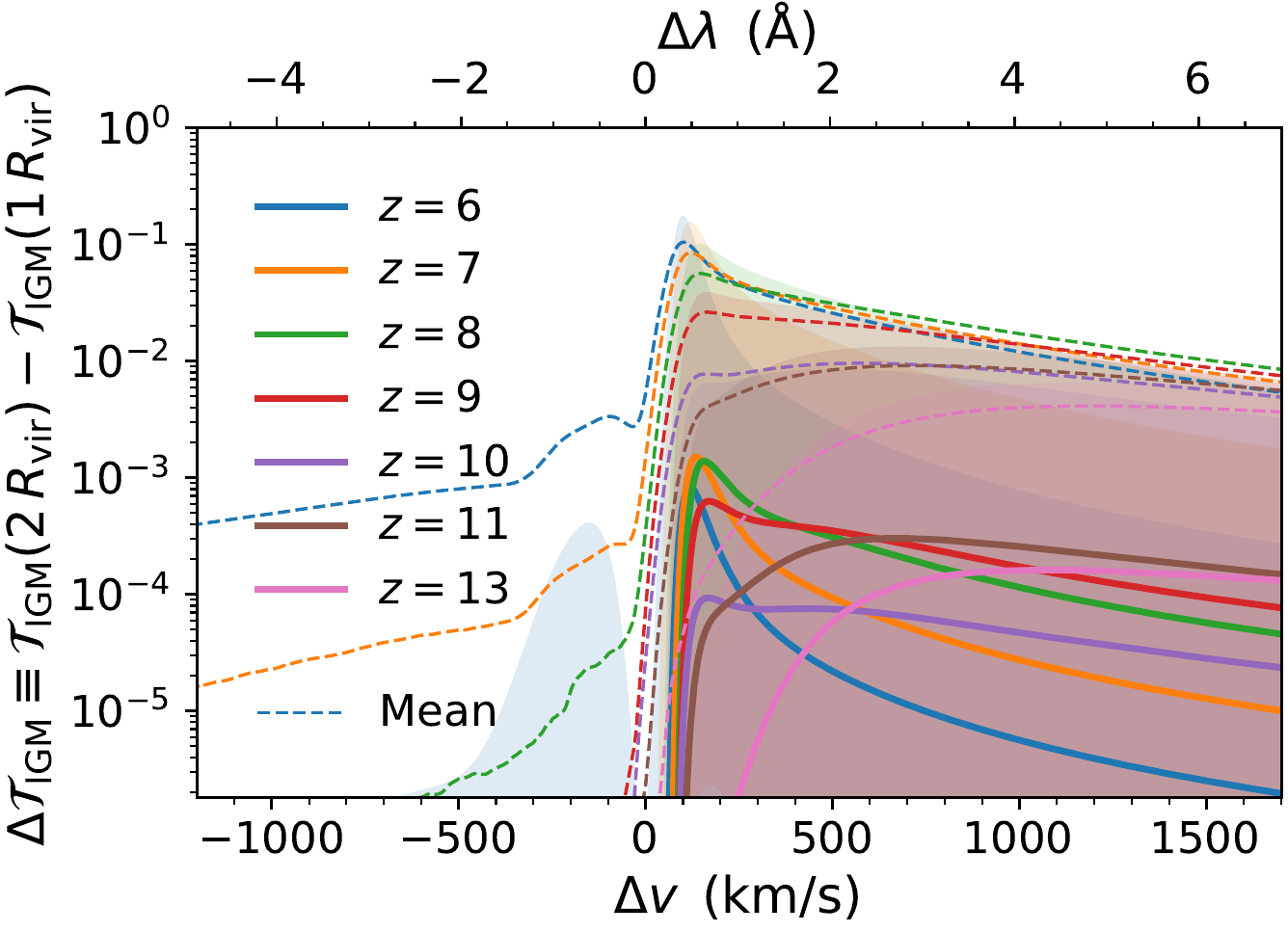}
  \caption{Difference in IGM transmission $\Delta \mathcal{T}_\text{IGM}$ between $1\,R_\text{vir}$ and $2\,R_\text{vir}$ as a function of velocity offset $\Delta v$ and rest-frame wavelength offset $\Delta \lambda$ around the Ly$\alpha$ line at each redshift. The solid (dashed) curves show the catalogue median (mean) statistics and shaded regions give the $1\sigma$ confidence levels. Overall, the choice of starting integration radius mainly affects results on the red side of line centre as blue photons are already highly suppressed. The mean values ($\Delta \mathcal{T}_\text{IGM} \sim 1\%$) are higher than the median ones ($\Delta \mathcal{T}_\text{IGM} \sim 0.01\%$) due to outliers that experience significant local absorption.}
  \label{fig:T_diff}
\end{figure}

As a start of integration we use a fixed value of $1\,R_\text{vir}$ (i.e. $R_{200}$ of the entire group). Beyond this we assume that Ly$\alpha$ scatterings back into the aperture line-of-sight are negligible. While this choice is both physically motivated and informed by previous studies \citep[e.g.][]{Laursen2011,Byrohl2020}, it is important to evaluate the impact for the present simulation and redshift range. Therefore, in Fig.~\ref{fig:T_diff} we show the difference in IGM transmission $\Delta \mathcal{T}_\text{IGM}$ between $2\,R_\text{vir}$ and $1\,R_\text{vir}$, which demonstrates that changes in the median (solid) and mean (dashed) curves are quite small. The peaks of the differences are around $\Delta v \approx 100\,\text{km\,s}^{-1}$ with plateaus on the red side of line centre, while the blue side is essentially unchanged due to already being highly suppressed. Interestingly, the mean variations ($\Delta \mathcal{T}_\text{IGM} \sim 1\%$) are higher than the median ones ($\Delta \mathcal{T}_\text{IGM} \sim 0.01\%$) due to outlier sightlines with significant localized absorption. In the extreme case of maximally bimodal distributions the mean variation can be viewed as the fraction of sightlines that are impacted by the starting radius value, especially in the context of describing IGM transmission with a covering fraction model that increases sightline variability as reionization progresses. We expect such variations to be mirrored in end-to-end radiative transfer calculations, so our main conclusions are largely unaffected by our choice of initial integration radius.

\bsp 
\label{lastpage}
\end{document}